\documentclass[12pt]{article}
\usepackage{xspace}
\usepackage[table,dvipsnames]{xcolor}
\usepackage{amsmath}
\usepackage{amssymb}
\usepackage{graphicx}
\usepackage{hyperref}
\usepackage{underscore}
\usepackage{ulem}
\usepackage{mathrsfs}
\newcommand{\Agama}{\textsc{Agama}\xspace}
\newcommand{\Amuse}{\textsc{Amuse}\xspace}
\newcommand{\Galpy}{\textsc{Galpy}\xspace}
\newcommand{\Nemo} {\textsc{Nemo}\xspace}
\newcommand{\Gsl}  {\textsc{Gsl}\xspace}
\newcommand{\Eigen}{\textsc{Eigen}\xspace}
\newcommand{\Nbody}{\textsl{N}-body\xspace}
\newcommand{\Cpp}  {\texttt{C++}\xspace}
\newcommand{\CppII}{\texttt{C++11}\xspace}
\newcommand{\Python}{\texttt{Python}\xspace}
\newcommand{\Fortran}{\texttt{Fortran}\xspace}
\definecolor{darkviolet}{rgb}{0.3,0.0,0.5}
\definecolor{darkolive} {rgb}{0.3,0.5,0.0}
\definecolor{darkcyan}  {rgb}{0.0,0.5,0.6}
\definecolor{darkblue}  {rgb}{0.0,0.0,0.8}
\newcommand{\ttt}[1]{\textcolor{darkviolet}{\texttt{#1}}}
\newcommand{\ppp}[1]{\textcolor{darkolive} {\texttt{#1}}}
\renewcommand{\d}{\mathrm{d}}
\newcommand{\D}{\partial}
\newcommand{\bA}{\boldsymbol{A}}
\newcommand{\bv}{\boldsymbol{v}}
\newcommand{\bx}{\boldsymbol{x}}
\newcommand{\by}{\boldsymbol{y}}
\newcommand{\bz}{\boldsymbol{z}}
\newcommand{\bX}{\boldsymbol{X}}
\newcommand{\bY}{\boldsymbol{Y}}
\newcommand{\bZ}{\boldsymbol{Z}}
\newcommand{\bJ}{\boldsymbol{J}}
\newcommand{\bt}{\boldsymbol{\theta}}
\newcommand{\scE}{\mathscr E}
\newcommand{\Beta}{\mathrm B}
\newcommand{\dvpar}{\langle \Delta v_\| \rangle}
\newcommand{\dvsqpar}{\langle \Delta v^2_\| \rangle}
\newcommand{\dvsqper}{\langle \Delta v^2_\bot \rangle}
\DeclareMathOperator{\trig}{trig}
\hypersetup{
    colorlinks,
    linkcolor={darkcyan},
    citecolor={darkblue},
    urlcolor ={darkblue}
}

\begin{document}
\title{\vspace*{-10mm}
\includegraphics[width=8cm]{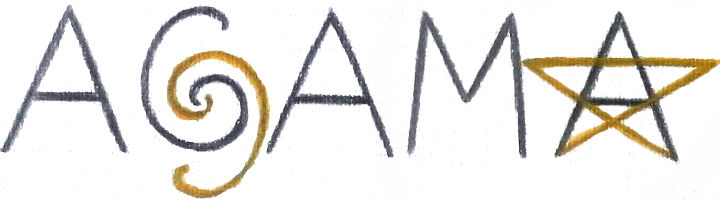}\protect\\[5mm]\Agama reference}
\author{Eugene Vasiliev\\
%\normalsize\textit{Lebedev Physical Institute, Moscow, Russia}\\
\normalsize\textit{University of Oxford \& Institute of Astronomy, Cambridge}\\
\normalsize\textrm{email: eugvas@lpi.ru} }

\maketitle
\tableofcontents
\newpage

%%%%%%%%%%%%%%%%%%
\section{Overview}

\Agama (Action-based Galaxy Modelling Architecture) is a software library intended for a broad range of tasks within the field of stellar dynamics. As the name suggests, it is centered around the use of action/angle formalism to describe the structure of stellar systems, but this is only one of its many facets. The library contains a powerful framework for dealing with arbitrary density/potential profiles and distribution functions (analytic, extracted from \Nbody models, or fitted to the data), a vast collection of general-purpose mathematical routines, and covers many aspects of galaxy dynamics up to the very high-level interface for constructing self-consistent galaxy models. It provides tools for analyzing \Nbody simulations, serves as a base for the Monte Carlo stellar-dynamical code \textsc{Raga} \cite{Vasiliev2015}, the Fokker--Planck code \textsc{PhaseFlow} \cite{Vasiliev2017}, and the Schwarzschild modelling code \textsc{Forstand} \cite{VasilievValluri2020} (in turn, derived from the earlier code \textsc{Smile} \cite{Vasiliev2013,VasilievAthanassoula2015}).

The core of the library is written in \Cpp and is organized into several modules, which are considered in turn in Section~\ref{sec:Structure}:
\begin{itemize}  \setlength{\parskip}{2pt} \setlength{\itemsep}{2pt}
\item Low-level interfaces and generic routines, which are not particularly tied to stellar dynamics: various mathematical tasks, coordinate systems, unit conversion, input/output of particle collections and configuration data, and other utilities.
\item Gravitational potential and density interface: the hierarchy of classes representing density and potential models, including two very general and powerful approximations of any user-defined profile, and associated utility functions.
\item Routines for numerical computation of orbits and their classification.
\item Action/angle interface: classes and routines for conversion between position/velocity and action/angle variables.
\item Distribution functions expressed in terms of actions.
\item Galaxy modelling framework: computation of moments of distribution functions, interface for creating gravitationally self-consistent multicomponent galaxy models, construction of \Nbody models and mock data catalogues.
\item Data handling interface, selection functions, etc.
\end{itemize}

A large part of this functionality is available in \Python through the eponymous extension module. Many high-level tasks are more conveniently expressed in \Python, e.g., finding best-fit parameters of potential and distribution function describing a set of data points, or constructing self-consistent models with arbitrary combination of components and constraints.
A more restricted subset of functionality is provided as plugins to several other stellar-dynamical software packages (Section~\ref{sec:Interfaces}).

The library comes with an extensive collection of test, demonstration programs and ready-to-use tools; some of them are internal tests that check the correctness of various code sections, others are example programs illustrating various applications and usage aspects of the library, and several programs that actually perform some useful tasks are also included in the distribution. There are both \Cpp and \Python programs, sometimes covering exactly the same topic; a brief review is provided in Section~\ref{sec:ExamplesTests}.

The main part of this document presents a comprehensive overview of various features of the library and a user's guide. The appendix contains a developer's guide and most technical aspects and mathematical details. The science paper describing the code is \cite{Vasiliev2019}.

The code can be downloaded from \url{http://agama.software}.

%%%%%%%%%%%%%%%%%%%%%%%%%%%%%%%%%%%%%%%%%%%%%%%%%%%%%%%%%%%%%%%%%%%%%
\section{Structure of the \Agama \Cpp library}  \label{sec:Structure}

%%%%%%%%%%%%%%%%%%%%%%%%%%%%%%%%%%
\subsection{Low-level foundations}

%%%%%%%%%%%%%%
\subsubsection{Math routines}  \label{sec:Math}
\Agama contains an extensive mathematical subsystem covering many basic and advanced tasks. Some of the methods are implemented in external libraries (\Gsl, \Eigen) and have wrappers in \Agama that isolate the details of implementation, so that the back-end may be switched without any changes in the higher-level code; other parts of this subsystem are self-contained developments.
All classes and routines in this section belong to the \ttt{math::} namespace.

\paragraph{Fundamental objects}
throughout the entire library are functions of one or many variables, vectors and matrices.
Any class derived from the \ttt{IFunction} interface should provide a method for computing the value and up to two derivatives of a function of one variable $f(x)$; \ttt{IFunctionNdim} represents the interface for a vector of functions of many variables $\boldsymbol{f}(\bx)$, and \ttt{IFunctionNdimDeriv} additionally provides the Jacobian of this function (the matrix $\D f_i /\D x_k$). Many mathematical routines operate on instances of classes derived from one of these interfaces.

For one-dimensional vectors we use \ttt{std::vector} when a dynamically-sized array is needed; some routines take input arguments of type \ttt{const double[]} or store the output in \ttt{double[]} variables which may be also statically-sized arrays (for instance, allocated on the stack, which is more efficient in tight loops).

For two-dimensional matrices there is a dedicated \ttt{math::Matrix} class, which provides a simple fixed interface to an implementation-dependent structure (either the \Eigen matrix type, or a custom-coded flattened array with 2d indexing, if \Eigen is not available).
Matrices may be dense and sparse; the former provide full read-write access, while the latter are constructed from the list of non-zero elements and provide read-only access. Sparse matrices are implemented in \Eigen or, in its absense, in \Gsl starting from version 2.0; for older versions we substitute them internally with dense matrices (which, of course, defeats the purpose of having a separate sparse matrix interface, but at least allows the code to compile without any modifications).

\paragraph{Numerical linear algebra}  routines in \Agama are wrappers for either \Eigen (considerably more efficient) or \Gsl library.
There are a few standard BLAS functions (matrix-vector and matrix-matrix multiplication for both dense and sparse matrices) and several matrix decomposition classes (\ttt{LUDecomp}, \ttt{CholeskyDecomp}, \ttt{SVDecomp}) that can be used to solve systems of linear equations $\mathsf{A}\bx=\boldsymbol{b}$.

\textsl{LU} decomposition of a non-degenerate square matrix $\mathsf{A}$ (dense or sparse) into a product of lower and upper triangular matrices is the standard tool for solving full-rank systems of linear equations. Once a decomposition is created, it may be used several times with different r.h.s.\ vectors $\boldsymbol{b}$.

Cholesky decomposition of a symmetric positive-definite dense matrix $\mathsf{A} = \mathsf{L}\mathsf{L}^T$ serves the same purpose in this more specialized case (being twice more efficient). It is informally known as ``taking the square root of a matrix'': for instance, a quadratic form $\bx^T \mathsf{A} \bx$ may be written as $|\mathsf{L}^T\bx|^2$ -- this is used in the context of dealing with correlated random variables, where $\mathsf{A}$ would represent the correlation matrix.

Singular-value decomposition (SVD) represents a generic $M\times N$ matrix ($M$ rows, $N$ columns; here $M\ge N$) as $\mathsf{A}=\mathsf{U}\,\mathrm{diag}(\boldsymbol{S})\,\mathsf{V}^T$, where $\mathsf{U}$ is a $M\times N$ orthogonal matrix (i.e., $\mathsf{U}\mathsf{U}^T=\mathsf{I}$), $\mathsf{V}$ is a $N\times N$ orthogonal matrix, and the vector $\boldsymbol{S}$ contains singular values, sorted in descending order. In the case of a symmetric positive definite matrix $\mathsf{A}$, SVD is identical to the eigenvalue decomposition, and $\mathsf{U}=\mathsf{V}$.
SVD is considerably more costly than the other two decompositions, but it is a more powerful tool that may be applied for solving over-determined and/or rank-deficient linear systems while maintaining numerical stability.
If $M>N$, there are more equations than variables, and the solution is obtained in the least-square sense; if the nullspace of the system is non-trivial (i.e., $\mathsf{A}\bx=\boldsymbol{0}$ for a non-zero $\bx$), the solution with the lowest possible norm is returned.
%The ratio of maximal to minimal singular values is called the condition number of the matrix $\mathsf{A}$, and the larger it is, the more degenerate is the matrix.

\paragraph{Root-finding}  is handled differently in one or many dimensions.
\ttt{findRoot} searches for a root of a continuous one-dimensional function $f(x)$ on an interval $[a..b]$, which may be finite or infinite, provided that $f(a)\,f(b) \le 0$ (i.e., the interval encloses the root). It uses a combination of Brent's method with an optional Hermite interpolation in the case that the function provides derivatives.
\ttt{findRootNdim} searches for zeros of an $N$-dimensional function of $N$ variables, which must provide the Jacobian, using a hybrid Newton-type method.

\paragraph{Integration}  of one-dimensional functions can be performed in several ways. \ttt{integrateGL} uses fixed-order Gauss--Legendre quadrature without error estimate. \ttt{integrate} uses variable-order Gauss--Kronrod scheme with the order of quadrature doubled each time until it attains the required accuracy or reaches the maximum; it is a good balance between fixed-order and fully adaptive methods, and is very accurate for smooth analytic functions. \ttt{integrateAdaptive} handles more sophisticated integrands, possibly with singularities, using a fully adaptive recursive scheme to reach the required accuracy, but is also more expensive.

Multidimensional integration over an $N$-dimensional hypercube is performed by the \ttt{integrateNdim} routine, which serves as a unified interface to either \textsc{Cubature} or \textsc{Cuba} library \cite{Cuba}; the former is actually included into the \Agama codebase. Both methods are fully adaptive and have similar performance (either one is better on certain classes of functions). The input function may provide $M\ge 1$ values, i.e., several functions may be integrated simultaneously over the same domain.

\paragraph{Sampling} \label{sec:Sampling}  from a probability distribution (\ttt{sampleNdim}) serves the following task: given a $N$-dimensional function $f(\bx)\ge 0$ over a hypercube domain, construct an array of $M$ random sample points $\bx_k$ such that the density of samples in the neighborhood of any point is proportional to the value of $f$ at that point. Obviously, the function $f$ must have a finite integral over the entire domain, and in fact the integral may be estimated from these samples (however it is not as accurate as the deterministic cubature routines, which are allowed to attribute different weights to each sampled point). This routine uses a multidimensional variant of rejection algorithm with adaptive subdivision of the entire domain into smaller regions, and performing the rejection sampling in each region (a more detailed description is given in Section~\ref{sec:MathSamplingDetails}).

\paragraph{Optimization methods}
A broad range of tasks may be loosely named ``optimization problems'', i.e., finding a minimum of a certain function (objective) of one or many variables under certain constraints.

For a function of one variable, there is a straightforward minimization routine \ttt{findMin} that can operate on any finite or (semi-)infinite interval $[a..b]$, and finds $\min f(x)$ on this interval (including endpoints); if there are multiple minima, then one of them will be found (not necessarily the global one), depending on the initial guess. The starting point $x_0$ such that  $f(x_0)<f(a), f(x_0)<f(b)$ may be optionally be provided by the caller; in its absense the routine will try to come up with a guess itself. Only the function values are needed by the algorithm.

For a function of $N$ variables $\bx$, there are several possibilities. If only the values of the function $f(\bx)$ are available, then the Nelder--Mead (simplex, or amoeba) algorithm provided by the routine \ttt{findMinNdim} may be used. 
If the partial derivatives $\D f/\D \bx$ are available, they may be used in a more efficient quasi-Newton BFGS algorithm provided by the routine \ttt{findMinNdimDeriv}.

A special case of optimization problem is a non-linear least-square fit: given a function $g(\bx; \boldsymbol{d})$, where $x_i$ are $N$ parameters that are being optimized, and $d_k$ are $M$ data points, minimize the sum of squared differences between the values of $g$ at these points and target values $v_k$: $\min f(\bx) = \sum_{k=1}^M [g(\bx; d_k) - v_k]^2$. This task is solved by the Levenberg--Marquardt algorithm, which needs the Jacobian matrix of partial derivatives of $g$ w.r.t.\ its parameters $\boldsymbol x$ at each data point $d_k$. It is provided by the routine \ttt{nonlinearMultiFit}.
Of course, if the function $g$ is linear w.r.t.\ its parameters, this reduces to a simpler linear algebra problem, solved by the routine \ttt{linearMultiFit}. And if there is only one or two parameters (i.e., a linear regression with or without a constant term), this is solved by the routines \ttt{linearFit} and \ttt{linearFitZero}.

In the above sequence, more specialized problems require more knowledge about the function, but generally converge faster, although all of them may be recast in terms of a general (unconstrained) minimization problem, as demonstrated in \texttt{test_math_core.cpp}.
All of them (except the linear regression routines) need a starting point or a $N$-dimensional neighborhood, but may move away from it in the direction of (one of possible) minima; again there is no guarantee to find the global minimum.

If there are restrictions on the values of $\bx$ in the form of a matrix $\mathsf{A}$ of element-wise linear inequality constraints $\mathsf{A}\bx \preccurlyeq \boldsymbol{b}$, and if the objective function $f$ is linear or quadratic in the input variables, these cases are handled by the routines \ttt{linearOptimizationSolve} and \ttt{quadraticOptimizationSolve}. They depend on external libraries (GLPK and/or CVXOPT; the former can only handle linear optimization problems). 

\paragraph{Interpolation} \label{sec:SplineInterpolation}
There are various classes for performing interpolation in one, two or three dimensions.
All methods are based on the concept of piecewise-polynomial functions defined by the nodes of a grid $x_0<x_1<\dots<x_{N_x-1}$; in the case of multidimensional interpolation the grid is rectangular, i.e., aligned with the coordinate lines in each dimension. The advantages of this approach are locality (the function value depends only on the adjacent grid points), adaptivity (grid nodes need not be uniformly spaced and may be concentrated in the region of interest) and efficiency (the cost of evaluation scales as $\log(N_x)$ -- time needed to locate the grid segment containing the point $x$, plus a constant additional cost to evaluate the interpolating polynomial on this segment).

There are linear, cubic and quintic (fifth-order) interpolation schemes in one, two and three dimensions (quintic -- only in 1d and 2d). The former two are defined by the values of the interpolant at grid nodes, and the last one additionally requires its (partial) derivatives w.r.t.\ each coordinate at grid nodes. All these classes compute the function value and up to two derivatives at any point inside the grid; 1d functions are linearly extrapolated outside the grid.

An alternative formulation of the piecewise-polynomial interpolation methods is in terms of B-splines -- $N_x+N-1$ basis functions defined by the grid nodes, which are polynomials of degree $N$ on each of at most $N+1$ consecutive segments of the grid, and are zero otherwise. The case $N=1$ corresponds to linear interpolation, $N=3$ -- to (clamped) cubic splines%
\footnote{A general cubic spline in 1d is defined by $N_x+2$ parameters: they may be taken to be the values of spline at $N_x$ grid nodes plus two endpoint derivatives, which is called a clamped spline. The more familiar case of a natural cubic spline instead has these two additional parameters defined implicitly, by requiring that the second derivative of the spline is zero at both ends.}.
The interpolating function is defined as $f(\bx) = \sum_\alpha\,A_\alpha\,B_\alpha(\bx)$, where $\alpha$ is a combined index in all dimensions, $A_\alpha$ are the amplitudes and $B_\alpha$ are the basis functions (in more than one dimension, they are formed as tensor products of 1d B-splines, i.e., $B_{ij}(x,y) = B_i(x)\,B_j(y)$). Again, the evaluation of interpolant only requires $O(\log(N_x)+N^2)$ operations per dimension to locate the grid segment and compute all $N$ possibly nonzero basis functions using a $N$-step recursion relation. This formulation is more suitable for constructing approximating splines from a large number of scattered points (see next paragraph), and the resulting B-splines may be subsequently converted to more efficient linear or cubic interpolators. This approach is currently implemented in 1 and 3 dimensions.

B-splines can also be used as basis functions in finite-element methods: any sufficiently smooth function can be approximated by a linear combination of B-splines on the given interval, and hence represented as a vector of expansion coefficients. Various mathematical operations on the original functions (sum, product, convolution) can then be translated into linear algebra operations on these vectors. The 1d finite-element approach is used in the Fokker--Planck code \textsc{PhaseFlow}, which is included in the library, and in a few other auxiliary tasks (e.g., solution of Jeans equations).

Spline interpolation is heavily used throughout the entire \Agama library as an efficient and accurate method for approximating various quantities that are expensive to evaluate directly. By performing suitable additional scaling transformations on the argument and/or value of the interpolator, it is possible to achieve exquisite accuracy (sometimes down to machine precision) with a moderate ($\mathcal O(10^2)$) number of nodes covering the region of interest; for one-dimensional splines a linear extrapolation beyond that region often remains quite accurate under a carefully chosen scaling (usually logarithmic). Quintic splines are employed when it is possible to compute analytically the derivatives (or partial derivatives in the 2d case) of the approximated function at grid nodes during the spline construction in addition to its values -- in this case the accuracy of approximation becomes $1-2$ orders of magnitude better than that of a cubic spline. (Of course, computing the derivatives by finite-differencing or from a cubic spline does not achieve the goal).
Mathematical foundations of splines are described in more detail in the Appendix (sections \ref{sec:MathBSplineDetails} and \ref{sec:MathSplineDetails}).

\paragraph{Penalized spline fitting}  \label{sec:SplineFitting}
There are two kinds of tasks that involve the construction of a spline curve from an irregular set of points (as opposed to the values of the curve at grid nodes, as in the previous section).

The first task is to create a smooth least-square approximation $f(x)$ to a set of points $\{x_i, y_i\}$: 
minimize $\sum_i [y_i-f(x_i)]^2 + \lambda \int [f''(x)]^2\,\d x$, where $\lambda$ is the smoothing parameter controlling the tradeoff between approximation error (the first term) and the curvature penalty (the second term). The solution is given by a cubic spline with grid nodes placed at all input points $\{x_i\}$ \cite{GreenSilverman}; however, it is not practical in the case of a large number of points. Instead, we approximate it with a cubic spline having a much smaller number of grid nodes $\{X_k\}$ specified by the user. The class \ttt{SplineApprox} is constructed for the given grid  $\{X_k\}$ and $x$-coordinates of input points; after preparing the ground, it may be used to find the amplitudes of B-splines for any $\{y_i\}$ and $\lambda$, and there is a method for automatically choosing the suitable amount of smoothing.

The second task is to determine a density function $P(x)$ from an array of samples $\{x_i\}$, possibly with individual weights $\{w_i\}$. It is also solved with the help of B-splines, this time for $\ln P(x)$, which is represented as a B-spline of degree $N$ defined by user-specified grid nodes $\{X_k\}$. The routine \ttt{splineLogDensity} constructs an approximation for $\ln P$ for the given grid nodes and samples, with adjustable smoothing parameter $\lambda$.

Both tasks are presently implemented only for the 1d case, but in the future may be generalized to multidimensional data represented by tensor-product B-splines. More details on the mathematical formulation are given in the Appendix (sections \ref{sec:MathSplineApproxDetails} and \ref{sec:MathSplineDensityDetails}).

%%%%%%%%%%%%%%
\subsubsection{Units}  \label{sec:Units}

Handling of units is a surprisingly difficult and error-prone task. \Agama adopts a somewhat clumsy but consistent approach to unit handling, which mandates a clear separation between internal units inside the library and external units used to import/export the data. This alone is a rather natural idea; what makes it peculiar is that we do not fix our internal units to any particular values. There are three independent physical base units -- mass, length, and time, or velocity instead of time. The only convention used throughout the library is that $G=1$, which is customary for any stellar-dynamical code. This leaves only two independent base units, and we mandate that the results of all calculations should be independent of the choice of base units (up to insignificant roundoff errors at the level $\sim 10^{-4}\div 10^{-6}$ -- typical values for root-finder or integration tolerance parameters). This places heavier demand on the implementation -- in particular, all dimensional quantities should generally be converted to logarithms before being used in a scale-free context such as finding a root on the interval $[0..\infty)$. But the reward is greater robustness in various applications.

In practice, the \ttt{units::} namespace defines \textit{two} separate unit classes. The first is \ttt{InternalUnits}, defining the two independent physical scales (taken to be length and time) used as the internal units of the library. Typically, a single instance of this class (let's call it \texttt{intUnit}) is created for the entire program. It does not provide any methods -- only conversion constants such as \texttt{from_xxx} and \texttt{to_xxx}, where xxx stands for some physical quantity. For instance, to obtain the value of potential expressed in (km/s)${}^2$ at the galactocentric radius of 8~kpc, one needs to write something like \\
\texttt{double E = myPotential.value(coord::PosCyl( 8 * intUnit.from_Kpc, 0, 0 ));}\\
\texttt{std::cout << E * pow_2(intUnit.to_kms);}

The second is \ttt{ExternalUnits}, which is used to convert physical quantities between the external datasets and internal variables. External units, of course, do not need to follow the convention $G=1$, thus they are defined by three fundamental physical scales (length, velocity and mass) plus an instance of \ttt{InternalUnits} class that describes the working units of the library. An instance of unit converter is supplied as an argument to all functions that interface with external data: read/write potential and distribution function parameters, \Nbody snapshots, and any other kinds of data. Thus the dimensional quantities ingested by the library are always in internal units, and are converted back to physical units on output.

When the external data follows the convention $G=1$ in whatever units, no conversion is necessary, thus one may provide an \ttt{ExternalUnits} object with a default constructor wherever required (it is usually a default value for this argument); in this case also no \ttt{InternalUnits} need to be defined. The reason for existence of two classes is that neither of them can fulfill both roles: to serve as an arbitrary internal ruler for testing the scale-invariance of calculations, and to have three independent fundamental physical scales (possibly different for various external data sources). In practice, one may create a single global instance of \ttt{ExternalUnits} with a temporary instance of arbitrary \ttt{InternalUnits} as an argument; however, having a separate global instance of the latter class is handy because its conversion constants indicate the direction (to or from physical units).

The \Python interface supports the unit conversion internally: the user may set up a global instance of \ttt{ExternalUnits}, and all dimensional quantities passed to the library will be converted to internal library units and then back to physical units on output. Or, if no such conversion has been set up, all data is assumed to follow the convention $G=1$. In the future, we may adopt an alternative unit handling approach that would be seamlessly integrated with the units subsystem of the \textsc{Astropy} library \cite{Astropy}.

%%%%%%%%%%%%%%
\subsubsection{Coordinates}  \label{sec:Coords}
The \ttt{coords::} namespace contains classes and routines for representing various mathematical objects in several coordinate systems in three-dimensional space.

There are several built-in coordinate systems: \ttt{Car}tesian, \ttt{Cyl}indrical, \ttt{Sph}erical, and \ttt{ProlSph} -- prolate spheroidal. Their names are used as tags in other templated classes and conversion routines; only the last one has an adjustable parameter (focal distance).

Templated classes include position, velocity, a combination of the two, an abstract interface \ttt{IScalarFunction} for a scalar function evaluated in a particular coordinate system, gradient and hessian of a scalar function, and coefficients for coordinate transformations from one system to the other. Templated functions convert these objects from one coordinate system to the other: for instance, \ttt{toPosVelCyl} converts the position and velocity from any source coordinate system into cylindrical coordinates; these routines should be called explicitly, to make the code self-documenting. An even more powerful family of functions \ttt{evalAndConvert} take the position in one (output) coordinate system and a scalar function defined in the other (evaluation) system, calls the function with transformed coordinates, and perform the transformation of gradient and hessian back to the output system. The primary use of these routines is in the potential framework (Section~\ref{sec:Potential}) -- each potential defines a method for computing it in the optimal system, and uses the conversion routines to provide the remaining ones. Another use is for transformation of probability distributions, which involve Jacobian matrices of coordinate conversions. In the future, we may add other coordinate systems (e.g., heliocentric) into the same framework.

%%%%%%%%%%%%%%
\subsubsection{Particles}  \label{sec:Particles}
A particle is an object with phase-space coordinates and mass; the latter is just a single number, and the former may be either just the position or the position and velocity in any coordinate system. Particles are grouped in arrays (templated struct \ttt{ParticleArray<ParticleT>}).
Particle arrays in different coordinate systems can be implicitly converted to each other, to simplify the calling convention of routines that use one particular kind of coordinate system, but accept all other ones with the same syntax.

\Agama provides routines for storing and loading particle arrays in files (\ttt{readSnapshot} and \ttt{writeSnapshot}), with several file formats available, depending on compilation options. Text files are built-in, and support for \Nemo and \textsc{Gadget} binary formats is provided through the \textsc{Unsio} library (optional).

Particle arrays are also used in constructing a potential expansion (\ttt{Multipole} or \ttt{CylSpline}) from an \Nbody snapshot, and created by routines from the \texttt{galaxymodel} module (Section~\ref{sec:GalaxyModel}), e.g., by sampling from a distribution function.

The particle array type and input/output routines belong to the \ttt{particles::} name\-space.

%%%%%%%%%%%%%%
\subsubsection{Utilities}  \label{sec:Utilities}

There are quite a few general-purpose utility functions that do not belong to any other module, and are grouped in the \ttt{utils::} namespace. 
Apart from several routines for string manipulation (e.g., converting between numbers and strings), and logging, there is a self-sufficient mechanism for dealing with configuration files. These files have a standard INI format, i.e., each line contains \ppp{name=value}, and parameters belonging to the same subject domain may be grouped in sections, with a preceding line \ppp{[section name]}. Values may be strings or numbers, names are case-insensitive, and lines starting with a comment symbol \texttt{\#} or \texttt{;} are ignored.

The class \ttt{KeyValueMap} is responsible for a list of values belonging to a single section; this list may be read from an INI file, or created by parsing a single string like \ppp{"param1=value1 param2=1.0"}, or from an array of command-line arguments. Various methods return the values converted to a particular type (number, string or boolean) or set/replace values.
The class \ttt{ConfigFile} operates with a collection of sections, each represented by its own \ttt{KeyValueMap}; it can read and write INI files.

%%%%%%%%%%%%%%%%%%%%%%%%%%%%%%%%%%%%%%%%%%%%%%
\subsection{Potentials}  \label{sec:Potential}

\Agama provides a versatile collection of density and potential models, including two very general and efficient approximations that can represent almost any well-behaved profile of an isolated stellar system. All classes and routines in this section are located in the \ttt{potential::} namespace.

All density models are derived from the \ttt{BaseDensity} class, which defines methods for computing the density in three standard coordinate systems (derived classes choose the most convenient one to implement directly, and the two other ones use coordinate transformations), a function returning the symmetry properties of the model, and two convenience methods for computing mass within a given radius and the total mass (by default they integrate the density over volume, but derived classes may provide a cheaper alternative).

All potential models are derived from the \ttt{BasePotential} class, which itself descends from \ttt{BaseDensity}. It defines methods for computing the potential, its first derivative (gradient vector) and second derivative (hessian tensor) in three standard coordinate systems. By default, density is computed from the hessian, but derived classes may override this behaviour.
Furthermore there are several derived abstract classes serving as bases for potentials that are easier to evaluate in a particular coordinate system (Section~\ref{sec:Coords}): the function \ttt{eval()} for this system remains to be implemented in descendant classes, and the other two functions use coordinate and derivative transformations to convert the computed value to the target coordinate system.
For instance, a triaxial harmonic potential is easier to evaluate in Cartesian coordinates, while the St\"ackel potential is naturally expressed in a prolate spheroidal coordinate system.

Any number of density components may be combined into a single \ttt{CompositeDensity} class, and similarly for potential components.

%%%%%%%%%%%%%%
\subsubsection{Analytic potentials}  \label{sec:PotentialAnalytic}

There are several commonly used models with known expressions for the potential and its derivatives. 

Spherical models include the \ttt{Plummer}, \ttt{Isochrone}, \ttt{NFW} (Navarro--Frenk--White) potentials, and a generalized \ttt{King} (lowered isothermal) model which is specified by its distribution function $f(E)$, as given by Equation~1 in \cite{GielesZocchi2015}.
Moreover there is a wrapper class that turns any user-provided function $\Phi(r)$ with two known derivatives into a form compatible with the potential interface. A point mass (Kepler) potential is obtained by constructing a \ttt{Plummer} potential with zero scale radius.

Axisymmetric models include the \ttt{MiyamotoNagai} and \ttt{OblatePerfectEllipsoid} potentials (the latter belongs to a more general class of St\"ackel potentials \cite{deZeeuw1985}, but is the only one implemented at present).
There is another type of axisymmetric models that have a dedicated potential class, namely a separable \ttt{Disk} profile with $\rho(R,z) = \Sigma(R)\, h(z)$. A direct evaluation of potential requires 2d numerical quadrature, or 1d in special cases such as the exponential radial profile, which is still too costly. Instead, we use the \textsc{GalPot} approach introduced in \cite{KuijkenDubinski1995, DehnenBinney1998}: the potential is split into two parts, \ttt{DiskAnsatz} that has an analytic expression for the potential of the strongly flattened component, and the residual part that is represented with the \ttt{Multipole} expansion.

Triaxial models include the \ttt{Logarithmic}, \ttt{Harmonic}, \ttt{Dehnen} \cite{Dehnen1993} and \ttt{Ferrers} potentials. The first two have infinite extent and are usable only in certain contexts (such as orbit integration), because most routines expect the potential to vanish at infinity.
Dehnen models may have any symmetry from spherical to triaxial; in non-spherical cases, the potential and its derivatives are computed using a 1d numerical quadrature \cite{MerrittFridman1996}, so this is rather costly (and also inaccurate at large distances). A preferred way of using an axisymmetric or triaxial Dehnen model is through the \ttt{Multipole} expansion. Ferrers ($n=2$) models are strictly triaxial, and have analytic expressions for the potential and its derivatives \cite{Pfenniger1984}.
There is also a \ttt{Spheroid} class that describes general triaxial two-power-law ($\alpha\beta\gamma$) density profiles \cite{Zhao1996} with an optional exponential cutoff. Dehnen, Plummer, Isochrone and NFW profiles are all special cases of this model; however, this class only provides the density profile and not the potential. \ttt{Sersic} represents another commonly used density model, which can also be triaxial. 
Generalized \ttt{King} models (with an adjustable strength of the outer cutoff, as in \cite{GielesZocchi2015}) provide both the density and the potential.

%%%%%%%%%%%%%%
\subsubsection{Multipole expansion}  \label{sec:PotentialMultipole}

\ttt{Multipole} is a general-purpose potential approximation that delivers highly accurate results for density profiles with axis ratio not very different from unity (say, at most a factor of few). It represents the potential as a sum of spherical-harmonic functions of angles multiplied by arbitrary functions of radius: $\Phi(r,\theta,\phi) = \sum_{l,m}\, \Phi_{l,m}(r)\, Y_l^m(\theta,\phi)$. The radial dependence of each term is given by a quintic spline, defined by a rather small number of grid nodes ($N_r\sim 20\div 50$), typically spaced equally in $\log r$ over a range $r_\mathrm{max}/r_\mathrm{min} \gtrsim 10^6$; the suitable order of angular expansion $l_\mathrm{max}$ depends on the shape of the density profile, and is usually $\lesssim 10$.

The potential approximation may be constructed in several ways:
\begin{itemize} \setlength{\parskip}{0pt} \setlength{\itemsep}{2pt}
\item from another potential (makes sense if the latter is expensive to compute, e.g., a triaxial \ttt{Dehnen} model);
\item from a smooth density profile, thereby solving the Poisson equation in spherical coordinates;
\item from an \Nbody model (an array of particle coordinates and masses) -- in this case a temporary smooth density model is created and used in the same way as in the second scenario;
\item by loading a previously computed array of coefficients from a text file.
\end{itemize}

This type of potential is rather inexpensive to initialize, very efficient to compute, provides an accurate extrapolation to small and large radii beyond the extent of its radial grid, and is the right choice for ``spheroidal'' density models -- from spherical to mildly triaxial, and even beyond (i.e., a model may have a twist in the direction of principal axes, or contain an off-centered odd-$m$ mode).

As a side note, a related class of potential approximations is based on expanding the radial dependence of spherical-harmonic terms $\Phi_{l,m}(r)$ into a sum over functions from a suitable basis set \cite{HernquistOstriker1992,Zhao1996}. For several reasons, this approach is less efficient: the choice of the family of basis functions implies certain biases in the approximation, and the need to compute a full set of them (involving rather expensive algebraic operations) at each radius is contrasted with a much faster evaluation of a spline (essentially using only a few adjacent grid points). \cite{Vasiliev2013} demonstrated the superiority of a previous implementation of spline-interpolated spherical-harmonic expansion over the basis-set approach, and \ttt{Multipole} is improved even further.

%%%%%%%%%%%%%%
\subsubsection{Azimuthal harmonic expansion}  \label{sec:PotentialCylSpline}

\ttt{CylSpline}\footnote{an improved version of the method presented in \cite{VasilievAthanassoula2015}} is another general-purpose potential approximation that is more effective for strongly flattened (disky) systems, whether axisymmetric or not. It represents the potential as a sum of Fourier terms in the azimuthal angle ($\phi$), with coefficients of each term interpolated via a 2d quintic spline spanning a finite region in the $R,z$ plane. The accuracy of approximation is determined by the number and extent of the grid nodes in $R$ and $z$ (also scaled logarithmically to achieve a high dynamic range) and the order $m_\mathrm{max}$ of angular expansion; in the axisymmetric case only one term is used, but generally it may represent any geometry, e.g., spiral arms and a triaxial bar.

This potential may also be constructed in the same four ways as \ttt{Multipole}, but the solution of Poisson equation is much more expensive in this case; still, for typical grid sizes of a few dozen in each direction, it takes between a few seconds and minutes on a single CPU core (and is almost ideally parallelized). After initialization, the computation of potential and forces is as efficient as \ttt{Multipole}. In many cases, it delivers comparable or better accuracy than the latter, but is not suitable for cuspy density profiles and for extended tails of density at large radii, since it may only represent it over a finite region (the potential and its first derivative is still quite accurately extrapolated outside the grid, but the density is identically zero there). Its main advantage is the ability to handle disky systems which are not suitable for a spherical-harmonic expansion%
\footnote{Potential of separable axisymmetric disk density profiles can be efficiently computed using a combination of \ttt{DiskAnsatz} and \ttt{Multipole} (the \textsc{GalPot} approach), but this applies only to this restricted class of systems, and is comparable to \ttt{CylSpline} in both speed and accuracy.}.

To summarize, both potential approximations have wide, partially overlapping range of applicability, are equally efficient in evaluation (but not construction), and deliver good accuracy (see Figures~\ref{fig:PotentialAccuracy1},~\ref{fig:PotentialAccuracy2} in the Appendix, with more technical details given in Section~\ref{sec:PotentialDetails}).
We note that application of these methods to represent the potential of a galaxy like the Milky Way is computationally more demanding than simple models based e.g.\ on a combination of Miyamoto--Nagai disks and spherically-symmetric two-power-law profiles, but only moderately (by a factor of 2--3), and allows much greater flexibility and realism (especially if non-axisymmetric features are required).

%%%%%%%%%%%%%%
\subsubsection{Potential factory}  \label{sec:PotentialFactory}

\begin{table}
\caption{Available density and potential models and their parameters}  \label{tab:PotentialParams}
\begin{tabular}{l m{5cm} >{\raggedright\arraybackslash}m{7cm}}
Name & Formula & Parameters \\
\hline
\multicolumn{3}{c}{Density-only models} \\[2mm]

\ttt{Disk} & $\rho = \Sigma_0\,\exp\big(-\big[\frac{R}{R_\mathrm{d}}\big]^{\frac1n} - \frac{R_\mathrm{cut}}{R}\big)$
$\times\left\{ \begin{array}{ll} \delta(z)\qquad\qquad\mbox{if} & h=0 \\[1mm]
\frac{1}{2h} \exp\big(-\big|\frac{z}{h}\big|\big) & h>0 \\[1mm]
\frac{1}{4|h|}\, \mathrm{sech}^2\big(\big|\frac{z}{2h}\big|\big) & h<0 \end{array} \right. $ &
\ppp{surfaceDensity}~($\Sigma_0$) or \ppp{mass}, \ppp{scaleRadius}~($R_\mathrm{d}$), \ppp{scaleHeight}~($h$), \ppp{innerCutoffRadius}~($R_\mathrm{cut}$), \ppp{sersicIndex}~($n$)\\

\ttt{Spheroid} & $\rho = \rho_0  \left(\frac{\tilde r}{a}\right)^{-\gamma} \Big[ 1 + \big(\frac{\tilde r}{a}\big)^\alpha \Big]^{\frac{\gamma-\beta}{\alpha}}$ $\times \exp\Big[ -\big(\frac{\tilde r}{r_\mathrm{cut}}\big)^\xi\Big] $ &
\ppp{densityNorm}~($\rho_0$) or \ppp{mass}, \ppp{alpha}~($\alpha$), \ppp{beta}~($\beta$), \ppp{gamma}~($\gamma$), \ppp{scaleRadius}~($a$), \ppp{axisRatioY}~($p$), \ppp{axisRatioZ}~($q$), \ppp{outerCutoffRadius}~($r_\mathrm{cut}$), \ppp{cutoffStrength}~($\xi$) \\[2mm]

\ttt{Sersic} & \mbox{deprojection of} \mbox{$\Sigma = \Sigma_0 \exp\big[-b_n\,(R/a)^{1/n}\big]$} & \ppp{surfaceDensity}~($\Sigma_0$) or \ppp{mass}, \ppp{scaleRadius}~($a$), \ppp{sersicIndex}~($n$), \ppp{axisRatioY}~($p$), \ppp{axisRatioZ}~($q$) \\[2mm]

\multicolumn{3}{c}{Density/potential models} \\[2mm]

\ttt{Plummer} & $\Phi = -\frac{M}{\sqrt{a^2+r^2}}$ & \ppp{mass}~($M$), \ppp{scaleRadius}~($a$) \\[2mm]

\ttt{Isochrone} & $\Phi = - \frac{M}{a + \sqrt{r^2 + a^2}}$ & \ppp{mass}~($M$), \ppp{scaleRadius}~($a$) \\[2mm]

\ttt{NFW} & $\Phi = -\frac{M}{r} \ln\left(1 + \frac{r}{a}\right)$ & \ppp{mass}~($M$ {\footnotesize is the mass enclosed in $\sim5.3a$, the total mass is $\infty$}), \ppp{scaleRadius}~($a$) \\[2mm]

\ttt{MiyamotoNagai} & $\Phi = -\frac{M}{\sqrt{R^2 + \left(a + \sqrt{z^2+b^2}\right)^2}}$ & \ppp{mass}~($M$), \ppp{scaleRadius}~($a$), \ppp{scaleRadius2} or \ppp{scaleHeight}~($b$) \\[2mm]

\ttt{PerfectEllipsoid}\!\! & $\rho = \frac{M}{\pi^2\,q\,a^3} \left[ 1 + \frac{R^2+(z/q)^2}{a^2} \right]^{-2}$ &  \ppp{mass}~($M$), \ppp{scaleRadius}~($a$), \ppp{axisRatioZ}~($q$) \\[2mm]

\ttt{Dehnen} & $\rho = \frac{M\,(3-\gamma)}{4\pi\,p\,q\,a^3} \left(\frac{\tilde r}a\right)^{-\gamma} \left(1+\frac{\tilde r}a\right)^{\gamma-4}$\!\! &  \ppp{mass}~($M$), \ppp{gamma}~($\gamma$), \ppp{axisRatioY}~($p$), \ppp{axisRatioZ}~($q$), \ppp{scaleRadius}~($a$) \\[2mm]

\ttt{Ferrers} & $\rho = \frac{105\,M}{32\pi\,p\,q\,a^3} \left[1 - \left(\frac{\tilde r}a\right)^2\right]^2$ & \ppp{mass}~($M$), \ppp{scaleRadius}~($a$), \ppp{axisRatioY}~($p$), \ppp{axisRatioZ}~($q$) \\[2mm]

\ttt{King} & specified by $f(E)$, see text & \ppp{mass}, \ppp{scaleRadius}~($r_\mathrm{c}$), \ppp{W0}~($W_0$), \ppp{trunc}~($g$) \\[2mm]

\ttt{Logarithmic} & $\Phi = \frac{1}{2} v_0^2\,\ln(r_\mathrm{core}^2 + \tilde r^2)$ & \ppp{v0}~($v_0$), \ppp{scaleRadius}~($r_\mathrm{core}$), \ppp{axisRatioY}~($p$), \ppp{axisRatioZ}~($q$) \\[2mm]

\ttt{Harmonic} & $\Phi = \frac{1}{2} \Omega^2\,\tilde r^2$ & \ppp{Omega}~($\Omega$), \ppp{axisRatioY}~($p$), \ppp{axisRatioZ}~($q$) \\[2mm]

\multicolumn{3}{l}{\footnotesize $R=\sqrt{x^2+y^2}$ is the cylindrical radius and $\tilde r=\sqrt{x^2+(y/p)^2+(z/q)^2}$ is the ellipsoidal radius}
\end{tabular}
\end{table}

\begin{table}
\caption{Symmetry types and their implications}  \label{tab:Symmetry}
\renewcommand{\arraystretch}{1.25}
\begin{tabular}{l m{9.2cm} m{3.5cm}}
Name & Invariant transformations & \mbox{Sph.-harm.~coefs} identically zero \\
\hline
\ttt{None} & --- & --- \\[2mm]
\ttt{Reflection} & \mbox{$\{x,y,z\} \to \{-x,-y,-z\}$} \mbox{(twofold discrete symmetry)} & odd $l$ \\
\ttt{Bisymmetric} & \mbox{same or $z \to -z$, also implies $\{x,y\} \to \{-x,-y\}$} \mbox{(fourfold discrete symmetry}, e.g., a two-arm spiral) & same + odd $m$ \\
\ttt{Triaxial} & \mbox{same or $x \to -x$ or $y \to -y$} \mbox{(eightfold discrete symmetry, e.g., a bar)} & same + negative $m$ \\
\ttt{Axisymmetric} & \mbox{same or rotation about $z$ axis by any angle} \mbox{(continuous symmetry in $\phi$)} & same + any $m \ne 0$ \\
\ttt{Spherical} & \mbox{same or rotation about origin by any angle} \mbox{(continuous symmetry in both $\theta$ and $\phi$)} & same + any $l \ne 0$
\end{tabular}
\end{table}

All density and potential classes may be constructed using a universal ``factory'' interface -- several routines \ttt{createDensity} and \ttt{createPotential} that return new instances of \ttt{PtrDensity} or \ttt{PtrPotential} according to the provided parameters.
The parameters can be supplied in several ways. One is an INI file with one or several components of the potential described in separate sections \ppp{[Potential]}, \ppp{[Potential2]}, \ppp{[Potential disk]}, etc. (all section names should start with ``Potential''). Another possibility is to provide a \ttt{KeyValueMap} object (Section~\ref{sec:Utilities}) corresponding to a single section from an INI file (it may be read from the file, or constructed manually, e.g., from named arguments in the \Python interface, or from command-line parameters for console programs, or from a single string like \ppp{"key1=value1 key2=value2"}). These parameters may describe the potential completely (e.g., if this is one of the known analytical models), or define the parameters of Multipole or CylSpline potential expansions to be constructed from the user-provided density or potential object, or from an array of particles -- in the latter case these objects are also passed to the factory routine. Finally, the coefficients of a potential or density expansion may be stored into a text file and subsequently used to load and construct a new object, using \ttt{writePotential}/\ttt{readPotential} routines.

Below follows the list of possible parameters of a single potential or density component for the factory routines (not all of them make sense for all models, but unknown or irrelevant parameters will simply be ignored); see Table~\ref{tab:PotentialParams} for complete information:
\begin{itemize}
\item \ppp{type} -- determines the type of potential used; should be the name of a class derived from \ttt{BasePotential} -- either an analytic potential listed in the first column of Table~\ref{tab:PotentialParams}, or an expansion (\ttt{Multipole} or \ttt{CylSpline}). It is usually required, except if the potential is loaded from a coefficients file -- in that case the name of the potential appears in the first line of this text file, so is determined automatically.
\item  \ppp{density} -- if \ppp{type} is a potential expansion, this parameter determines the density model to be used; should be the name of a class derived from \ttt{BaseDensity} (or, by consequence, the name of an analytic potential), except that it cannot be a model with unbound potential (Logarithmic or Harmonic) or another potential expansion.\\
\phantomsection\label{sec:PotentialGalpot}%
There is one exception to the rule that \ppp{type} must encode a potential class: it may also contain the names of the density profiles originally used in \textsc{GalPot} -- \ttt{Disk}, \ttt{Spheroid} or \ttt{Sersic}. All such components are collected first, and used to construct a \textit{single} instance of \ttt{Multipole} potential with default parameters, plus zero or more instances of \ttt{DiskAnsatz} potentials (according to the number of disk profiles). The source density for this Multipole potential contains all Spheroid, S\'ersic and Disk components, plus \textit{negative} contributions of DiskAnsatz potentials (i.e., with inverted sign of their masses). Of course, one may use them also as regular \ppp{density} components (e.g., \ppp{type=CylSpline density=Disk}, which yields comparable accuracy), but in that case each one would create a separate potential expansion, which is of course not efficient. In order to lift this limitation, one may construct all density components individually, manually combine them into a single \ttt{CompositeDensity} model, and pass it to the constructor of a potential expansion (this approach is used for self-consistent multicomponent models, Section~\ref{sec:SCM}).
\item \ppp{symmetry} -- defines the symmetry properties of the density model passed to the potential expansion. All built-in models report this property automatically; this parameter is useful if the input is given by an array of particles, or by a user-defined routine returning the density or potential in \Python and \Fortran interfaces. It could be either a text string with one of the standard choices from Table~\ref{tab:Symmetry} (only the first letter is used), or a number encoding a more complicated symmetry (see the definitions in \texttt{coord.h}).
\item \ppp{file} -- the name of a file with potential expansion coefficients, or with an \Nbody snapshot to be used for creating a potential expansion. In the former case the type of potential expansion is stored in the first line of the file, so the \ppp{type} parameter is not required.
\end{itemize}
Parameters defining an analytic density or potential model (if \ppp{type} is a potential expansion, they refer to the \ppp{density} argument, otherwise to \ppp{type}); default values are given in brackets:
\begin{itemize}
\item \ppp{mass} [1] -- total mass of an analytic model.
\item \ppp{scaleRadius} [1] -- the first (sometimes the only) parameter with the dimension of length that defines the profile.
\item \ppp{scaleHeight} [1] or \ppp{scaleRadius2} -- the second such parameter (e.g., for Miyamoto--Nagai or exponential disk models).
\item \ppp{outerCutoffRadius} [0] -- another length-scale parameter defining the radius of exponential truncation, used for \ttt{Spheroid} models (0 means no cutoff).
\item \ppp{innerCutoffRadius} [0] -- similar parameter for \ttt{Disk} that defines the radius of an inner hole.
\item \ppp{surfaceDensity} [0] -- value of surface density at $R=0$ for the exponential \ttt{Disk} profile or for the \ttt{Sersic} profile.
\item \ppp{densityNorm} [0] -- value that defines the volume density at the scale radius for the \ttt{Spheroid} profile. Alternatively, instead of this or the previous parameter, one may provide the total mass of the corresponding model (these two parameters have a priority over mass), but this can't be done for infinite-mass models, so the density normalization remains the only option.
\item \ppp{alpha} [1] -- parameter controlling the steepness of transition between two asymptotic power-law slopes for \ttt{Spheroid}.
\item \ppp{beta} [4] -- power-law index of the outer density profile for \ttt{Spheroid}; should be $>2$ except when there is an outer cutoff, otherwise the potential is unbound.
\item \ppp{gamma} [1] -- power-law index of the inner density profile $\rho \propto r^{-\gamma}$ as $r\to 0$ for \ttt{Dehnen} (should be $0\le\gamma\le 2$) or \ttt{Spheroid} models (should be $\gamma<3$).
\item \ppp{cutoffStrength} [2] -- parameter controlling the steepness of the exponential cutoff in \ttt{Spheroid}.
\item \ppp{sersicIndex} -- shape parameter of the \ttt{Sersic} profile (larger values correspond to a models with steeper inner and shallower outer profiles, default is the de Vaucouleur's value of 4), or the same parameter for the \ttt{Disk} profile (default is 1 corresponding to the exponential disk).
\item \ppp{p} or \ppp{axisRatioY} [1] -- the axis ratio $y/x$ of equidensity surfaces of constant ellipticity for \ttt{Dehnen}, \ttt{Spheroid}, \ttt{Sersic} or \ttt{Ferrers} models, or the analogous quantity for the \ttt{Logarithmic} or \ttt{Harmonic} potentials.
\item \ppp{q} or \ppp{axisRatioZ} [1] -- the same parameter for $z/x$.
\item \ppp{W0} -- dimensionless potential depth of generalized King (lowered isothermal) models: $W_0 = [ \Phi(r_t) - \Phi(0) ] / \sigma^2$; larger values correspond to more extended envelopes (larger ratio between the outer truncation radius $r_t$ and the scale radius). In the above expression, the velocity dispersion $\sigma$ is not an independent parameter: the model in dimensionless units is specified by $W_0$ and the truncation strength parameter $g$; the potential, the truncation radius, and the total mass in dimensionless units are all determined by integrating a second-order ODE, and then the length and mass units are rescaled to match the given total mass $M$ and the scale radius (also called King radius or core radius).
\item \ppp{trunc} [1] -- truncation strength parameter of lowered isothermal models (denoted by $g$ in \cite{GielesZocchi2015}); should be between 0 and 3.5 (0 corresponds to Woolley, 1 -- to King, 2 -- to Wilson models), larger values result in softer density fall-off near the truncation radius.
\item \ppp{Omega} [1] -- the frequency of oscillation in the \ttt{Harmonic} potential.
\item \ppp{v0} [1] -- the asymptotic circular velocity for the \ttt{Logarithmic} potential.
\end{itemize}
Parameters defining the potential expansions (default values in brackets are all sensible and only occasionally need to be changed):
\begin{itemize}
\item \ppp{gridSizeR} [25] -- the number of grid nodes in spherical (\ttt{Multipole}) or cylindrical (\ttt{CylSpline}) radius; in the latter case this includes the 0th node at $R=0$. 
\item \ppp{gridSizeZ} [25] -- same for the grid in $z$ direction in \ttt{CylSpline}, including the $z=0$ node.
\item \ppp{rmin} [0] -- the radius of the innermost nonzero node in the radial grid (for both potential expansions); zero means automatic determination.
\item \ppp{rmax} [0] -- same for the outermost node; zero values mean automatic determination.
\item \ppp{zmin} [0], \ppp{zmax} [0] -- same for the vertical grid in \ttt{CylSpline}; zero values mean take them from the radial grid. Note that the grid auto-setup mechanism is currently less optimal in \ttt{CylSpline} than in \ttt{Multipole}, so a sensibly chosen manual grid extent may be beneficial for accuracy.
\item \ppp{lmax} [6] -- the order of \ttt{Multipole} expansion in $\cos\theta$; 0 means spherical symmetry. 
\item \ppp{mmax} [lmax] -- the order of azimuthal Fourier expansion in $\phi$ for both  \ttt{CylSpline} and \ttt{Multipole}; 0 means axisymmetry, and $m_\mathrm{max}$ should be $\le l_\mathrm{max}$. Of course, the actual order of expansion in all cases is also determined by the symmetry properties of the input density model -- if it reports to be axisymmetric, no $m\ne 0$ terms will be used anyway.
\item \ppp{smoothing} [1] -- the amount of smoothing applied to the non-spherical harmonics during the construction of the \ttt{Multipole} potential from an array of particles.
\end{itemize}

These keywords, with some modifications, are also used in potential construction routines in \Python and \Fortran interfaces and in the \Amuse and \Galpy plugins (Sections~\ref{sec:Python}, \ref{sec:Fortran}, \ref{sec:Amuse}, \ref{sec:Galpy}). For instance, \Python interface allows to provide a user-defined function specifying the density profile in the \ppp{density=} argument, or an array of particles in the \ppp{particles=} argument.

All dimensional values in the potential factory routines can optionally be specified in physical units and converted into internal units by providing an extra unit conversion parameter (Section~\ref{sec:Units}). For instance, masses and radii in the INI file may be given in solar masses and parsecs. This conversion also applies during write/read of density or potential coefficients to/from text files. Of course, if all data is given in the same units and follows the convention $G=1$, no conversion is needed.

%%%%%%%%%%%%%%
\subsubsection{Utility functions}  \label{sec:PotentialUtility}

\texttt{potential_utils.h} contains several frequently used functions that operate on any potential object: determination of the radius that encloses a given mass; conversion between energy $E$, angular momentum of a circular orbit $L_\mathrm{circ}$, and radius; epicyclic frequencies $\kappa,\nu,\Omega$ as functions of radius%
\footnote{defined as $\displaystyle \kappa^2\equiv \frac{\D ^2\Phi}{\D R^2} + \frac 3 R \frac{\d\Phi}{\D R},\;\; \nu^2\equiv \frac{\D ^2\Phi}{\D z^2},\;\; \Omega^2\equiv \frac 1 R \frac{\d\Phi}{\D R} = \left(\frac{L_\mathrm{circ}}{R^2}\right)^2,\;$ evaluated at $z=0$.};
peri- and apocenter radii of a planar orbit with given $E,L$ (in the $z=0$ plane of an axisymmetric potential), etc. They are implemented as standalone functions (generally using a root-finding routine to solve equations such as $\Phi(r)=E$ for $r$), and as two interpolator classes that pre-compute these values on a 1d or 2d grid in $E$ or $E,L$, and provide a faster (but still very accurate) alternative to the standalone functions. These interpolators are used, e.g., in the spherical action finder/mapper class (Section~\ref{sec:ActionsSpherical}).

%%%%%%%%%%%%%%%%%%%%%%%%%%%%%%%%%%%%%%%%%%%%%%%%%%%%%%%%%%%%%%%
\subsection{Orbit integration and analysis}  \label{sec:Orbits}

Orbits of particles in the smooth time-independent potential are computed using the routine \ttt{orbit::integrate} in any of the three standard coordinate systems, plus optionally a rotating reference frame. It solves the coupled system of ordinary differential equations (ODEs) for time derivatives of position and velocity, using one of the available methods derived from \ttt{math::BaseOdeSolver}; currently we provide only the 8th order Runge--Kutta with adaptive timestep \cite{DOP853}. Other possibilities previously implemented in \cite{Vasiliev2013} include 15th order Gauss--Radau scheme \cite{IAS15}, 4th order Hermite method \cite{Hermite}, and several methods from \textsc{Odeint} package \cite{odeint}, including Bulirsch--Stoer and various Runge--Kutta schemes. However, in practice all of them have rather similar performance in the appropriate range of tolerance parameters, thus we have only kept one at the moment.

There are various tasks that can be performed during orbit integration, using classes derived from \ttt{orbit::BaseRuntimeFnc}. The simplest one (\ttt{orbit::RuntimeTrajectory}) is the recording of the trajectory at regular intervals of time, which are unrelated to the internal timestep of ODE solver (that is, the  position/\-velocity at any time is obtained by interpolation provided by the solver -- so-called dense output feature). More complicated tasks involve storage of some other kind of information, e.g., in the context of Schwarzschild modelling, or in some cases, even modifying the orbit itself (random perturbations mimicking the effect of two-body relaxation in the Monte Carlo code \textsc{Raga}).

Orbit analysis refers to the determination of orbit class (box, tube, resonant boxlet, etc.) and degree of chaoticity. This is performed using a Fourier transform of position as a function of time and detecting the most prominent ``spectral lines''; the ratio between their frequencies is an indicator of orbit type \cite{BinneySpergel1984, CarpinteroAguilar1998}, and their rate of change with time is a measure of chaos \cite{ValluriMerritt1998}. These methods were implemented in \cite{Vasiliev2013}, but as the focus of \Agama in galaxy modelling is shifted from discrete orbits to smooth distribution functions, we have not yet included them in the library.

A finite-time estimate of Lyapunov exponent $\lambda$ is another measure of stochasticity (see \cite{Carpintero2014, Skokos2010} for reviews of methods based on variational equations). It may be estimated by following the time evolution of a deviation vector, which depends on the second derivatives of potential evaluated along the orbit. For a regular orbit, its magnitude grows at most linearly with time, while for a chaotic orbit it eventually starts to grow exponentially. The class \ttt{orbit::RuntimeLyapunov} implements the method described in Section~4.3 and illustrated in Figure~4 of \cite{Vasiliev2013}: if no exponential growth has been detected, it returns $\lambda=0$, otherwise a median value of $\lambda$ on the interval of exponential growth, normalized to the characteristic orbital time (so that orbits at different energies can be more directly compared).

%%%%%%%%%%%%%%%%%%%%%%%%%%%%%%%%%%%%%%%%%%%%%%%%%%%%%%%%%%%%
\subsection{Action/angle variables}  \label{sec:ActionAngle}

As the name implies, \Agama deals with models of stellar system described in terms of action/angle variables. They are defined, e.g., in Section 3.5 of \cite{BinneyTremaine}.

In a spherical or axisymmetric potential, the most convenient choice for actions is the triplet $\{J_r, J_z, J_\phi\}$, where $J_r\ge 0$ (radial action) describes the motion in cylindrical radius, $J_z\ge 0$ (vertical action) describes the motion in $z$ direction, and $J_\phi \equiv R v_\phi$ (azimuthal action) is the conserved component $L_z$ of angular momentum (it may have any sign). In a spherical potential, the sum $J_z + |J_\phi|$ is the total angular momentum $L$. Actions are only defined for a bound orbit -- if the energy is positive, they will be reported as \texttt{NAN} (except $L_z$ which can always be computed).
%The corresponding angles are defined such that $\theta_r=0$ corresponds to the pericenter, $\theta_z=0$ -- to the passage through the $x-y$ plane with positive vertical velocity, and $\theta_\phi=0$ -- to [?]. 

The \ttt{actions::} namespace introduces several concepts: \ttt{Actions} and \ttt{Angles} are the triplet of action and angle variables, \ttt{ActionAngles} is their combination, \ttt{Frequencies} is the triplet of frequencies $\boldsymbol{\Omega}\equiv \D H/\D \bJ$ (derivatives of Hamiltonian w.r.t.\ actions). 
The transformation from $\{\bx,\bv\}$ to $\{\bJ,\bt\}$ is provided by action finders, and the inverse transformation -- by action mappers. There are several distinct methods discussed later in this section, and they may exist as standalone routines and/or instances of classes derived from the \ttt{BaseActionFinder} and \ttt{BaseActionMapper} classes. The action finder routines exist in two variants: computing only the actions (the most common usage), or in addition the angles and frequencies (more expensive).

The following sections describe the methods suitable for specific cases of spherical or axisymmetric potentials (see \cite{SandersBinney2016} for a review and comparison of various approaches).
At present, \Agama does not contain any methods for action/angle computation in non\--axi\-sym\-met\-ric potentials, but they may be added in the future within the same general framework.

%%%%%%%%%%%%%%
\subsubsection{Isochrone mapping}  \label{sec:ActionsIsochrone}

The spherical isochrone potential, specified by two parameters (mass $M$ and scale radius $b$) admits analytic expressions for the transformation between $\{\bx,\bv\}$ and $\{\bJ,\bt\}$ in both directions. These expression are given, e.g., in Eqs.~3.225--3.241 of \cite{BinneyTremaine}.
The standalone routines providing these transformations, optionally with partial derivatives of $\{\bx,\bv\}$ w.r.t.\ $\bJ, M, b$, are located in \texttt{actions_isochrone.h}.

%%%%%%%%%%%%%%
\subsubsection{Spherical potentials}  \label{sec:ActionsSpherical}

In a more general case of an arbitrary spherical potential, the radial action is given by 
\begin{align*}
J_r = \frac{1}{\pi} \int_{r_\mathrm{min}}^{r_\mathrm{max}} \sqrt{2[E-\Phi(r)] - L^2/r^2}\;\d r,
\end{align*}
where $r_\mathrm{min,max}(E,L)$ are the roots of the expression under the radical.
The standalone routines in \texttt{actions_spherical.h} perform the action/angle transformation in both directions, using numerical root-finding and integration functions in each invocation. If one needs to compute actions for many points ($\gtrsim 10^3$) in the same potential, it is more efficient to construct an instance of \ttt{ActionFinderSpherical} class that provides high-accuracy interpolation from the pre-computed 2d tables for $r_\mathrm{min,max}(E,L)$ (using the helper class \ttt{potential::Interpolator2d}) and $J_r(E,L)$, the inverse mapping $E(J_r,L)$ also provided via an interpolation table, and the complete inverse mapping $\{\bJ,\bt\} \Rightarrow \{\bx,\bv\}$.

%%%%%%%%%%%%%%
\subsubsection{St\"ackel approximation}  \label{sec:ActionsStaeckel}

In a still more general axisymmetric case, the action/angle variables can be exactly computed for a special class of St\"ackel potentials, in which the motion is fully integrable and separable in a prolate spheroidal coordinate system. This computation is performed by the standalone routines \ttt{actionsAxisymStaeckel} and \ttt{actionAnglesAxisymStaeckel} in \mbox{\texttt{actions_staeckel.h}}, which operate on an instance of \ttt{potential::OblatePerfectEllipsoid} class (the only example of a St\"ackel potential in \Agama). The procedure consists of several steps: numerically find the extent of oscillations in the meridional plane in both coordinates $\lambda, \nu$ of the prolate spheroidal system; numerically compute the 1d integrals for $J_\lambda, J_\nu$ (which correspond to $J_r, J_z$); and if necessary, find the frequencies and angles (again by 1d numerical integration).

For the most interesting practical case of a non-St\"ackel axisymmetric potential, the actions can only be approximated under the assumption that the motion is integrable and is locally well described by a St\"ackel potential. This is the essence of the ``St\"ackel fudge'' approach \cite{Binney2012}. In a nutshell, it pretends that the potential \textit{is} of a St\"ackel form (without explicitly constructing it), computes the would-be integrals of motion in this presumed potential, and then performs essentially the same steps as the routines for the genuine St\"ackel potential. Actions computed in this way are approximate, in the sense that even for a regular (non-chaotic) motion, they are not exactly conserved along the orbit; the variation of $\bJ$ is smallest for nearly-circular orbits close to the equatorial plane, but typically remains $\lesssim 1-10\%$ even for rather eccentric orbits that stray far from the plane (note that the method does not provide any error estimate). However, if the actual orbit is chaotic or belongs to one of minor resonant families, the variation of estimated actions along the orbit is rather large because the method does not account for resonant motion.

In order to proceed, the St\"ackel approximation requires the parameter of the prolate spheroidal coordinate system -- the focal distance $\Delta$; the accuracy (variation of estimated actions along the orbit) strongly depends on its value. Importantly, we do not need to have a single value of $\Delta$ for the entire system, but may use the most suitable value for the given set of integrals of motion (depending on $\{\bx,\bv\}$).
The \ttt{ActionFinderAxisymFudge} class pre-computes a table of best-fit values of $\Delta$ as a function of $E,L_z$ (this takes a couple of seconds) and uses interpolation to obtain a suitable value at any point, which is then fed into the routines that compute the actions.
This is the main workhorse for many higher-level tasks in the \Agama library.

A variation of this approach is to pre-compute the actions $J_r,J_z$ as functions of three integrals of motion (one of them being approximate) on a suitable grid, and then use a 3d interpolation to obtain the values of actions at any point. The construction of such interpolation table takes another couple of seconds, and the evaluation of actions through interpolation is $\sim 10\times$ faster than using the St\"ackel approximation directly. However, the accuracy of this approach is somewhat worse (not because of interpolation, but due to the approximate nature of the third integral); nevertheless, it is still sufficient in many contexts.

More technical details are provided in Section~\ref{sec:ActionsStaeckelDetails}.

%%%%%%%%%%%%%%
\subsubsection{Torus mapping}  \label{sec:ActionsTorus}

The transformation from $\{\bJ, \bt\}$ to $\{\bx,\bv\}$ in an arbitrary axisymmetric potential is performed using the Torus mapping approach \cite{BinneyMcMillan2016}. An instance of \ttt{ActionMapperTorus} class is constructed for any choice of $\bJ$ and allows to perform this mapping for multiple values of $\bt$; however, the cost of torus construction is rather high, and it may not always succeed (depending on the properties of potential and required accuracy). The code is adapted from the original \textsc{TorusMapper} package, with several modifications enabling the use of an arbitrary potential and a more efficient angle mapping approach; however, it does not quite comply to the coding standards adopted in \Agama (Section~\ref{sec:DeveloperGuide}) and in the future will be replaced by a fresh implementation.

%%%%%%%%%%%%%%%%%%%%%%%%%%%%%%%%%%%%%%%%%%%%%%%%%%%
\subsection{Distribution functions}  \label{sec:DF}

By Jeans' theorem, a steady-state distribution of stars or other species in a stationary potential may depend only on integrals of motion, taken here to be the actions $\bJ$. The \ttt{df::} namespace contains the classes and methods for working with such distribution functions (DFs) formulated in terms of actions. They are derived from the \ttt{BaseDistributionFunction} class, which provides a single method for computing the value $f(\bJ)$ at the given triplet of actions. All physically valid DFs must have a finite mass $M = (2\pi)^3 \iiint f(\bJ)\,\d ^3J$, computed by numerical integration (the pre-factor comes from a trivial integration over angles) and returned by the \ttt{totalMass()} method of the DF instance.
The same DF corresponds to different density profiles in different potentials (Section~\ref{sec:Moments}), but the total mass of the density profile is always the same.

\Agama provides several DFs suitable for various components of a galaxy, described in the following sections. In addition there is a concept of a multi-component DF: since computing the actions -- arguments of the DF -- is a non-negligible cost, it is often advantageous to evaluate several DFs at the same set of actions at once.
There is also a ``DF factory'' routine \ttt{createDistributionFunction} for constructing various DF classes from a set of named parameters described by a \ttt{KeyValueMap} object (Section~\ref{sec:Utilities}); the choice of model is set by \ppp{type=...}, and model-specific parameters are described in the following sections.

Importantly, the DF formulated in terms of actions does not depend on the potential. However, some models use the concept of epicyclic frequencies to compute the value of $f(\bJ)$. These frequencies are represented by a special proxy class \ttt{potential::Interpolator}, which is constructed from a given potential, but then serves as an independent entity (essentially an array of arbitrary functions of one variable), so that $f(\bJ)$ has the same value in any other potential. This is important in the context of iterative construction of self-consistent models (Section~\ref{sec:SCM}).

%%%%%%%%%%%%%%
\subsubsection{Disky components}  \label{sec:DFdisk}

There are two classes of disk DFs in \Agama: the first, \ttt{QuasiIsothermal}, expresses the DF in terms of auxiliary functions that are related to a particular potential, while the second, \ttt{Exponential}, is written in an entirely self-contained form. We describe them in turn.

Stars on nearly-circular (cold) orbits in a disk are often described by a Schwarzschild or Shu DF, which have Maxwellian velocity distribution with different dispersions in each direction. A generalization for warm disks \cite{Dehnen1999} expressed in terms of actions \cite{BinneyMcMillan2011} is provided by the \ttt{QuasiIsothermal} class. The DF for a single population is given by
\begin{align*}
f(\bJ) &= \frac{\tilde\Sigma\,\Omega}{2\pi^2\,\kappa^2} \times
\frac{\kappa}{\tilde\sigma_r^2} \exp\left(-\frac{\kappa\,J_r}{\tilde\sigma_r^2}\right) \times
\frac{\nu}   {\tilde\sigma_z^2} \exp\left(-\frac{\nu\,   J_z}{\tilde\sigma_z^2}\right) \times
\left\{ \begin{array}{ll}  1 & \mbox{if }J_\phi\ge 0, \\
\exp\left( \frac{2\Omega\,J_\phi}{\tilde\sigma_r^2} \right) & \mbox{if }J_\phi<0, \end{array} \right. \\
\tilde\Sigma(R_c)  &\equiv \Sigma_0 \exp( -R_c / R_\mathrm{disk} ) , \qquad\quad
\tilde\sigma_r^2(R_c) \equiv \sigma_{r,0}^2 \exp( -2R_c / R_{\sigma,r} ) + \sigma_\mathrm{min}^2,\\
\tilde\sigma_z^2(R_c)&\equiv 2\,h_\mathrm{disk}^2\,\nu^2(R_c)  + \sigma_\mathrm{min}^2
\quad\mbox{or}\quad
\tilde\sigma_z^2(R_c) \equiv \sigma_{z,0}^2 \exp( -2R_c / R_{\sigma,z} )  + \sigma_\mathrm{min}^2.
\end{align*}

To construct such a DF, one needs to provide an instance of potential, which is used to initalize the mappings between actions and the radius of a circular orbit $R_c$ as a function of $z$-component of angular momentum, and the epicyclic frequencies $\kappa, \nu, \Omega$ as functions of radius. Other parameters listed in the  \ttt{QuasiIsothermalParam} structure are:
\begin{itemize}
\item \ppp{coefJr} [1], \ppp{coefJz} [0.25] are the dimensionless coefficients $k_r,k_z\sim \mathcal{O}(1)$ in the linear combination of the actions $\tilde J \equiv |J_\phi| + k_r J_r + k_z J_z$, which is used as the argument of $R_c$ instead of the angular momentum. The reason for this is that $\tilde J$ better corresponds to the average radius of a given orbit than $J_\phi$ alone: for instance, a star with $J_\phi\approx 0$ does not in reality stay close to origin if the other two actions are large.
Actually, we use $R_c(\hat J)$, where $\hat J \equiv (\tilde J^2 + J_\mathrm{min}^2)^{1/2}$, and $J_\mathrm{min}$ is introduced to prevent a pathological behaviour of DF in the case of a cuspy potential, when epicyclic frequencies tend to infinity as $R_c \to 0$.
\item \ppp{Sigma0} is the overall normalization of the surface density profile $\Sigma_0$.
\item \ppp{Rdisk} sets the scale length of the disk $R_\mathrm{disk}$.
\item \ppp{Hdisk} if provided, determines the scale height of the disk $h_\mathrm{disk}$ and the vertical velocity dispersion, which are related through the vertical epicyclic frequency $\nu$. 
\item Alternatively, \ppp{sigmaz0} and \ppp{Rsigmaz} can be used instead of \ppp{Hdisk} to make the vertical velocity dispersion profile close to an exponential function of radius, with central value $\sigma_{z,0}$ and radial scale length $R_{\sigma,z}$; this choice has been used historically for quasi-isothermal DFs \cite{BinneyMcMillan2011,Bovy2015}, but it does not generally produce constant-scaleheight disk profiles.
\item \ppp{sigmar0} and \ppp{Rsigmar} control the radial velocity dispersion profile, which is nearly exponental with central value $\sigma_{r,0}$ and radial scale $R_{\sigma,r}$. Typically $R_{\sigma,r}, R_{\sigma,z} \sim 2\,R_\mathrm{disk}$.
\item \ppp{sigmamin} [0] is the minimum value of velocity dispersion $\sigma_\mathrm{min}$, added in quadrature to both $\tilde\sigma_r$ and $\tilde\sigma_z$; it is introduced to avoid the pathological situation when the velocity dispersion drop so rapidly with radius that the value of DF at $J_r=J_z=0$ actually increases indefinitely at large $J_\phi$. A reasonable lower limit could be $(0.02\, .. \, 0.05)\sigma_0$.
\item \ppp{Jmin} [0] is the lower limit $J_\mathrm{min}$ imposed on the argument of the circular radius function $R_c(\hat J)$.
\end{itemize}
As stressed above, both epicyclic frequencies and $R_c(\hat J)$ are merely one-dimensional functions that are once initialized from an actual potential, but no longer need to be related to the potential in which the DF is later used. If these two potentials are close enough, then this DF produces density profiles and velocity dispersions that approximately correspond to the exponential disks: the surface density $\Sigma(R)$ is close to exponentially declining with central value $\Sigma_0$ and radial scale length $R_\mathrm{disk}$, the vertical profile is roughly isothermal with scale height $h_\mathrm{disk}$, and the radial velocity dispersion is similar to $\tilde\sigma_r$, althogh the actual profiles are somewhat different from the tilded functions.

If the potential has a nearly flat rotation curve with circular velocity $v_\circ$, then $R_c(\tilde J) \approx \tilde J/v_\circ$, and epicyclic frequencies are $\propto v_\circ^2 / \tilde J$. This motivates the introduction of another family of disk-like DFs, which has a similar functional form, but does not itself contain any reference to the potential, simplifying the construction of self-consistent models (Section~\ref{sec:SCM}). It is represented by the \ttt{Exponential} class:
\begin{align*}
f(\bJ) &= \frac{M}{(2\pi)^3}\, \frac{J_d}{J_{\phi,0}^2}\, \exp\bigg(-\frac{J_d}{J_{\phi,0}}\bigg) \times
\frac{J_v}{J_{r,0}^2} \exp\bigg(-\frac{J_v\,J_r}{J_{r,0}^2}\bigg) \times
\frac{J_v}{J_{z,0}^2} \exp\bigg(-\frac{J_v\,J_z}{J_{z,0}^2}\bigg) \\
&\times \left\{ \begin{array}{ll}  1 & \mbox{if }J_\phi\ge 0, \\
\exp\left(\frac{J_v\,J_\phi}{J_{r,0}^2} \right) & \mbox{if }J_\phi<0, \end{array} \right. \qquad
J_d \equiv \sqrt{\tilde J^2 + J_{d,0}^2}, \quad
J_v \equiv \sqrt{\tilde J^2 + J_{v,0}^2}
\end{align*}
Its parameters listed in the \ttt{ExponentialParam} structure are:
\begin{itemize}
\item \ppp{norm} is the overall scaling with the dimension of mass ($M$).
\item \ppp{Jphi0} determines the scale length of the disk: $J_{\phi,0} \sim R_\mathrm{disk}v_\circ$.
\item \ppp{Jz0} controls the scale height and vertical velocity dispersion: $J_{z,0} \sim h_\mathrm{disk}v_\circ \sim R\,\sigma_z(R)$.
\item \ppp{Jr0} plays a similar role for the radial velocity dispersion and the extent of radial excursion $\Delta R$ of a typical orbit: $J_{r,0} \sim \Delta R\,v_\circ \sim R\,\sigma_r(R)$.
\item \ppp{coefJr} [1], \ppp{coefJz} [0.25] are the coefficients $k_r$, $k_z$ in the linear combination of actions $\tilde J \equiv |J_\phi| + k_r J_r + k_z J_z$.
\item \ppp{addJden} [0] and \ppp{addJvel} [0] are the lower limits $J_{d,0}$ and $J_{v,0}$ for the arguments of the exponential functions, which are introduced to tweak the behaviour of density and velocity dispersion profiles, correspondingly, in the limit of small actions. Their role is similar to $J_\mathrm{min}$ for the QuasiIsothermal model: since the rotation curves of realistic potentials cannot be flat all the way down to the center, the unmodified DF would produce too low density and too high velocity dispersions at small radii, which is compensated by these additional parameters. Their values should typically be of order $J_{d,0} \sim J_{r,0}$ and $J_{v,0} \sim J_{\phi,0}$.
\end{itemize}

One can generalize either of these models to the continuous distribution of populations with different velocity dispersions, by introducing three additional dimensionless parameters. It is commonly assumed that the velocity dispersion of a coeval population of stars increases with time. Let $\tau\in[0..1]$ be the age of the star normalized to the galaxy age (youngest stars have $\tau=0$ and oldest -- $\tau=1$). Let $\sigma_0$, $\sigma_1$ be the radial or vertical velocity dispersion of youngest and oldest stars, correspondingly, and denote their ratio as $\xi\equiv \sigma_0/\sigma_1 \le 1$ (\ppp{sigmabirth}). The commonly adopted relation between age and velocity dispersion \cite{AumerBinney2009} is $\sigma(\tau) = \sigma_1 \big[\tau + (1-\tau)\xi^{1/\beta} \big]^\beta$, with $\beta\simeq 0.33$ (\ppp{beta}). Assume further that the star formation rate declines exponentially with characteristic timescale $\tau_\mathrm{SFR}$ (\ppp{Tsfr}, again normalized to the galaxy age), so that the number of stars increases with look-back time $\tau$ as $\exp(\tau/\tau_\mathrm{SFR})$. Then the value of the DF needs to be integrated over stars of all ages, weighted by the number of stars at each age and substituting $\tilde\sigma_r, \tilde\sigma_z$ by the age-scaled values:
\begin{align*}
\bar f(\bJ) = \frac{\int_0^1 \d \tau\, \exp(\tau/\tau_\mathrm{SFR})\,
f\big[ \bJ \,|\, \sigma(\tau) \big] }{\int_0^1 \d \tau\, \exp(\tau/\tau_\mathrm{SFR})}
\end{align*}
The \ttt{QuasiIsothermal} DF is then computed as
\begin{align*}
\bar f(\bJ) &= \frac{\tilde\Sigma\,\Omega\,\nu}
{2\pi^2\,\kappa\,\tilde\sigma_{r,1}^2\,\tilde\sigma_{z,1}^2}
\frac{ \displaystyle \int_0^1 \d \tau\, \exp(\tau/\tau_\mathrm{SFR})\,
\frac{1}{\lambda^4} \exp\left(-\frac{1}{\lambda^2}
\left[ \frac{\kappa\,J_r}{\tilde\sigma_{r,1}^2} + \frac{\nu\,J_z}{\tilde\sigma_{z,1}^2} +
\mathrm{max}\bigg\{0, -\frac{2\Omega\,J_\phi}{\tilde\sigma_{r,1}^2}\bigg\} \right]\right)}
{ \int_0^1 \d \tau\, \exp(\tau/\tau_\mathrm{SFR}) } ,\\
\lambda &\equiv \sigma(\tau)/\sigma_1 = \big[\tau + (1-\tau)\xi^{1/\beta}\big]^\beta ,
\end{align*}
where the auxiliary functions $\tilde\sigma_r,\tilde\sigma_z$ now refer to the oldest stars. A similar modification applies to the \ttt{Exponential} DF, except that the age-dependent parameters are the scale actions $J_{r,0}, J_{z,0}$, not the velocity dispersions.

%%%%%%%%%%%%%%
\subsubsection{Spheroidal components}  \label{sec:DFspheroid}

A suitable choice for DFs of elliptical galaxies, bulges or haloes is the \ttt{DoublePowerLaw} model, which is similar to the ones presented in \cite{Binney2014, Posti2015}, with a different notation:
\begin{align*}
f(\bJ) &= \frac{M}{(2\pi\, J_0)^3}
%\left(\frac{h(\bJ)}{J_0}\right)^{-\Gamma}
\left[1 + \left(\frac{J_0}{h(\bJ)}\right)^\eta \right]^{\Gamma / \eta} \;
\left[1 + \left(\frac{g(\bJ)}{J_0}\right)^\eta \right]^{(\Gamma-\Beta) / \eta} \\
&\times
\left[1 - \beta\left(\frac{J_\mathrm{core}}{h(\bJ)}\right) + 
\left(\frac{J_\mathrm{core}}{h(\bJ)}\right)^2 \right]^{-\Gamma/2}
\exp\bigg[-\left(\frac{g(\bJ)}{J_\mathrm{cutoff}}\right)^\zeta\bigg] \;
\bigg(1 + \varkappa\tanh \frac{J_\phi}{J_{\phi,0}} \bigg), \\
g(\bJ) &\equiv g_r J_r + g_z J_z\, + (3-g_r-g_z)\, |J_\phi|, \\
h(\bJ) &\equiv h_r J_r + h_z J_z   + (3-h_r-h_z)   |J_\phi|.
\end{align*}
The parameters of this model are listed in the structure \ttt{DoublePowerLawParams} and have the following meaning (default values in brackets):
\begin{itemize}
\item \ppp{norm} is the overall normalization $M$ with the dimension of mass; the actual total mass differs from $M$ by a numerical factor ranging from 1 to a few tens, which depends on other parameters of the model. 
\item \ppp{J0} is the characteristic action $J_0$ corresponding to the break in the double-power-law profile; it sets the overall scale of the model.
\item \ppp{slopeIn} is the power-law index $\Gamma$ in the inner part of the model, below the characteristic action; must be $<3$.
\item \ppp{slopeOut} is the power-law index $\Beta$ in the outer part of the model; must be $>3$.
\item \ppp{steepness} [1] is the parameter $\eta$ controlling the steepness of the transition between the two asymptotic regimes.
\item \ppp{coefJrIn} [1], \ppp{coefJzIn} [1] are the coefficients $h_r, h_z$ in the linear combination of actions, controlling the flattening and anisotropy in the inner part of the model; the third coefficient is implied to be $3-h_r-h_z$, and all three must be non-negative.
\item \ppp{coefJrOut} [1], \ppp{coefJzOut} [1] are the similar coefficients $g_r, g_z$ for the outer part of the model.
\item \ppp{Jcore} [0] introduces an optional central core, forcing the DF to have a finite central value even when the inner slope $\Gamma$ is nonzero \cite{ColeBinney2017}; in this case, the auxiliary coefficient $\beta$ is assigned automatically from the condition that the total normalization of the DF remains the same (although this is only true when $J_\mathrm{core}\ll \mathrm{min}(J_0, J_\mathrm{cutoff})$ or when the coefficients $g_r=h_r, g_z=h_z$, otherwise it still changes somewhat).
\item \ppp{Jcutoff} [0] additionally suppresses the DF at large actions (beyond $J_\mathrm{cut}$), 0  means disable.
\item \ppp{cutoffStrength} [2] sets the steepness $\zeta$ of this exponential cutoff at large $J$.
\item \ppp{rotFrac} [0] controls the amount of streaming motion by setting the odd-$J_\phi$ part of DF to be $\varkappa$ times the even-$J_\phi$ part; $\varkappa=0$ disables rotation and $\varkappa=\pm 1$ correspond to models with maximum rotation.
\item \ppp{Jphi0} [0] sets the extent $J_{\phi,0}$ the central core with suppressed rotation.
\end{itemize}
This DF roughly corresponds to the $\alpha\beta\gamma$ \ttt{Spheroid} model, with the asymptotic power-law indices $\Beta=2\beta-3$ and $\Gamma=(6-\gamma)/(4-\gamma)$ in the self-consistent case. An example program \texttt{example_doublepowerlaw.cpp} helps to find the values of these parameters that best correspond to the given spherical isotropic model.

%%%%%%%%%%%%%%
%\subsubsection{Nonparametric interpolated models}  \label{sec:DFinterpolated}
%
%A general way of representing an arbitrary DF in a three-dimensional action space is through an  \hyperref[sec:SplineInterpolation]{interpolating spline} in suitably scaled coordinates. The formulation in terms of B-splines is also an example of a multi-component DF, where each basis function, formed as a tensor product of one-dimensional B-splines, is a separate component. These DFs are used as building blocks in self-consistent models (work in preparation).

%%%%%%%%%%%%%%
\subsubsection{Spherical DFs constructed from a density profile}  \label{sec:DFspherical}

The previously described types of DFs were defined in terms of an analytic functions of actions, and the density profiles that they generate in a particular potential (Section~\ref{sec:Moments}) are not available in a closed form. The alternative approach is to start from a given density profile and a potential, and determine a DF that generates this density profile, using some sort of inversion formula. So far this approach has been mostly used in spherical systems, with the most well-known case being the Eddington inversion formula for a spherical isotropic DF. We implement a more general version of this formula \cite{Cuddeford1991}, which produces a DF with the following velocity anisotropy profile:
\begin{align*}
\beta(r) \equiv 1 - \frac{\sigma_t^2}{2\sigma_r^2} = \frac{\beta_0 + (r/r_a)^2}{1 + (r/r_a)^2}.
\end{align*}
$\beta_0$ is the limiting value of anisotropy in the center, and if $r_a<\infty$, the anisotropy coefficient tends to 1 at large $r$ (Osipkov--Merritt profile), otherwise stays equal to $\beta_0$ everywhere. The usual isotropic case is obtained by setting $\beta_0=0, r_a=\infty$.

The DF, expressed in terms of energy $E$ and angular momentum $L$, has the following form:
\begin{align*}
f(E,L) = \hat f(Q) \; L^{-2\beta_0}, \quad Q\equiv E + L^2 / (2 r_a^2) .
\end{align*}
The function $\hat f(Q)$ is computed numerically for the given choice of parameters, density and potential, as explained in Section~\ref{sec:DFsphericalDetails}), and represented in an interpolated form on a suitable grid in $Q$.
For some combinations of parameters, this produces a DF which is negative in some range of $Q$, in particular, when the central slopes of the density profile $\gamma \equiv -d\ln\rho/d\ln r$, the potential $\delta \equiv -d\ln(-\Phi)/d\ln r$, and the coefficient $\beta_0$ violate the so-called slope--anisotropy theorem \cite{AnEvans2006}: $\gamma \ge 2\beta + (1/2-\beta)\delta$. For the self-gravitating case, $\delta=0$ if the potential is finite at origin, or $\delta=\beta-2$ otherwise. If the computed DF is negative, it is replaced by zero.

This type of DF has the following parameters:
\begin{itemize}
\item \ppp{beta0} [0]  is the central value of velocity anisotropy, must be in the range $-0.5 \le \beta_0 < 1$.
\item \ppp{r_a} [$\infty$]  is the anisotropy radius, must be positive (in particular, infinity means a constant-anisotropy model).
\end{itemize}
In addition, one needs to provide one-dimensional functions representing the radial profile of the density and the potential (if they are taken from the same model, only the latter is needed). Any spherically-symmetric instances of \ttt{Density} and \ttt{Potential} classes can be used.

The DF is traditionally expressed in terms of $E,L$, and in this form can only be used in the same potential $\Phi$ as it was constructed in. However, it can be put on equal grounds with other action-based DFs, which are manifestly independent of potential, using the following approach.
An instance of \ttt{ActionFinderSpherical} class, constructed for the original potential $\Phi$, is attached to the instance of a \ttt{QuasiSpherical} DF. To compute the value of DF at the given triplet of actions $\bJ$, we map them to $E,L$ using this original spherical action finder, and then feed these values to the DF. Crucially, in this form the actions $\bJ(\bx,\bv \;|\;\tilde\Phi)$ may be computed in any other potential $\tilde\Phi$ (not necessarily spherical), not just the original $\Phi$. This corresponds to the DF being adiabatically transformed from the original potential to the new one, without changing its dependence on actions.
This is especially convenient for iterative construction of multicomponent self-consistent models (Section~\ref{sec:SCM}): the DFs of spheroidal components (bulge, halo) may be obtained using the Eddington inversion formula or its anisotropic generalization in the initial spherically-symmetric approximation of the total potential, and then expressed as functions of actions $\bJ$. The resulting DF is then viewed as a function of actions only. In subsequent iterations, the potential is no longer spherical, but the density and anisotropiy profiles of these components, obtained by integration of their DFs over velocity, are nevertheless quite close to the initial ones.

%%%%%%%%%%%%%%
\subsubsection{Spherical isotropic models}  \label{sec:DFsphericalIsotropic}

In a special case of spherical isotropic models, the DF has the form $f(E)$. There is an alternative formulation which makes it invariant with respect to the potential, retaining the convenience of action formalism without the need to compute actions explicitly. Namely, we use the phase volume $h$ instead of $E$ as the argument of the DF. It is defined as the volume of phase space enclosed by the given energy hypersurface:
\begin{align*}
h(E) &\equiv \iiint \d ^3x \iiint \d ^3v\, H\Big[E - \big(\Phi(|\bx|)+|\bv|^2/2\big)\Big] \;,\quad
\mbox{where $H$ is the step function,} \\
&= \int_0^{r_\mathrm{max}(E)} 4\pi\,r^2\,\d r \int_0^{v_\mathrm{max}(E,r)} 4\pi\, v^2\,\d v =
\frac{16\pi^2}{3} \int_0^{r_\mathrm{max}(E)} r^2\, \Big[2\big(E-\Phi(r)\big)\Big]^{3/2}\;\d r.
\end{align*}

The advantages of using $h$ instead of $E$ are that the total mass of the model is simply $M=\int_0^\infty f(h)\,\d h$, that the same DF may be used in different potentials, etc. The bi-directional correspondence between $E$ and $h$ is provided by a helper class \ttt{PhaseVolume}, constructed for a given potential. The derivative $\d h(E)/\d E \equiv g(E)$ is called the density of states (\cite{BinneyTremaine}, eq.~4.56), and is given by
\begin{align*}
g(E) \equiv 16\pi^2 \int_0^{r_\mathrm{max}(E)} r^2\, \sqrt{2\big(E-\Phi(r)\big)}\;\d r 
= 4\pi^2\,L^2_\mathrm{circ}(E)\,T_\mathrm{rad}(E).
\end{align*}

Any non-negative function of one variable (the phase volume $h$) may serve as an isotropic distribution function in a spherically-symmetric potential, provided that it satisfies the condition that $\int_0^\infty f(h)\, \d h$ is finite. 
One possible way of computing such a DF is through the Eddington inversion formula for any density profile in any given potential (not necessarily related), implemented in the routine \ttt{createSphericalIsotropicDF}. The other is to construct an approximating DF from an array of particles sampled from it, using the log-density estimation approach (Section~\ref{sec:MathSplineDensityDetails}), provided by the routine \ttt{fitSphericalIsotropicDF}.
More information on these models is given in section~\ref{sec:DFsphericalIsotropicDetails}.

These one-dimensional DFs may also be put on equal grounds with other action-based DFs, using a proxy class \ttt{QuasiSphericalIsotropic}. It provides the mapping $\bJ \to E \to h$ via \ttt{ActionFinderSpherical} and \ttt{PhaseVolume} classes, both constructed in a given potential; the value $h$ is then used as the argument to an arbitrary function $f(h)$ provided by the user. Similarly to the case of disky DFs (Section~\ref{sec:DFdisk}), the potential is only needed to construct the intermediate mappings between actions and the arguments of the DF; the resulting object is then viewed as a function of actions only, and could be used in any other potential.
When the original DF is constructed using the Eddington inversion formula, it is easier to use the \ttt{QuasiSpherical} class from the previous section directly, avoiding the additional transformation $E\leftrightarrow h$.

%%%%%%%%%%%%%%%%%%%%%%%%%%%%%%%%%%%%%%%%%%%%%%%%%%%%%%%%%%%%%%%%
\subsection{Galaxy modelling framework}  \label{sec:GalaxyModel}

This module (namespace \ttt{galaxymodel::}) broadly encompasses all tasks that involve both a DF and a potential, and additionally an action finder constructed for the given potential and used for transforming $\{\bx,\bv\}$ to $\bJ$.
As stressed previously, using $\bJ$ as the argument of $f$ has the advantage that the DF may be used with an arbitrary potenial without any modifications (because the possible range of actions does not depend on the potential, unlike, e.g., the possible range of energy).

%%%%%%%%%%%%%%
\subsubsection{Moments of distribution functions}  \label{sec:Moments}

The most basic task is the computation of DF moments (density, velocity dispersion, etc.), defined as
\begin{align*}
\rho(\bx) &\equiv \iiint \d ^3v\, f\big(\bJ(\bx,\bv)\big), \\
\overline{\bv} &\equiv \rho^{-1} \iiint \d ^3v \,\bv\, f\big(\bJ(\bx,\bv)\big), \\
\overline{v_i v_j} &\equiv \rho^{-1} \iiint \d ^3v \,v_i v_j\, f\big(\bJ(\bx,\bv)\big).
\end{align*}
The routine \ttt{computeMoments} calculates any combination of these quantities at the given point $\bx$ by numerically integrating $f$ over $\bv$; the DF may be single- or multi-component. 
This is not a cheap operation, as the integration requires $\gtrsim 10^3$ evaluation of DF and hence calls to the action finder; the computation of density is the major cost in self-consistent modelling (Section~\ref{sec:SCM}). The first moment of velocity may have only one non-zero component ($\overline{v_\phi}$); the tensor of second velocity moments is computed in cylindrical coordinates ($R,z,\phi$), and in axisymmetric systems may have at most 4 non-zero components ($\overline{\strut v_R^2}, \overline{\strut v_z^2}, \overline{\strut v_\phi^2}, \overline{\strut v_R v_z}$, the latter is related to the tilt of velocity ellipsoid in the meridional plane).

The routine \ttt{computeProjectedMoments} calculates the surface density and the line-of-sight velocity dispersion at the given cylindrical radius $R$ (currently for axisymmetric systems only) -- this involves an additional integration over $z$:
\begin{align*}
\Sigma(R) \equiv \int_{z=-\infty}^\infty \d z\, \rho(R,z), \qquad
\sigma_\mathrm{los} \equiv \frac{1}{\Sigma(R)} \int_{z=-\infty}^\infty \d z\, \rho(R,z)\, v_z^2 .
\end{align*}

There are analogous routines for spherical isotropic DFs of the form $f(h)$, defined in \texttt{galaxymodel_spherical.h} and used by the \texttt{mkspherical} tool (Section~\ref{sec:mkspherical}).

%%%%%%%%%%%%%%
\subsubsection{Velocity distribution functions}  \label{sec:VDF}

Instead of just a few DF moments at a given point, one may consider one-dimensional velocity distribution functions (VDFs):
\begin{align*}
\mathfrak{f}(\bx;v_1) &\equiv \frac{1}{\rho(\bx)} \iint \d v_2\,\d v_3\, f\big(\bJ(\bx,\bv)\big) ,\\
\mathfrak{f}_\mathrm{proj}(x_1,x_2;v_k) &\equiv \frac{1}{\Sigma(x_1,x_2)} \int \d x_3\, \rho(x_1,x_2,x_3)\,\mathfrak{f}(x_1,x_2,x_3;v_k) .
\end{align*}
Currently this is only implemented in cylindrical coordinates: $\bx=\{R,z,\phi\}, \bv=\{v_R,v_z,v_\phi\}$.
VDFs in each dimension are represented as B-splines of degree $N$: $\mathfrak{f}(\bx, v_k) = \sum_{\alpha=1}^{N_\mathrm{basis}} A_\alpha B_\alpha(v_k)$, where $B_\alpha$ are defined by the nodes of the grid in velocity space (provided by the user, but typically covering the entire available range of velocity with $\sim 100$ points). To compute the coefficients of expansion $A_\alpha$, we follow the finite-element approach by integrating the DF weighted with each basis function to obtain $f(\dots,v_k)\,B_\alpha(v_k)\,dv_k$, and solving the resulting linear system (Section~\ref{sec:MathSplineDetails}). Thus all three VDFs are computed at once in the course of a single 3-dimensional integration (or 4-dimensional for projected VDFs), which is, however, rather expensive (typically $\sim 10^6$ function evaluations). The simplest case $N=0$ corresponds to a familiar velocity histogram, but a more accurate one is given by $N=1$ (linear interpolation) or $N=3$ (cubic spline); note that in the latter case, the interpolated $\mathfrak{f}(v)$ may attain negative values, but on average better approximates the true VDF.
The VDFs or projected VDFs are computed by the routine \ttt{computeVelocityDistribution<N>}.

%%%%%%%%%%%%%%
\subsubsection{Conversion to/from \Nbody models}  \label{sec:Nbody}

As the DF is a probability distribution function (PDF), it can be sampled with a large number of points to create an \Nbody model of the system. There are two possible ways of doing this:
\begin{itemize}  \setlength{\parskip}{2pt} \setlength{\itemsep}{2pt}
\item Draw samples of actions from $f(\bJ)$, used as a three-dimensional PDF. Then create (possibly several) $\{\bx,\bv\}$ points for each value of actions with a random choice of angles, using the torus mapping approach (Section~\ref{sec:ActionsTorus}). This is performed by the routine \ttt{sampleActions}.
\item Draw samples directly from the six-dimensional $\{\bx,\bv\}$ space, evaluating $f\big(\bJ(\{\bx,\bv\})\big)$ with the help of an action finder. This is performed by the routine \ttt{samplePosVel}.
\end{itemize}
Both approaches should in principle deliver an equivalent discrete representation of the model, but may have a different cost; generally, the second one is preferred. It also has a separate, more efficient implementation for spherical isotropic DFs $f(h)$ in spherical potentials $\Phi(r)$; it is used by the \texttt{mkspherical} tool (Section~\ref{sec:mkspherical}).

There is also a related task for sampling just the density profile $\rho(\bx)$ with particles, without assigning any velocity to them; this may be used to visualize the density model, and is performed by the routine \ttt{sampleDensity} (of course, it does not use any action finder). All these tasks employ the adaptive multidimensional rejection method implemented in \ttt{math::sampleNdim}.

The inverse procedure for constructing a DF from a given \Nbody model is less well defined. In the case of a spherical isotropic system (Section~\ref{sec:DFsphericalIsotropic}), the one-dimensional function of phase volume $f(h)$ is estimated non-parametrically with the \hyperref[sec:SplineFitting]{penalized density fitting method} and represented as a spline in scaled coordinate (the routine \ttt{fitSphericalDF}). In principle this may be generalized for the case of a three-dimensional $f(\bJ)$, but this has not been implemented yet. The alternative is to fit a parametric DF to the array of actions, computed for the \Nbody particles in the given potential (of course, a suitable self-consistent \ttt{Multipole} or \ttt{CylSpline} potential itself may also be constructed from the same \Nbody model). This approach is demonstrated by one of the example programs (Section~\ref{sec:ExamplesTests}).

%%%%%%%%%%%%%%
\subsubsection{Iterative self-consistent modelling}  \label{sec:SCM}

As explained above, the same DF gives rise to a different density profile in each potential. A natural question is whether there always exists a unique potential-density pair $\rho,\Phi$ such that $\rho(\bx)=\int \d ^3v\,f\big(\bJ(\bx,\bv\;|\Phi)\big)$ corresponds to $\Phi(\bx)$ via the Poisson equation, with the mapping $\{\bx,\bv\}\implies \bJ$ constructed for the same potential.
While we are not aware of a strict mathematical proof, in most practical cases the answer is positive, and such potential may be constructed by the iterative self-consistent modelling approach \cite{Binney2014,Piffl2015}. 
In a more general formulation, one may have several DF components $f_c(\bJ), c=1..N_\mathrm{comp}$ and optionally several additional (external) density or potential components. The procedure consists of several steps, which use various pieces of machinery described previously:
\begin{enumerate}  \setlength{\parskip}{2pt} \setlength{\itemsep}{2pt}
\item Create a plausible initial guess for the total potential $\Phi(\bx)$.
\item Construct the action finder for this potential (Section~\ref{sec:ActionAngle}).
\item Compute the density profiles $\rho_c(\bx)$ of all components with DFs (Section~\ref{sec:Moments}).
\item Calculate the updated potential by solving the Poisson equation $\nabla^2\Phi = 4\pi\sum_c\rho_c$ for the combined density of all components (plus any external density or potential components such as a central supermassive black hole (SMBH), which are called static since they are not updated throughout the iterative procedure), using one or both general-purpose potential expansions (Sections~\ref{sec:PotentialMultipole},~\ref{sec:PotentialCylSpline}).
\item If desired, add new components or replace a static component with a DF-based one.
\item Repeat from step 2, until the potential changes negligibly between iterations. This typically requires $\mathcal{O}(10)$ steps.
\end{enumerate}

The only non-trivial aspect of this procedure is to choose whether the density of a given component is better described as spheroidal (not strongly flattened, possibly with a central cusp or an extended envelope) or disky (possibly strongly flattened, but with a finite-density core and a finite extent, or at least sharply declining at large radii). In the first case it will contribute to the potential represented by the \ttt{Multipole} expansion, and in the second -- by the \ttt{CylSpline} expansion. This applies to both DF-based and static density components; in addition there may be static components with already known potentials (e.g., a Plummer potential with a very small scale radius representing the central SMBH), which will be added directly to the total potential. Importantly, all disky components will be represented by a single \ttt{CylSpline} object, and similarly all spheroidal components by a single \ttt{Multipole} object.
The density of each DF-based component is first computed on a suitable grid of $\mathcal{O}(10^2-10^3)$ points, and a corresponding density interpolator (\ttt{DensitySphericalHarmonic} -- Section~\ref{sec:PotentialMultipoleDetails}, or \ttt{DensityAzimuthalHarmonic} -- Section~\ref{sec:PotentialCylSplineDetails}) is created that will be used in solving the Poisson equation. 
Presently this method is restricted to axisymmetric models, due to the lack of more general action finders.

This approach is implemented with the help of several classes derived from \ttt{BaseComponent}, the \ttt{SelfConsistentModel} structure which binds together the array of components, the potential, the action finder, and the parameters of potential expansions, and finally the routine \ttt{doIteration}, all defined in \texttt{galaxymodel_selfconsistent.h}. All these concepts are also available in the \Python wrapper (Section~\ref{sec:Python}), and a complete annotated example illustrating the entire workflow is presented both in the \Cpp and \Python variants.

%%%%%%%%%%%%%%
\subsubsection{Schwarzschild orbit-superposition modelling}  \label{sec:Schwarzschild}

Schwarzschild modelling approach \cite{Schwarzschild1979} is an alternative method for creating self-consistent models, in which the DF is determined numerically as a weighted superposition of individual building blocks. In the original approach, these blocks are numerically computed orbits (hence the alternative name ``orbit-superposition method''), which are essentially $\delta$-functions in the space of integrals of motion, but a more general definition might use finite-size building blocks, or bunches of individual orbits. In what follows, we use the original formulation.

There are several ingredients in these models:
\begin{itemize}  \setlength{\parskip}{2pt} \setlength{\itemsep}{2pt}
\item The total gravitational potential $\Phi(\bx)$ used to integrate the orbits.
\item One or more \ttt{Target} objects, which define various kinds of constraints: 3d density distribution, intrinsic (3d) or projected (line-of-sight) kinematics, etc. The required values of these constraints are denoted as $U_n^{(t)}$, with the index $t=1..N_\mathrm{tar}$ enumerating \ttt{Target}s, and $n=1..N_{\mathrm{cons},t}$ -- constraints of each target.
\item The orbit library (collection of $N_\mathrm{orb}$ orbits and their weights $w_i$ in the model), and associated arrays $u_{i,n}^{(t)}$ of contributions of $i$-th orbit to $n$-th constraint of $t$-th target.
\end{itemize}

The modelling workflow is split into several steps:
\begin{enumerate}  \setlength{\parskip}{2pt} \setlength{\itemsep}{2pt}
\item Initialize the potential $\Phi$ and $N_\mathrm{tar}$ \ttt{Target} objects.
\item Create initial conditions (IC) for the orbit library.
\item Integrate orbits while recording the arrays $u_{i,n}^{(t)}$.
\item Determine the orbit weights that satisfy the constraints.
\item (Optional) create an $N$-body representation of the model.
\end{enumerate}

The potential for Schwarzschild models needs to be specified in advance, in contrast to the DF-based iterative self-consistent models (Section~\ref{sec:SCM}). Often it will be computed from the deprojected surface density profile, e.g., parametrized by a Multi-Gaussian Expansion (MGE); there are several \Python routines for manipulating these parametrizations.

Target objects provide an abstract interface for discretizing both the data and the model into an array of constraints. There are 3 types (5 variants) of 3d density discretization schemes, described in the Appendix~\ref{sec:SchwarzschildDetails}, one target (4 variants) for representing the spherically averaged 3d kinematic profiles, and one target (4 variants) for recording the line-of-sight velocity distributions (LOSVD). All these schemes use $B$-splines (Section~\ref{sec:MathBSplineDetails}) for defining the discretization elements, and variants of the same scheme differ by the order of the $B$-spline basis.
A \ttt{Target} object does not contain any data itself, it only provides the methods for computing discretized representations of various other entities and storing them in external arrays.
For instance, a density target acting on a \ttt{Density} model produces the array of masses associated with each discretization element (in the simplest case, the mass contained in each cell of the 3d density grid), while a LOSVD target acting on a \ttt{Density} model computes the integrals of the PSF-convolved surface density profile over each spatial region (aperture) on the sky plane. The same LOSVD target applied to a \ttt{GalaxyModel} object computes the PSF-convolved LOSVDs produced by the combination of the DF and the potential in each aperture. Any target applied to an $N$-body snapshot computes the relevant quantities $U_n^{(t)}$ from the array of particles, weighted by their masses. And finally, any target can be attached to the orbit integrator to construct the discretized representaion of the $i$-th orbit $u_{i,n}^{(t)}$ (this is conceptually similar to recording the orbit as a collection of points sampled from the trajectory, with weights proportional to the time intervals between adjacent points, and then applying the target to this $N$-body snapshot, although in practice it is implemented on-the-fly, without actually storing the trajectory). The LOSVD target is used to constrain the model by observed kinematics, but this involves an additional step to convert the internal $B$-spline representation of the datacube into observable quantities (for details, see Appendix~\ref{sec:SchwarzschildDetails}).

The IC for the orbit library may be generated by one of the complementary approaches for constructing dynamical models. This is achieved by first sampling positions from the actual 3d density profile of the galaxy or one of its components, then assigning velocities drawn from a suitable DF or from a Jeans model. For spheroidal systems or galaxy components, the Eddington inversion or its anisotropic generalization (Section~\ref{sec:DFspherical}) provide a suitable DF, while for strongly flattened and rotating disk components (including bars), velocities may be drawn from a Gaussian distribution with the dispersions computed from the axisymmetric anisotropic Jeans equations. In either case, the resulting IC are not necessarily in equilibrium, but merely provide a convenient starting point for the orbit-based modelling. Moreover, one may stack together several sets of IC created with different parameters (e.g., to provide a denser sampling of orbits at high binding energies near the galactic center).

The orbit weights $w_i\ge 0$ that satisfy the constraints are determined from the system of linear equations
\begin{equation*}
\sum\nolimits_{i=1}^{N_\mathrm{orb}} w_i\, u_{i,n}^{(t)} = U_n^{(t)},\quad
t=1..N_\mathrm{tar}, \; n=1..N_{\mathrm{cons},t} .
\end{equation*}
In practice, the constraints may not be satisfied exactly, especially if they come from noisy observations. In this case, the best solution is obtained by minimizing the $L_2$-norm of the residual in the above equation system, weighted by the observational uncertainties $\epsilon_{U_n^{(t)}}$. Additionally, one may employ some sort of regularization to make the solution more well-behaved. In practice, this is achieved by adding a regularization term proportional to the sum of squared orbit weights to the objective function to be minimized, which encourages a more uniform distribution of orbit weights in the solution. The objective function is thus written as
\begin{equation*}
\mathcal Q \equiv \sum_{t=1}^{N_\mathrm{tar}}\, \sum_{n=1}^{N_{\mathrm{cons},t}} 
\left(
\frac{ \sum_{i=1}^{N_\mathrm{orb}} w_i\, u_{i,n}^{(t)} - U_n^{(t)} }{ \epsilon_{U_n^{(t)}} }
\right)^2 +\,
\lambda\,\sum_{i=1}^{N_\mathrm{orb}} \left(\frac{w_i}{w_{i,0}}\right)^2.
\end{equation*}
Here $\lambda$ is the regularization coefficient, $w_{i,0}$ are the priors on orbit weights (in the simplest case, uniform). If some constraints need to be satisfied exactly, the corresponding uncertainties should be set to zero; these constraints will not contribute to $\mathcal Q$. The minimization of the objective function, subject to the non-negativity constraints $w_i\ge 0$, is performed by the quadratic optimization solver CVXOPT.

In the case of constructing models constrained by observed LOSVD kinematics, one may use the same orbit library multiple times to represent systems with different mass normalizations $\Upsilon$, rescaling the velocities by $\sqrt{\Upsilon}$ as described in the Appendix~\ref{sec:SchwarzschildDetails}).

Finally, the orbit-superposition model can be converted into an $N$-body representation, e.g., to test its stability or provide initial conditions for a simulation. In order to do this, one needs to record the trajectories during orbit integration along with the other arrays $u_{i,n}^{(t)}$, and then select a certain number of points from each orbit's trajectory in proportion to its weight. In the case that the number of points recorded during orbit integration was insufficient to represent an orbit with a particularly high weight, this orbit needs to be re-integrated while storing the points more frequently. There is a \Python routine \ttt{sampleOrbitLibrary} that performs this task in a transparent way.

All computationally heavy operations of this approach are implemented in the \Cpp library, but the top-level modelling workflow is more conveniently expressed in \Python.

%%%%%%%%%%%%%%%%%%%%%%%%%%
%\subsection{Data handling}
%\subsubsection{Selection functions}

%%%%%%%%%%%%%%%%%%%%%%%%%%%%%%%%%%%%%%%%%%%%%%%%%%%%%%%%%%%%%%%%%%%%%%%%%%%%%%%%
\section{Interfaces with other languages and frameworks}  \label{sec:Interfaces}

%%%%%%%%%%%
\subsection{\Python interface}  \label{sec:Python}

The \Python interface provides a large subset of \Agama functionality expressed as Python classes and routines. Presently, this includes:
\begin{itemize}  \setlength{\parskip}{2pt} \setlength{\itemsep}{2pt}
\item A few mathematical tasks such as multidimensional integration and sampling, penalized spline fitting and density estimate, linear/quadratic optimization solvers.
\item Unit handling.
\item Potential and density classes.
\item Orbit integration.
\item Action finders (both classes and standalone routines).
\item Distribution functions.
\item Galaxy modelling framework: computation of DF moments, drawing samples from density profiles and DFs, iterative self-consistent and Schwarzschild orbit-superposition modelling.
\end{itemize}

The shared library \texttt{agama.so} can be used directly as a \Python extension module%
\footnote{for the most common implementation of \Python interpreter, named \texttt{CPython} -- not to be confused with \texttt{Cython}, which is a distinct compiled language. The \Python interface layer in \Agama relies solely on the \Python C API, and does not use any third-party libraries such as \texttt{swig} or \texttt{boost::python}.}. Both \Python 2.6--2.7 and \Python 3.x are supported, but the library must be compiled separately for either version. In addition, there are a few auxiliary routines written in \Python (they reside in \texttt{py/pygama.py}). To simplify the usage, the package initialization (\texttt{__init__.py}) imports everything from both \texttt{agama.so} and \texttt{py/pygama.py} into a single namespace, which is what you get by writing \texttt{import agama} in your \texttt{.py} file.

The \ttt{Density}, \ttt{Potential}, \ttt{DistributionFunction}, \ttt{Target} classes serve as universal proxies to the underlying hierarchy of \Cpp classes, and their constructors take a variety of named arguments covering all possible variants (including those requiring a more complicated setup in the \Cpp code).
Additionally, density and DF object may also be represented by an arbitrary user-defined Python function -- this can be used in all contexts where a corresponding interface is needed, e.g., in constructing a potential expansion from a density profile, or in computing DF moments, which greatly increases the flexibility of the \Python interface. Most routines or methods that operate on individual points in \Cpp (such as action finders or potentials) can accept \texttt{numpy} arrays in \Python, which again leads to a more concise code with nearly the same efficiency as a pure \Cpp implementation.Moreover, operations on such arrays are internally OpenMP-parallelized in the \Python extension module, except if they include callbacks to user-defined \Python functions%
\footnote{this restriction is due to the global interpreter lock mechanism in \texttt{CPython}, precluding its simultaneous access from multiple threads.}.

Below follows a brief overview of the classes and routines provided by the \Python interface.
As usual, writing \texttt{help(agama.Whatever)} brings up a complete description of the class or routine and its arguments. Moreover, there are several test and example programs demonstrating various aspects of usage of the \Python interface to \Agama; some of them have exact \Cpp equivalents.

\paragraph{Density} class instances can be constructed with the following syntax: \\
\texttt{d = agama.Density(type="...", mass=..., scaleRadius=..., otherParameters=...)}\\
using the parameters listed in Section~\ref{sec:PotentialFactory}; argument names are case-insensitive, as are the names of density models. 
Alternatively, to combine several density objects \texttt{d1}, \texttt{d2}, etc., one may list them as the unnamed arguments of the constructor (these could be proper Density objects, dictionaries with density parameters, or user-defined functions providing the Density interface, see below):\\
\texttt{comp = agama.Density(d1, dict(type='Plummer', mass=42), d2)}\\[2mm]
Elements of such composite density objects can be accessed by index or iterated over:\\
\texttt{for i in range(len(comp)): print(comp[i])}\\
\texttt{for d in comp: print d}\\[2mm]
Another possibility is to construct a spherically-symmetric density model from a cumulative mass profile -- a 2d array with two columns: $r$, $M(<r)$. The following example corresponds to a $\gamma=1$ Dehnen (aka Hernquist) profile:\\
\texttt{r = numpy.logspace(-3,3)}\\
\texttt{M = (r / (r+1))**2}\\
\texttt{d = agama.Density(cumulMass=numpy.column_stack((r, M)) )}\\[2mm]
In this case the \ttt{Density} object is internally represented by a \ttt{DensitySphericalHarmonic} \Cpp class. Both this class and \ttt{DensityAzimuthalHarmonic} are also utilized in the iterative self-consistent modelling framework (Section~\ref{sec:SCM}); such density can be written out to a text file and loaded back by\\
\texttt{d.export("hernquist_model.txt")}\\
\texttt{d = agama.Density("hernquist_model.txt")}\\[2mm]
The \ttt{Density} class provides only a couple of methods: first of all, the computation of the density itself -- the argument is a single point or a $N\times3$ array of $N$ points (the \Python interface always deals with cartesian coordinates to avoid confusion):\\
\texttt{print(d.density(1,0,0))}\\
\texttt{print(d.density([[1,0,0],[2,3,4]]))}\\[2mm]
\texttt{surfaceDensity} computes the integral of density along the line of sight in an arbitrarily rotated coordinate system, whose orientation is specified by Euler angles (Section~\ref{sec:CoordinateDetails}).\\[1mm]
Another useful method is \texttt{totalMass()}, with an obvious role; the mass may well be infinite, e.g., for a \ppp{type="NFW"} model.\\[1mm]
Finally, the density profile may be sampled with particles, producing two arrays -- coordinates ($N\times3$) and masses ($N$): \\
\texttt{pos, mass = d.sample(10000)}\\[2mm]
The density framework can be augmented with user-defined \Python functions that return the density at a given point (or, rather, an array of points). Such a function must be a free function (not a class method) or a lambda expression, accepting one argument, which should be treated as a $N\times3$ array of points (even when $N=1$). Such function would typically be called from the \Cpp code with more than one input point, to reduce overhead from switching between \Cpp and \Python. The following two equivalent examples define a spherical Plummer model:\\
\texttt{fnc1~~=~~lambda~~x:  3/(4*numpy.pi) * (1 + numpy.sum(x**2, axis=1))**-2.5}\\
\texttt{def fnc2(x):  return 3/(4*numpy.pi) * (1 + numpy.sum(x**2, axis=1))**-2.5}\\
These functions can be provided as an argument to the \ttt{Density} class constructor:\\
\texttt{print(agama.Density(fnc1).totalMass())}\\
Or they could be supplied directly in other places where a \ttt{Density} object is expected, such as the \ttt{Potential} and \ttt{DistributionFunction} constructors.

\paragraph{Potential} class is a further development of \ttt{Density} and provides access to the entire hierarchy of \Cpp potentials. It can be constructed in a variety of ways: \\
\texttt{p = agama.Potential(type="...", mass=..., otherParameters=...)}\\
i.e., in the same way as a \ttt{Density} object, using named arguments listed in Section~\ref{sec:PotentialFactory};\\
\texttt{p = agama.Potential("MWPotential2014.ini")}\\
reads the potential parameters from an INI file, where each section \ppp{[Potential1]}, \ppp{[Potential disk]}, etc., corresponds to a single component (see below for special rules about the construction of composite potentials).\\
\texttt{p = agama.Potential(file="potential_disk.coef_cyl")}\\
loads the coefficients of potential expansion previously stored in a text file by the \texttt{export()} method.\\
If \ppp{type="Multipole"} or \ppp{"CylSpline"}, then a potential expansion is constructed from the density profile provided in the \ppp{density=...} argument. This could be the name of a density model (e.g., \ppp{"Spheroid"}), or the instance of a \ttt{Density} class, or a user-defined function providing the Density interface as described above, or an $N$-body snapshot provided as a tuple of two arrays (coordinates and masses), or a file containing such \hyperref[sec:PythonSnapshot]{snapshot}:\\
\texttt{p1 = agama.Potential(type="Multipole", density="Plummer", axisRatioZ=0.5)}\\
\texttt{p2 = agama.Potential(type="Multipole", density=fnc1)}\\
\texttt{p3 = agama.Potential(type="Multipole", particles=(pos,mass))}\\
\texttt{p4 = agama.Potential(type="Multipole", file="nbody_snapshot.txt")}\\[2mm]
A composite potential can also be created from several other \ttt{Potential} objects:\\
\texttt{p5 = agama.Potential(p1, p2, p3)}\\[2mm]
Like a composite \ttt{Density}, elements of such composite potential can be accessed by index or iterated over.
Another way of creating a composite potential is to provide a list of \texttt{dict} instances containing the parameters for each potential to the constructor:\\
\texttt{disk_par = dict(type="Disk", mass=5, scaleRadius=3, scaleHeight=0.4)}\\
\texttt{bulge_par= dict(type="Sersic", mass=1, scaleRadius=1, axisRatioZ=0.6)}\\
\texttt{halo_par = dict(type="NFW", mass=70, scaleRadius=20, axisRatioZ=0.8)}\\
\texttt{potgal~~~= agama.Potential(disk_par, bulge_par, halo_par)}\\[2mm]
If we examine the potential created in the last line,\\
\texttt{print(pot_gal)}  \textit{\color{Sepia} \ \ \# CompositePotential: DiskAnsatz, Multipole}\\
it becomes apparent that some rearrangement took place behind the stage. Indeed, in the case when the potential is constructed from several sets of parameters (but \textit{not} from several existing potential instances), the code attempts to optimize the efficiency by using the \hyperref[sec:PotentialGalpot]{\textsc{GalPot}} approach. In this example, the \ppp{Disk} density profile was split into two parts -- the \ttt{DiskAnsatz} potential class and the residual density profile; other spheroidal density components (\ppp{Sersic} and \ppp{NFW}) were combined with this residual profile, and used to initialize a single instance of \ttt{Multipole} potential. This is advantageous if one needs to evaluate the potential many times (e.g., in action computation), but makes it difficult to examine the contribution of each mass component separately. In order to do so, we may instead create another potential used only for visualization:\\
\texttt{potvis~= agama.Potential(agama.Potential(disk_par), }\\
\texttt{\mbox{}~~~~agama.Potential(bulge_par), agama.Potential(halo_par))}\\[2mm]
An instance of \ttt{Potential} class provides the same methods as the \ttt{Density} class (and may be used in all places where a density instance is needed), plus the following ones:\\[1mm]
\texttt{p.potential(points)} evaluates the potential at one or several input points (which should be either a 1d list/array of length 3, or a 2d array of size $N\times3$);\\[1mm]
\texttt{p.force(points)} computes the acceleration (force per unit mass), i.e., \textit{minus} the derivative of potential, returning a $N\times3$ array;\\[1mm]
\texttt{p.forceDeriv(points)} computes the forces and force derivatives:\\
\texttt{r = numpy.linspace(0,20)}\\
\texttt{points = numpy.column_stack((r, r*0, r*0))}
\textit{\color{Sepia} \ \ \# a $N\times3$ array}\\
\texttt{force,deriv = potgal.forceDeriv(points)}
\textit{\color{Sepia} \ \ \# $\Rightarrow N\times3$ and $N\times6$ arrays} \\
\texttt{kappa = numpy.sqrt(-deriv[:,0] - 3*force[:,0]/r)}
\textit{\color{Sepia} \ \ \# radial epicyclic frequency $\kappa$} \\
\texttt{nu~~~~= numpy.sqrt(-deriv[:,2])}
\textit{\color{Sepia} \ \ \# vertical epicyclic frequency $\nu$} \\[1mm]
Plotting the rotation curve of the above constructed potential and all its components:\\
\texttt{plt.plot(r, numpy.sqrt(-r*potvis.force(points)[:,0]))}\\
\texttt{for pot in potvis: plt.plot(r, numpy.sqrt(-r*pot.force(points)[:,0]))}\\[1mm]
\texttt{p.Tcirc(E)} computes the characteristic time (period of a circular orbit with the given energy); for convenience it may also be called with an $N\times6$ input array of position/velocity coordinates: \texttt{p.Tcirc(points)}.\\[1mm]
\texttt{p.Rcirc(E=...)} or \texttt{p.Rcirc(L=...)} return the radius of a circular orbit in the equatorial plane corresponding to the given energy or $z$-component of the angular momentum;\\[1mm]
\texttt{p.Rmax(E)} returns the maximum radius accessible with the given energy, i.e., the root of $\Phi(R_\mathrm{max})=E$;\\[1mm]
\texttt{p.Rperiapo(E, L)} returns the pericenter and apocenter radii for an orbit with the given energy and angular momentum in the equatorial plane of an axisymmetric potential; for convenience, it may also be called with an $N\times6$ input array of position/velocity coordinates, and of course, with a $N\times2$ array of $E,L$ values for multiple input points (all methods can accept vectorized input).

\paragraph{ActionFinder} is the \Python class constructed for the given potential:\\
\texttt{af = agama.ActionFinder(pot, [interp=True|False])}\\
where the optional second argument chooses between the interpolated (faster, less accurate) and non-interpolated version of St\"ackel fudge. If the potential is spherical, the underlying \Cpp implementation will use the \ttt{ActionFinderSpherical} class, otherwise the \ttt{ActionFinderAxisymFudge} class (interpolated or not).\\
This class has only one method \texttt{__call__()}, computing actions, and optionally angles and frequencies, for the given 6d position/velocity point or an $N\times6$ array of points:\\
\texttt{act = af(points)}
\textit{\color{Sepia}\ \ \# 0th column is $J_r$, 1st -- $J_z$, 2nd -- $J_\phi$} \\
\texttt{act, ang, freq = af(points, angles=True)}\\[2mm]
There is also a standalone routine \texttt{actions()}, which may be used without constructing an instance of action finder:\\
\texttt{agama.actions(points, potential [, fd=focal_distance] [, angles=True])}\\
for a non-spherical potential one needs to provide the focal distance $\Delta$ (the \ttt{ActionFinder} class retrieves it from an internally constructed interpolation table).

\paragraph{ActionMapper} class performs the inverse operation -- transform from actions/angles to position/velocity coordinates. Currently it uses the Torus machinery (Section~\ref{sec:ActionsTorus}), and needs to be constructed for a given potential \textit{and} the triplet of actions:\\
\texttt{am = agama.ActionMapper(pot, (Jr, Jz, Jphi))}\\
When applied to one or more triplets of angles, it returns the corresponding $\bx,\bv$:\\
\texttt{xv = am([[theta_r1, theta_z1, theta_phi1], [theta_r2, theta_z2, theta_phi2]])}

\paragraph{DistributionFunction} class provides the \Python interface to the hierarchy of \Cpp DF classes. It is constructed either from a list of keyword arguments,\\
\texttt{df = agama.DistributionFunction(type="...", [norm=..., other params])}\\
where \ppp{type} may be one of the following: \ppp{DoublePowerLaw}, \ppp{QuasiIsothermal}, \ppp{Exponential}, \ppp{QuasiSpherical}, and the other parameters are specific to each type of DF (Section~\ref{sec:DF}), \\
or from a list of existing \ttt{DistributionFunction} objects, creating a composite DF:\\
\texttt{dfcomp = agama.DistributionFunction(df1, df2, myfnc)}\\[2mm]
Similarly to the \ttt{Density} class, one may provide a custom \Python function \texttt{myfnc} which returns the DF value at the given triplet of actions $\{J_r,J_z,J_\phi\}$ (again the calling convention is to process a 2d array of such triplets, even if with one row).\\[2mm]
The \ppp{QuasiIsothermal} DF (Section~\ref{sec:DFdisk}) additionally needs an instance of \ttt{Potential} to initialize the auxiliary functions $R_c(J)$, $\kappa,\nu,\Omega,\Sigma,\sigma$.\\[2mm]
The \ppp{QuasiSpherical} DF (Section~\ref{sec:DFspherical}) is constructed from the provided instances of \ttt{Density} and \ttt{Potential}, using the generalized Eddington inversion formula to create $f(E,L)$ and then a spherical action finder to convert it to an action-based form.

\paragraph{GalaxyModel} is the combination of a potential, an action finder, and a DF:\\
\texttt{gm = agama.GalaxyModel(pot, df)}
\textit{\color{Sepia}\ \ \# action finder is constructed automatically} \\[2mm]
This class provides methods for computing DF moments and drawing position/velocity samples from it. These operations are rather expensive, and if the input consists of several points, the computation is internally parallelized using \texttt{OpenMP} (except when the DF is a user-defined Python function).\\[2mm]
\texttt{dens, meanvel, vel2 = gm.moments(points, dens=True, vel=True, vel2=True)}\\
computes the density, mean $v_\phi$, and six second moments of the velocity at the provided point(s) (Section~\ref{sec:Moments}); one may choose which of these quantities are needed, eliminating unnecessary computations.\\[2mm]
\texttt{Sigma, rmsheight, sigmaz = gm.projectedMoments(r)}\\
computes the projected quantites in the equatorial plane (surface density, r.m.s. height, and vertical velocity dispersion).\\[2mm]
\texttt{fvr, fvz, fvphi = gm.vdf(points [, gridv])}\\
constructs 1d velocity distribution functions at the given point(s) (Section~\ref{sec:VDF}), represented by cubic spline interpolators. The optional second argument specifies the grid in velocity space used to represent the spline; if not provided, the default is to cover the range $\pm v_\mathrm{escape}$ with 50 points. The grid needs not be too dense -- a 3rd degree interpolating spline provides a substantially higher level of detail than an equivalently spaced histogram.
One may plot the velocity distribution as a smooth function on a denser grid:\\
\texttt{v_esc = numpy.sqrt(-2 * pot.potential(point))\\
gridv = numpy.linspace(-v_esc, v_esc, 200)\\
plt.plot(gridv, fvr(gridv))} \\[2mm]
\texttt{posvel, mass = gm.sample(N)}\\
draws $N$ equal-mass samples from the DF (Section~\ref{sec:Nbody}); the result is an $N\times6$ array of position/velocity points and a 1d array of masses (its values are approximately \texttt{df.totalMass()/N}, but here the total mass is computed by the sampling routine). This routine is used for constructing an $N$-body representation of the DF-based model.\\[2mm]
One may also sample positions from a given density profile, and then assign velocities using either the spherical anisotropic DF (Section~\ref{sec:DFspherical}) or axisymmetric Jeans equations. This is currently achieved by the \texttt{sample} method of a \ttt{Density} object, which additionally takes an instance of the \ttt{Potential} and optionally the parameters of Jeans equations. Be cautioned that the API will be changed in the future, to harmonize the usage conventions with those used for action-based \ttt{GalaxyModel}s.

\paragraph{SelfConsistentModel} is the driver class for performing iterative self-consistent modelling (Section~\ref{sec:SCM}). It is initialized with a list of arguments determining the parameters of the two potential expansions (\ttt{Multipole} and \ttt{CylSpline}) that are constructed in the process. To run the model, one needs to add one or more components; the list of components may be changed between iterations.\\
\texttt{params = dict(rminSph=1e-3, rmaxSph=1e3, sizeRadialSph=40, lmaxAngularSph=0)}\\
\texttt{scm = agama.SelfConsistentModel(**params)}

\paragraph{Component} class is a single component of this model; it could either represent a static density or potential profile, or provide a DF which will contribute to the density used to compute either of the two potential expansions.\\[2mm]
\texttt{comp = agama.Component(df=df, density=initdens, disklike=False, **params)}\\
creates the component with a spheroidal DF and an initial guess for the density profile;\\[2mm]
\texttt{scm.components.append(comp)}\\
adds the component to the model (\texttt{scm.components} is a simple \Python list);\\[2mm]
\texttt{scm.iterate()}\\
performs one iteration of the modelling procedure, recomputing the density of all components and then reinitializing the total potential.\\[2mm]
\texttt{comp.getDensity()}\\
returns the density of this component; \\[2mm]
\texttt{scm.potential}\\ is the instance of the total potential, which may be combined with the DF of each component into a \ttt{GalaxyModel} object, and used to compute other DF moments or construct an $N$-body model:\\
\texttt{posvel, mass = agama.GalaxyModel(scm.potential, df).sample(10000)}

\paragraph{Target} class represents one of several possible targets in Schwarzschild models. It is initialized by providing \ppp{type="..."} and other parameters depending on the target type (see Section~\ref{sec:SchwarzschildDetails}). This object can be used in two contexts: either as an argument for the \texttt{orbit} routine, collecting the contribution of each orbit to each constraint in the target, or as a function applied to a \ttt{Density} or \ttt{GalaxyModel} object or an $N$-body snapshot, returning the array of constraints computed from this object.

\paragraph{Orbit} integration is performed by the following routine:\\
\texttt{result = agama.orbit(potential=pot, ic=posvel, time=int_time, ...)}\\
here \ppp{posvel} contains initial conditions for one or several orbits (a $N\times6$ array), integration \ppp{time} for each orbit may be different, but typically is a multiple of the dynamical time returned by the \texttt{Tcirc} method of \ttt{Potential} (e.g., \texttt{100*pot.Tcirc(posvel)}), and dots indicate additional parameters (at least some must be provided to produce a result): \ppp{Omega} specifies the pattern speed of a rotating coordinate system, \ppp{trajsize} is the number of output points $M_k$ recorded from the trajectory for each orbit $k$ (may be a single number or vary between orbits), \ppp{lyapunov=True} additionally estimates the Lyapunov exponent for each orbit (an indicator of chaos), and \ppp{targets=...} optionally lists \ttt{Target} objects for Schwarzschild modelling.
The \texttt{result} returned by this routine is one or several arrays, depending on the requested output. For each \ttt{Target}, a $N\times K$ array is produced with the contribution of each orbit to each of $K$ constraints in the given target. When the output trajectory is requested by providing a nonzero \ppp{trajsize}, it is returned as either a tuple of length 2 (for a single orbit) or an $N\times 2$ array (for $N$ orbits), with the first element of each row $k$ containing a 1d array of times $t_i |_{i=0}^{M_k-1}$ at which the trajectory is stored, and the second element -- a 2d array ($M_k\times6$) containing the trajectory itself. Note that the entire ensemble of orbit is not a single \texttt{numpy} array, but an array of arrays, because each orbit may have a different size.
For instance, to plot the time evolution of $z$ coordinate of each orbit, one may use\\
\texttt{for times,trj in result: plt.plot(times, trj[:,2])}\\
and to plot the meridional $(R-z)$ cross-section of each orbit, one may use\\
\texttt{for trj in result[:,1]: plt.plot((trj[:,0]**2 + trj[:,1]**2)**0.5, trj[:,2])}\\
Finally, when \texttt{lyapunov=True}, another array of length $N$ is returned, containing the estimates of Lyapunov exponents for each orbit.

\paragraph{sampleOrbitLibrary} routine constructs an $N$-body snapshot from the orbit library of a Schwarzschild model, in which each orbit has a weight assigned by the \ttt{solveOpt} routine (see Section~\ref{sec:SchwarzschildDetails} for details):\\
\texttt{
matrix, traj = agama.orbit(potential=pot, ic=posvel, time=int_time, \\
\mbox{}~~~~trajsize=1000, targets=[target])\\
weights = agama.solveOpt(matrix, rhs)\\
nbody = 1000000\\
status, result = agama.sampleOrbitLibrary(nbody, traj, weights)}\\
The previously recorded trajectories returned by the \ttt{orbit} routine contain a certain number of points each, which may not be sufficient to sample orbits with particularly high weights. In this case, the function fails (returns \texttt{status=False}), and \texttt{result} contains the list of orbit indices which did not have enough points in previously recorded trajectories, and corresponding required numbers of samples for each orbit in this list. Then one should rerun the \ttt{orbit} routine with these parameters:\\
\texttt{
if not status:\\
\mbox{}~~~~indices, trajsizes = result\\
\mbox{}~~~~traj[indices] = agama.orbit(potential=pot, ic=posvel[indices], \\
\mbox{}~~~~~~~~time=int_time[indices], trajsize=trajsizes)\\
\mbox{}~~~~status, result = agama.sampleOrbitLibrary(nbody, traj, weights)}\\
In case of success (\texttt{status=True}), the \texttt{result} array contains a tuple of two arrays: \texttt{nbody}$\times6$ coordinates/velocities and \texttt{nbody} masses of particles in the $N$-body snapshot.

\paragraph{N-body snapshot handling} \label{sec:PythonSnapshot} is very rudimentary; the routines\\
\texttt{agama.writeSnapshot(filename, particles[, format])} and\\ \texttt{agama.readSnapshot(filename)} can deal with text files (7 columns -- $x,y,z,vx,vy,vz,m$), and optionally \Nemo or \textsc{Gadget} snapshots if the library was compiled with their support. Here \ppp{particles} is a tuple of two arrays: $N\times6$ position/velocity points and $N$ masses; the same convention is used to pass around snapshots in the rest of the \Python extension (e.g., in potential and sampling routines). A more powerful framework for dealing with $N$-body snapshots is provided, e.g., by the \textsc{Pynbody} library \cite{Pynbody}.

\paragraph{Unit handling} is optional: if nothing is specified explicitly, the library operates with the natural $N$-body units ($G=1$). However, the user may set up a unit system with three independent basic dimensional units (mass, length and velocity), and all dimensional quantities in Python will be expressed in these basic units and converted into natural units internally within the library:\\
\texttt{agama.setUnits(mass=1, length=1, velocity=1)}\\
instructs the library to work with the commonly used Galactic units: $1\,M_\odot$, 1\,kpc, 1\,km/s, with the derived unit of time being 0.98~Gyr (this is \textit{not} the same as using no units at all, because $G=4.3\times10^{-6}$ in these units). Importantly, this setup needs to be performed at the beginning of work, otherwise the values returned by the previously constructed classes and methods (e.g., potentials) would be incorrectly scaled.
At the moment there is no way to explicitly attach the units to the dimensional quantities passed to or returned from the library: for instance, \texttt{posvel} would be still a plain \texttt{numpy} array, with the implied convention that the first three columns are expressed in the length unit (e.g., 1~kpc) and the second three columns -- in velocity units (e.g., km/s), but it carries no attributes containing this information. In the future, the unit system may be integrated with the one from the \texttt{astropy} framework \cite{Astropy}.

\paragraph{Mathematical methods} provided by the \Python extension module include:\\
\texttt{integrateNdim} routine for multidimensional integration (an alternative interface to the \texttt{cubature} library which is included in the \Cpp code);\\
\texttt{sampleNdim} routine for sampling from a user-defined multidimensional function;\\
\texttt{splineApprox} and \texttt{splineLogDensity} routines for constructing a penalized cubic spline approximation or density estimate from discrete samples (Section~\ref{sec:SplineFitting});\\
\texttt{nonuniformGrid} and \texttt{symmetricGrid} routines for constructing one- and two-sided semi-exponentially-spaced arrays;\\
\texttt{bsplineInterp}, \texttt{bsplineMatrix}, \texttt{bsplineIntegrals} routines for dealing with B-splines;\\
\texttt{ghInterp}, \texttt{ghMoments} routines for dealing with Gauss--Hermite moments (see Section~\ref{sec:SchwarzschildDetails} for \hyperref[sec:SchwarzschildExample]{examples of usage} on the last two groups).

\begin{figure}
\begin{center}
\includegraphics{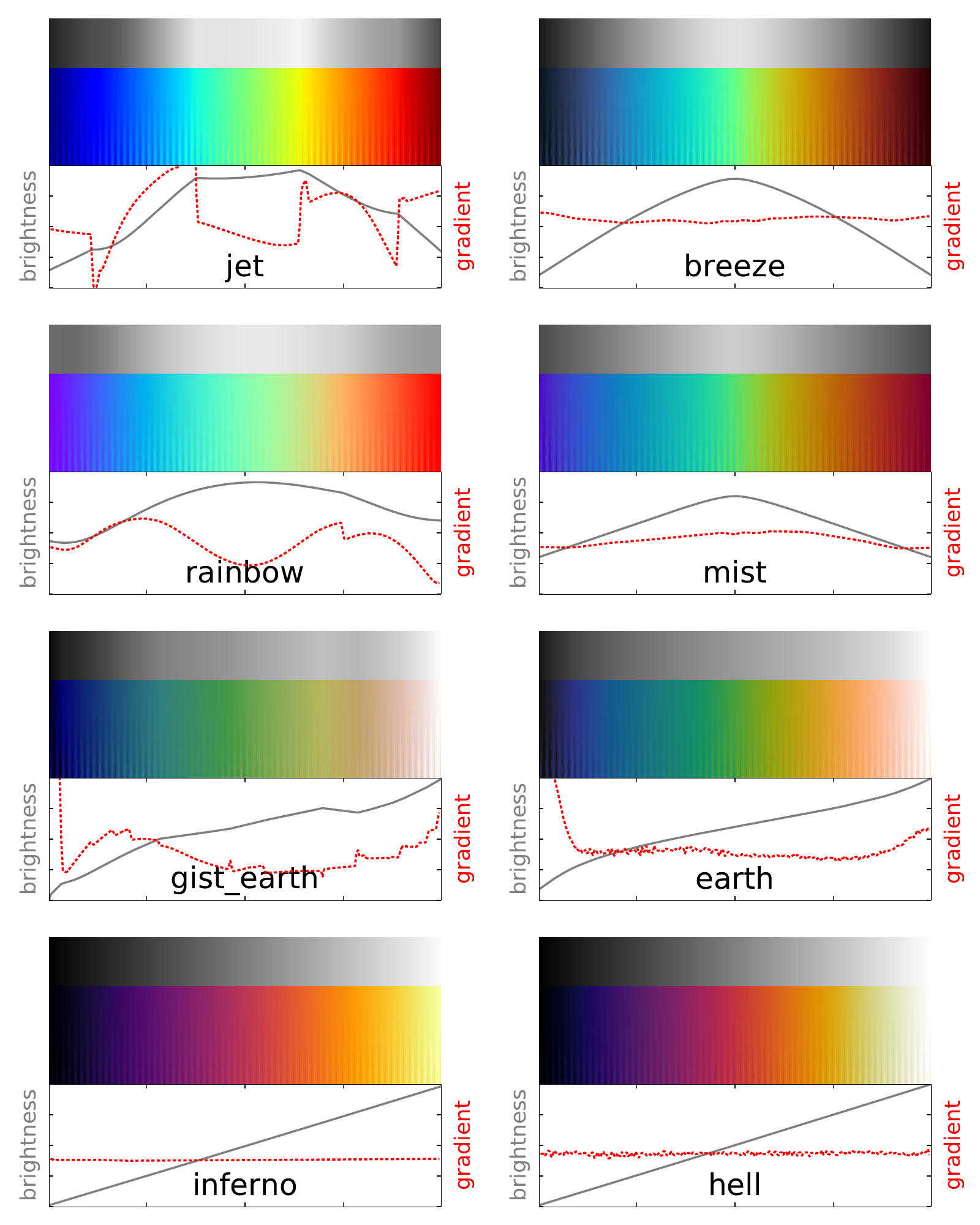}
\end{center}
\vspace*{-5mm}
\caption{Custom colormaps in \Agama (right) compared to the ones from \textsc{Matplotlib}.
}  \label{fig:colormaps}
\end{figure}

\paragraph{Colormaps} augment the rich collection of color schemes included in \texttt{matplotlib} with several custom-designed maps, which are better-balanced analogues of \texttt{jet}, \texttt{rainbow}, \texttt{gist_earth} and \texttt{inferno} maps (Figure~\ref{fig:colormaps}). They can be accessed from any \texttt{matplotlib} function under the names \ppp{breeze}, \ppp{mist}, \ppp{earth} and \ppp{hell}, respectively.

%%%%%%%%%%%
\subsection{\Fortran interface}  \label{sec:Fortran}

The \Fortran interface is much more limited compared to the \Python interface, and provides access to the potential solvers only. 

One may create a potential in several ways:
\begin{enumerate}  \setlength{\parskip}{2pt} \setlength{\itemsep}{2pt}
\item Load the parameters from an INI file (one or several potential components).
\item Pass the parameters for one component directly as a single string argument.
\item Provide a \Fortran routine that returns a density at a given point, and use it to create a potential approximation with the parameters provided in a text string.
\item Provide a \Fortran routine that returns potential and force at a given point, and create a potential approximation for it in the same way as above (this is useful if the original routine is expensive).
\end{enumerate}
Once the potential is constructed, the routines that compute the potential, force and its derivatives (including density) at any point can be called from the \Fortran code. No unit conversion is performed (i.e., $G=1$ is implied).
There is an example program showing all these modes of operation.

%%%%%%%%%%%
\subsection{\Amuse plugin}  \label{sec:Amuse}

\Amuse \cite{PortegiesZwart2013} is a heterogeneous framework for performing and analyzing \Nbody simulations using a uniform approach to a variety of third-party codes. The core of the framework and the user scripts are written in \Python, while the community modules are written in various programming languages and interact with each other using a standartized interface.

\Agama may be used to provide an external potential to any \Nbody simulation running within \Amuse. The plugin interface allows to construct a potential using either any of the built-in models, or a potential approximation constructed from an array of point masses provided from the \Amuse script. This potential presents a \ttt{GravityFieldInterface} allowing it to be used as a part of the \texttt{Bridge} coupling scheme in the simulation. For instance, one may study the evolution of a globular cluster that orbits a parent galaxy, by following the internal dynamics of stars in the cluster with an \Nbody code, while the galaxy is represented by a static potential using this plugin. This application is illustrated in the example script \texttt{py/example_amuse.py}.

%%%%%%%%%%%
\subsection{\Galpy plugin}  \label{sec:Galpy}

\Galpy \cite{Bovy2015} is a \Python-based framework for galaxy modelling, similar in scope to \Agama. It includes a collection of gravitational potentials, routines for orbit integration, action computation, distribution functions and more. 
The potential solvers and action finders from \Agama may be seamlessly integrated into the \Galpy framework with the help of a compatibility class \ttt{agama.GalpyPotential}, which is a subclass of both \ttt{agama.Potential} and \ttt{galpy.potential.Potential}. It is constructed in the same way as a regular \Agama potential and provides a \Galpy-compatible interface for any potential type from \Agama (note however that it cannot be used to wrap a native \Galpy potential into \Agama-compatible form, but in most cases it is possible to find an analogue among native \Agama types, or simply provide a user-defined density function wrapping a \Galpy density object to a \ppp{Multipole} or \ppp{CylSpline} potential solver from \Agama). This class provides methods from both interfaces (fortunately, their names do not coincide), and can be used in all relevant contexts in both frameworks. Of course, overlapping functionality (e.g., orbit integration or action computation) is much more efficiently performed using \Agama routines, avoiding in particular the overhead of repeatedly passing the control flow between \Python and \Cpp.

An example, comparing the native \Galpy action finder with that from \Agama, is provided in the file \texttt{py/example_galpy.py}. Overall, the potential approximations and action finders in \Agama are more versatile, accurate and computationally efficient, while \Galpy provides a convenient plotting interface.

%%%%%%%%%%%
\subsection{\Nemo plugin}  \label{sec:Nemo}

\Nemo \cite{Teuben1995} is a collection of programs for performing and analyzing \Nbody simulations, which use common data exchange format and UNIX-style pipeline approach to chain together several processing steps. The centerpiece of this framework is the \Nbody simulation code \textsc{gyrfalcON} \cite{Dehnen2000}. It computes the gravitational force between particles using the fast multipole method, and can optionally include an external potential.

The \Nemo plugin allows to use any \Agama potential as an external potential in \textsc{gyrfalcON} and other \Nemo programs (in a similar context as the \Amuse plugin). The potential may be specified either as a file with coefficients (for potential expansions), or more generally, as an INI file with parameters of possibly several components defined in groups \ppp{[Potential1]}, \ppp{[Potential whatever]}, \dots

To build the plugin, one needs to have \Nemo installed (obviously) and the environment variable \$NEMO defined; then \texttt{make nemo} will compile the plugin and place it in \texttt{\$NEMO/obj/acc} folder, where it can be found by \Nemo programs. For instance, this adds an extra potential in a \textsc{gyrfalcON}  simulation:\\
\texttt{\$ gyrfalcON infile outfile accname=agama accfile=mypot.ini [accpars=1.0] \dots}\\
where the last optional argument specifies the pattern speed $\Omega$ (frequency of rotation of the potential figure about $z$ axis). All units in the INI or coefs file here should follow the convention $G=1$.

%%%%%%%%%%%%%%%%%%%%%%%%%%%%%%%%%%%%%%%%%%%%%%%%%%%%%%%%%%%%%%%
\section{Tests and example programs}  \label{sec:ExamplesTests}

The \Agama library itself is indeed just a ``library'', not a ``program'', but it comes with a number of example programs and internal tests. The latter ones are intended to ensure the consistency of results as the development goes on, so that new or improved features do not break any existing code. All \texttt{test_***.cpp} and \texttt{test_***.py} programs are intended to run reasonably quickly and display either a \textcolor{Green}{PASS} or \textcolor{Red}{FAIL} message; they also illustrate some aspects of the code, or check the accuracy of various approximations on realistic data. Example programs, on the other hand, are more targeted towards the library users and demonstrate how to perform various tasks. Finally, there are several tools built on top of the library that perform some useful tasks themselves. Some of them are described below.

\paragraph{example_galpy} is a \Python program showing the use of \Agama plugin for \Galpy to construct a potential, integrate orbits, and compare the accuracy of action finders between the two libraries. 

\paragraph{example_amuse} illustrates the use of \Agama to provide an external potential in an $N$-body simulation performed within the \Amuse framework.

\paragraph{example_fortran} demonstrates how to create and use \Agama potentials in \Fortran, both for built-in density or potential models, or for user-defined \Fortran functions that provide the density or potential.

\paragraph{example_deprojection} is an interactive \Python plotting script illustrating the projection and deprojection of triaxial ellipsoidal bodies viewed at different orientations.

\paragraph{example_smoothing_spline} is a \Python program showing the use of penalized smoothing splines for fitting a curve to noisy data (\ttt{splineApprox}) and for estimating 1d probability distributions from discrete samples (\ttt{splineLogDensity}).

\paragraph{example_actions_nbody} shows how to determine the actions for particles from an \Nbody snapshot taken from a simulation of a disk+halo system. It first reads the snapshot and constructs two potential approximations -- \ttt{Multipole} for the halo component and \ttt{CylSpline} for the disk component -- from the particles themselves. Then it computes the actions for each particle and writes them to another file. This program exists both in \Cpp and \Python variants that perform the same task.

\paragraph{example_torus} is a \Python script illustrating the use of the \ttt{ActionMapper} class to construct an orbit, comparing it to a numerically integrated trajectory.

\paragraph{example_df_fit} shows how to find the parameters of a DF belonging to a particular family from a collection of points drawn from this DF in a known potential. It first computes the actions for these points, and then uses the multidimensional minimization routine \ttt{findMinNdim} to locate the parameters which maximize the likelihood of the DF given the data.
The \Python equivalent of this program additionally determines the confidence intervals on these parameters by running a MCMC algorithm starting around the best-fit parameters.

A more elaborate \Python program \texttt{gc_runfit} determines simultaneously the parameters of the spherically-symmetric potential and the DF that together describe the mock data points drawn from a certain (non-self-consistent) DF but with incomplete data (only the line-of-sight velocity and the projected distance from the origin). The goal is to determine the properties of the potential, treating the DF as nuisance parameters; it also uses the MCMC algorithm to determine uncertainties. 
This program is designed to work with the mock data from the \href{http://astrowiki.ph.surrey.ac.uk/dokuwiki/doku.php?id=tests:sphtri:spherical}{Gaia Challenge} test suite \cite{Read2020}, but can be easily adapted to other situations.

\paragraph{example_doublepowerlaw} performs a related task: given a spherically-symmetric density and potential profiles (not necessarily related through Poisson equation), numerically construct a \ttt{QuasiSpherical} DF and approximate it with a \ttt{DoublePowerLaw} DF which has an analytic form (see \cite{Jeffreson2017} for a similar approach).

\paragraph{example_self_consistent_model} illustrates various steps of the workflow for creating multicomponent self-consistent galaxy models determined by DFs. It begins with initial guesses for the density profiles of all components, and computes the total potential, which is used to construct a \ttt{QuasiIsothermal} DF for the disk; the \ttt{DoublePowerLaw} DFs of the halo and the bulge do not need a potential. Then it performs several iterations, updating the density of both disk and halo components and recomputing the total potential. Finally, it creates an \Nbody realization of the composite system by sampling particles from both DFs in the converged potential. It also demonstrates the use of INI files for keeping parameters of the model. This example is provided in equivalent \Cpp and \Python versions.

There are a couple of closely related \Python programs:\\
\texttt{example_self_consistent_model3} performs the same task for a slightly different three-component galactic model fully specified by DFs and plots several physical quantities in the model (velocity dispersion profiles, LOSVDs, etc.);\\
\texttt{example_self_consistent_model_simple} is a stripped-down version showing only the bare minimum of steps in the context of a spherical model.

\paragraph{example_schwarzschild_simple} is a \Python script illustrating the basic steps in Schwarzschild orbit-superposition  modelling. It creates a self-consistent triaxial model with a Dehnen density profile, and then converts it into an $N$-body snapshot.

\paragraph{schwarzschild} is a more general \Python program for constructing multicomponent Schwarzschild models and converting them into $N$-body models. It reads all model parameters from an \texttt{ini} file (an example of a three-component axisymmetric disk galaxy model is provided in \texttt{data/schwarzschild_axisym.ini}). This program can be used in the ``theoretical'' context, when the goal is to construct a single equilibrium model with given parameters, not to fit a model to some observational data -- that job is performed by the following program.

\paragraph{example_forstand} is a \Python program illustrating various aspects of observationally-constrained Schwarzschild modelling. It creates mock datasets from $N$-body models, runs a grid of models, and displays results in an interactive plot. This program can serve as a template for user scripts adapted to particular observational datasets.

\phantomsection\label{sec:mkspherical}%
\paragraph{mkspherical} is a \Cpp program for creating and analyzing spherical isotropic models. These models are defined by a potential $\Phi(r)$ and a distribution function $f(E)$, or rather, $f(h)$, where $h(E)$ is the phase volume (Section~\ref{sec:DFsphericalIsotropic}). These models may or may not be self-consistent, i.e., the potential may be generated by the density profile corresponding to $f(h)$, but also contain other external components (e.g., a central massive black hole). This tool can be used in two distinct modes: (a) creating a spherical model with prescribed properties (given by a built-in density profile, or interpolated from a user-provided table of enclosed mass within a range of radii), using the Eddington inversion formula to compute the distribution function, or (b) analyzing an \Nbody snapshot (constructing smooth spline approximations to the spherical potential, density and isotropic distribution function). The output consists of a text table with several variables ($\Phi$, $\rho$, $f$, etc.) as functions of radius or energy, and/or an \Nbody realization of the model, which may then serve as the initial conditions in a simulation.

\paragraph{phaseflow} is a \Cpp program for computing the dynamical evolution of a spherical isotropic stellar system driven by two-body relaxation \cite{Vasiliev2017}. It solves a coupled system of Fokker--Planck and Poisson equations for the joint evolution of $\Phi(r), \rho(r)$ and $f(h)$ discretized on a grid, using the formalism presented in Section~\ref{sec:DFsphericalIsotropicDetails}. It reads all input parameters from an \texttt{ini} file; a couple of examples are given in \texttt{data/phaseflow_corecollapse.ini} and \texttt{data/phaseflow_bahcallwolfcusp.ini}

\paragraph{raga} is a Monte Carlo stellar-dynamical code for simulating the evolution of non-spherical stellar systems \cite{Vasiliev2015}. It represents the system as a collection of particles that move in the smooth potential, represented as a \ttt{Multipole} expansion with coefficients being regularly recomputed from the particles themselves, and can include several additional dynamical processes: explicitly simulated two-body relaxation, loss-cone effects (capture of particles by a massive black hole), and interaction between a binary massive black hole and the stellar system. The code is described in a separate document (\texttt{readme_raga.pdf}), and an example input file is provided in \texttt{data/raga.ini}.

\newpage
%%%%%%%%%
\appendix

%%%%%%%%%%%%%%%%%%%%%%%%%%%
\section{Technical details}

%%%%%%%%%%%%%%%%%%%%%%%%%%%%%%%%%%%%%%%%%%%%%%%%%%%%%%%%%%
\subsection{Developer's guide}  \label{sec:DeveloperGuide}

Any large piece of software needs to follow a number of generic programming rules, which are well-known standards in commercial software development, but unfortunately are insufficiently widespread in the scientific community. Here we outline the most important guidelines adopted in the development of \Agama. Some of them are \Cpp-specific \cite{Meyers,SutterAlexandrescu}, others are more general \cite{Martin,McConnell}.
As a practical matter, we do not use any of \CppII features, except \hyperref[sec:SmartPointers]{smart pointers}, to keep compatibility with older compilers.

\paragraph{Code readability} is extremely important in long-term projects developed and used by several persons. All public classes, types and routines in \Agama are documented in-code, using the \textsc{Doxygen} syntax for comments that can be parsed and used to automatically generate a collection of HTML pages. These comments mostly describe the intent of each class and function and the meaning of each argument or variable, at least in their public interface -- in other words, a programmer's reference to the library. Sometimes a more technical description is also provided in these comments, but generally it is more likely to be presented in this document rather than in the code (with the inevitable risk of de-synchronizing as the code development progresses...)

\paragraph{Modularity} is an essential approach for keeping the overall complexity at a reasonable level. What this means in practice is that each unit of the code (a class or a function) should be responsible for a single well-defined task and provide a minimal and clean interface to it, isolating all internal details. The calling code should make no assumptions about the implementation of the task that this unit of code is promised to deliver. On many occasions, there are several interchangeable back-ends for the same interface -- this naturally applies to all class hierarchies descending from a base abstract class such as \ttt{BasePotential}, but also to the choice of back-end third-party libraries dictated by compilation options, with a single wrapper interface to all alternatives implementations.

Another facet of modularity is loose coupling, that is, instead of a single large object that manages many aspects of its internal state, it is better to create a number of smaller objects with minimal necessary interaction. For instance, composition (when one class has another class as a member variable) is preferred over inheritance (when the class has full access to the parent class's private members), as it reduces the strength of coupling.

\paragraph{Programming paradigm} throughout the library is a mixture of object-oriented and procedural, gently spiced with template metaprogramming.\\
Generally, when there is a need to provide a common interface to a variety of implementations, the choice between compile-time (templates) and run-time (virtual functions) polymorphism is dictated by the following considerations.\\
Templates are more efficient because the actual code path is hardwired at the compilation time, which allows for more optimizations and diagnoses more possible errors already at this stage. On the other hand, it is applicable when the actual workflow is syntactically the same, or the number of possible variants is known in advance -- for instance, conversion between all built-in coordinate systems (Section~\ref{sec:Coords}) is hard-coded in the library. Each function that uses a templated argument produces a separate compiled fragment; therefore it is impossible for the user to extend built-in library functions with a new variety of template parameter.\\
Abstract classes (or, rather, ``interfaces'') providing virtual functions that are fleshed out in descendant classes offer more flexibility, at the expense of a small overhead (negligible in all but the tighest loops) and impossibility to securely prevent some errors. This is the only way to provide a fully extensible mechanism for supplying a user-defined object (e.g., a mathematical function implementing a \ttt{IFunction} interface) into a pre-compiled library function such as \ttt{findRoot}.

The boundary between object-oriented and procedural paradigms is less well-defined. There are several possible ways of coupling the code and the data:
\begin{enumerate} \setlength{\parskip}{2pt} \setlength{\itemsep}{2pt}
\item data fields are encapsulated as private members of a class, and all operations are provided through public methods of that class;
\item a collection of assorted variables is kept in a structure or array, and there is a standalone function performing some operation on this data;
\item a standalone function takes an instance of a class and performs some operation using public member functions of this class;
\item a class contains a pointer, reference or a copy of another class or structure, and its member functions follow either of the two previous patterns.
\end{enumerate}

The first approach is used for most classes that provide nontrivial functionality and can be treated as \hyperref[sec:Const]{immutable objects}, or at least objects with full control on their internal state. If the data needs to be modified, it is usually kept in a structure with public member fields and no methods, so that it may be accessed by non-member functions (which need to check the correctness of data on each call); the \ttt{SelfConsistentModel} struct and assocated routines follow this pattern. The third approach is used mostly for classes that are derived from an abstract base class that declares only virtual methods; since any non-trivial operation on this class only uses this public interface, it does not need to be a part of the class itself, thus loosening the coupling strength. That's why we have many non-member functions operating on \ttt{BasePotential} descendants. Finally, the fourth scenario is the preferred way of creating layered and weakly coupled design.

\paragraph{Naming conventions}  are quite straightforward: class names start with a capital letter, variable names or function arguments -- with a lowercase, constants are in all capital with underscores, and other names are in CamelCase without underscores. Longer and more descriptive names are preferred -- as a matter of fact, we read the code much more than write, so it's better to aid reading than to spare a few keystrokes in writing.

We use several namespaces, roughly corresponding to the overall structure of the library as described in Section~\ref{sec:Structure}: this improves readability of the code and helps to avoid naming collisions, e.g., there could be two different \ttt{Isochrone} classes -- as a potential and as a concept in stellar evolution, living in separate namespaces. A feature of \Cpp called ``argument-dependent lookup'' allows to omit the namespace prefix if it can be deduced from the function arguments: for instance, if \texttt{pot} is an instance of class derived from \ttt{potential::BasePotential}, we may call \texttt{\sout{potential::}writePotential(fileName, pot)} without the prefix. This doesn't apply to name resolution of classes and templates, and to functions which operate on builtin types (e.g., in the \ttt{math::} namespace). We also do not use the \texttt{auto} keyword which is only available in \CppII.

When several different quantities need to be grouped together, we use \texttt{struct} with all public members and no methods (except possibly a constructor and a couple of trivial convenience functions). If something has an internal state that needs to be maintained consistently, and provides a nontrivial behaviour, this should be a \texttt{class} with private member variables. We prefer to have named fields in structures rather than arrays, e.g., a position in any coordinate system is specified by three numbers, but they are not just a \texttt{double pos[3]} -- rather, each case has its own dedicated type such as \texttt{struct PosCyl\{ double R,z,phi; \}} (Section~\ref{sec:Coords}). This eliminates ambiguity in ordering the fields (e.g., what is the 3rd coordinate in a cylindrical system -- $z$ or $\phi$? different codes may use different conventions, but naming is unique) and makes impossible to accidentally mis-use a variable of an incorrect type which has the same length.

\paragraph{Immutability} \label{sec:Const}  of objects is a very powerful paradigm that leads to simpler design and greater robustness of programs. We allow only ``primitive'' variables -- builtin types, \texttt{struct}s with all public member fields, or arrays (including vectors and matrices) -- to change their content. Almost all instances of non-trivial classes are read-only: once created, they may not be changed anymore; if any modification is needed, a new object should be constructed. All nontrivial work in setting up the internal state is done in the constructor, and all member functions are marked as \texttt{const}. This convention is a strong constraint that allows to pass around complex objects between different parts of the code and be sure that they always do the same thing, and that there are no side effects from calling a method of a class. This also simplifies the design of parallel programs: if an object needs a temporary workspace for some function to operate, it should not be allocated as a private variable in the class, but rather as a temporary variable on the stack in each function; thus concurrent calls to the same routine from different threads do not interfere, because each one has its own temporary variable, and only share constant member variables of the class instance.
There are, of course, some exceptions, for instance, in classes that manage the input/output, string collections (\ttt{utils::KeyValueMap}), and ``runtime functions'' that perform some data collection tasks during orbit integration (even in this case, sometimes the data is stored in an external non-member variable).

Another aspect of the same rule is \texttt{const} correctness of the entire code. Instances of read-only classes may safely be declared as \texttt{const} variables, and all input arguments for functions are also marked as \texttt{const} (see \hyperref[sec:CallingConventions]{below}).
These rules improve readability and allow many safety checks to be performed at compile time -- an incorrect usage scenario will not even compile, rather than produce an unexpected error at runtime.

\paragraph{Memory management}  is a non-trivial issue in \texttt{C} and a source of innumerable bugs in poorly written software. Fortunately, \Cpp has very powerful features that almost eliminate these problems, if followed consistently. The key element is the automatic management of object lifetimes for all classes or structures. Namely, if a variable of some class is created in a block of code, the destructor of this class is guaranteed to be called when the variable goes out of scope -- whether it occurs in a normal code path or after an exception has occurred (see \hyperref[sec:Exceptions]{below}). Thus if some memory allocation for a member variable was done in the constructor, it should be freed in the destructor and not in any other place. And of course the rule applies recursively, i.e., if a member variable of a class is a complex structure itself, its destructor is called automatically from the destructor of this class, and then the destructor of the parent class (if it exists) is invoked. In practice, it is almost never necessary to deal with these issues explicitly -- by using standard containers such as \texttt{string}s and \texttt{vector}s instead of \texttt{char*} or \texttt{double*} arrays, one transfers the hassle of dynamic memory management entirely to the standard library classes.

\phantomsection\label{sec:SmartPointers}
The picture gets more complicated if we have objects that represent a hierarchy of descendants of an abstract base class, and need to create and pass around instances of the derived types without knowing their actual type. In this case the object must be created dynamically, and a correct destructor will be automatically called when an object is deleted -- but the problem is that a raw pointer to a dynamically-allocated object is not an object and must be manually deallocated before it goes out of scope, which is precisely what we want to avoid. The solution is simple -- instead of raw pointers, use ``smart pointers'', which are proxy objects that manage the resource themselves. There are several kinds of smart pointers, but we generally use only one -- \texttt{shared_ptr}. Its main feature is automatic reference counting: if we dynamically create an object derived from \ttt{BasePotential} and wrap the pointer into \ttt{PtrPotential} (which is defined as \texttt{shared_ptr<const BasePotential>}), we may keep multiple copies of this shared pointer in different routines and objects, and the underlying potential object will stay alive as long as it is used in at least one place, and will be automatically deallocated once all shared pointers go out of scope and are destroyed. If a new value is assigned to the same shared pointer, the reference counter for the old object is also decreased, and it is deallocated if necessary. Thus we never need to care about the lifetime of our dynamically created objects; this semantics is similar to \Python.
Of course, if we know the actual type of potential that we only need locally, we may create it on the stack without dynamical allocation; most routines only need a (\texttt{const}) reference to a potential object -- does not matter whether it is an automatic local variable or a dereferenced pointer.

Finally, it's better to avoid dynamical memory allocation (including creation of \texttt{std::vector}s) in routines that are expected to be called frequently (such as \ttt{BasePotential::eval()}). All temporary variables should be created on the stack; if the size of an array is not known at compile time (e.g., it depends on the parameters of potential), we either reserve an array of some pre-defined maximum permitted size, or use \texttt{alloca()} routine which creates a variable-length array on the stack.

\paragraph{Calling conventions} \label{sec:CallingConventions}  refer to the way of passing and returning data between various parts of the code. Arguments of a function can be input, output, or both. All input arguments are either passed by value (for simple built-in types) or as a \texttt{const} reference (for more complex structures or classes); if an argument may be empty, then it is passed as a const pointer which may take the \texttt{NULL} value. Output and input/output arguments are passes as non-\texttt{const} references to existing objects. Thus the function signature unambiguously defines what is input and what is not, but does not indicate whether a mixed-intent argument must have any meaningful value on input -- this should be explained in the \textsc{Doxygen} comment accompanying the definition. Unfortunately there is no indication of the direction of arguments at the point where the function is called.

Usually the input arguments come first, followed by output arguments, except the cases of input arguments with default values, which must remain at the end of the list. (Unfortunately, \Cpp does not have named arguments, which would be more descriptive, but we encourage their use in the \Python interface).
When the function outputs a single entity (even if it is a complex object), it is usually a return type, not an output argument; in most contexts, there is no extra cost because temporary objects are not created (copy elision and return-value optimization rules). However, extra output information may be stored in output arguments (sometimes optional, i.e., they may be \texttt{NULL}, indicating that this extra information is not required by the caller). When the return value is not a copyable type (e.g., if a function creates a new instance of a class derived from an abstract base class), then it is returned as a smart pointer.

These conventions apply to ordinary non-member functions and class methods; for constructors they are somewhat different. If we create an object A which has a link to another object B, it usually should not be just a reference or a raw pointer -- because the lifetime of B may be shorter than the newly created A. In these cases, B is either copied by value (like a \texttt{std::vector}), or else it should be provided as a shared pointer to the actual object, and a copy of this pointer is stored in A, increasing its reference counter. This ensures that the actual instance of B continues to exist as long as it is used anywhere, and is automatically destroyed when it is no longer needed.
Thus, if a class constructor takes a shared pointer as an argument, this indicates that a copy of this pointer will be kept in the class instance during its lifetime; if it takes a reference, then it is only used within the constructor but not any longer.
This rule also has exceptions -- several wrapper classes used as proxy object for type conversion. For instance, when a certain routine (e.g., \ttt{math::integrate}) expects an argument of \ttt{const math::IFunction\&} type to perform some calculations on it without storing the object anywhere else, this argument could be a temporary instance of a wrapper class (e.g., \ttt{potential::DensityWrapper}) taking a reference to a \ttt{const potential::BaseDensity\&} object in the constructor. In other words, instances of these wrapper classes should only be created as unnamed temporary objects passed as an argument to another function, but not as stack- or heap-allocated variables -- even local ones.

\paragraph{Numerical issues} -- efficiency and accuracy -- are taken very seriously throughout the code. Floating-point operations often require great care in re-arranging expressions in a way that avoids catastrophic cancellation errors. A classic example is the formula for the roots of a quadratic equation: $x_{1,2} = (-b \pm\sqrt{b^2-4ac})/(2a)$. In the case of $ac\ll b^2$, one of the two roots is a difference between two very close numbers, thus it may suffer from the loss of precision. Another mathematically equivalent expression is $2c/(-b \mp\sqrt{b^2-4ac})$, and a numerically robust approach is to use both expressions -- each one for the root that has two numbers of the same sign \textit{added}, not subtracted. Going one step further, if the coefficients $a,b,c$ are themselves obtained from other expressions, it may be necessary to reformulate them in such a way as to avoid subtraction under the radical, etc. These details are necessary to ensure robust behaviour in all special and limiting cases; a good example are coordinate conversion routines.

Efficiency is also a prime goal: this includes a careful consideration of algorithmic complexity and minimization of computational effort. Some mathematically equivalent functions have rather different computational cost: for instance, generating two Gaussian random numbers with the Box--Muller algorithm is a few times faster than using the inverse error function; finding a given quantile (e.g., median) of an array can be done in $\mathcal{O}(N)$ operations without sorting the entire array (which costs $\mathcal{O}(N\,\log N)$ operations); and computing potential and three components of force simultaneously is faster than doing it separately.

For numerical integration, a suitable coordinate transformation may dramatically improve the accuracy -- or reduce the number of function calls in the adaptive integration routine; often a fixed-order Gauss--Legendre integration is enough in a particular case, with the degree of quadrature selected by extensive numerical experiments.
Consider, for instance, two one-dimensional integrals $I_a \equiv \int_{-1}^1 \sqrt{1-x^2}\,\d x$,
$I_b \equiv \int_{-1}^1 1/\sqrt{1-x^2}\,\d x$; their analytical values are respectively $\pi/2$ and $\pi$. Both integrands have integrable endpoint singularities (even though the first one tends to zero as $x\to \pm 1$, it is not analytic at these points), and a na\"ive $N$-point Gauss--Legendre quadrature rule has poor convergence: relative errors are 0.003 (0.0005) with $N=5$ (10) for $I_a$, and 0.1 (0.05) for $I_b$. However, if we apply transformation of the interval $[-1..1]$ onto itself, stretching it near the endpoints, the results are far better: for a substitution $x = (3-y^2)y/2$, the errors in 5- or 10-point rule are $\sim10^{-6}$ or $10^{-12}$ for both functions, while another similar substitution $x = \sin(y\,\pi/2)$ would even make the second integral exact. Such transformations, in exactly the same context, are used, e.g., in computing the actions and frequencies.

Multidimensional integration methods employ adaptive refinement of the integration domain, and in principle should be able to cope with strongly varying or non-analytic functions, again at the expence of accuracy (for a fixed upper limit on the number of function evaluations). However, there is another reason for applying non-trivial hand-crafted scaling transformations: if the function in question is very strongly localized in some small region of the entire domain, the adaptive integration routine may simply miss some part of the domain where none of the initially considered points yielded non-negligible function values, and hence this part was not refined at all. Such situation may arise, e.g., when computing the density $\rho(\bx) = \int f(\bx,\bv)\,\d ^3v$ from a DF of a cold disk, which is very narrowly peaked at $v_r=0, v_z=0, v_\phi=v_\mathrm{circ}(r)$; hence we employ a sophisticated transformation that stretches the region around $|v| = v_\mathrm{circ}$, facilitating the task of "hitting the right point".

Dimensional quantities are usually converted to logarithms before applying any scaling transformations, to ensure a (nearly-)scale-invariance. Consider, e.g., $\int_0^\infty f(x)\,\d x$ with a na\"ive transformation of the interval onto $[0..1]$: $x = y/(1-y)$. If the function $f(x)$ peaks between $x=10^6$ and $10^7$ (not uncommon when dealing with astronomical scales), no reasonable adaptive quadrature rule would be able to detect this peak crammed into a tiny interval $1-10^{-6} < y < 1-10^{-7}$! If, on the other hand, we employ a two-stage transformation, $x = \exp(y)$, $y = 1/(1-z)-1/z$, then the peak lies between $0.93<z<0.94$, which is far more likely to be found and handled correctly. Same considerations apply to root-finding and minimization routines; in these cases the added benefit is increased precision. If a root in the na\"ively scaled variable is at $1-10^{-10}$, it has only 5 significant digits; if our relative tolerance in root-finder was $10^{-12}$, the un-scaled variable would carry a relative error of $10^{-2}$! By contrast, in the two-stage transformation the root will be around 0.96 and we only lose one or two digits of precision.

Consistency and reproducibility of floating-point (FP) calculations is a painful issue, as anyone seriously working with them can attest. Theoretically, with no compiler optimizations and running in a single thread, one should be able to get identical results on all architectures conforming to the IEEE 754 FP standard (that is, almost all current ones). However, once we wish the program to be even mildly efficiently optimized, the results can differ between compilers and even between running the same executable on different processors. Adding the multi-threading computations surely makes things even less predictable. For instance, computing a sum of array elements in parallel will yield different results depending on the number of threads, because the sub-ranges summed by different threads will be different, and FP addition is not commutative. When dynamic load-balancing is allowed in runtime, these sub-ranges may be different between runs even with the same number of threads. These issues notwithstanding, a reasonable effort has been invested to keep as much reproducibility as possible when running on the same machine with the same number of threads, and keep the differences at the level of FP precision when running with different number of threads.

A few words about \texttt{INFINITY} and \texttt{NAN} values. Infinities are valid floating-point numbers and are quite useful in some contexts where a really large number is required (for instance, as the endpoint of the root-finder interval); they propagate correctly through most expressions and some functions (e.g., \texttt{exp}, \texttt{log}, comparison operators). The infamous \texttt{NAN} is a different story: it usually%
\footnote{In some functions, \texttt{NAN} it is used as a special, usually a default value of an input argument, indicating something like ``value is unknown''.}
indicates an incorrect result of some operation, and is infectious -- it propagates through all floating-point operations, including comparison operators (that is, \texttt{a>b} and \texttt{a<b} are \textit{both} false if \texttt{b} is \texttt{NAN}; however, \texttt{!(a<b)} and \texttt{b!=b} are true in this case). This feature is useful to pass the indication of an error to the upper-level routines, but it does not allow to tag the origin of the offensive operation. This brings us to the next topic:

\paragraph{Error handling} \label{sec:Exceptions}  is an indispensable part of any software. Whenever something goes wrong, this must be reported to the upper-level code -- unless there is a safe fallback value that may be returned without compromising the integrity of calculation. The standard approach in \Cpp is the mechanism of exceptions. They propagate through the entire call stack, until handled in a \texttt{catch} statement -- or terminate the program if no handler was found. They also carry upward any user-defined diagnostic information (e.g., a string with error description), and most importantly, they ensure a correct disposal of all temporary objects created in intermediate routines (of course, if these are real objects with destructors, not just raw pointers to dynamically-allocated memory -- which are bad anyway). 
Thus a routine does not need to care about a possible exception occurring at a lower level, if it cannot handle it in a meaningful way -- it should simply let it propagate upwards. Exceptions should be used to check that the input parameters are correct and consistent with the internal state of an object, or perhaps to signal an un-implemented special case. Within a constructor of a class, they are the only available mechanism for error signalling, because a constructor cannot return a value, and storing an error code as a class member variable doesn't make sense, since the object is not usable anyway. Instead, if a constructor fails, the object is immediately and correctly destroyed.

On the other hand, if a certain condition is never expected to occur, this may be expressed as an \texttt{assert} statement -- which terminates the program unconditionally if violated, and this clearly indicates some internal inconsistency in the code, e.g., a memory corruption. It also promotes a self-documenting code -- all assumptions on input parameters and (preferrably) results of calculation (pre- and post-conditions) are clearly visible. This mechanism should only be used within a single unit of code (e.g., a class), which has a full control on its internal state; if a function is part of public interface, it may not assume that the passed arguments are valid and should check them, but in the case of incorrect values should raise an exception rather than terminate the entire program.

Finally, it should be noted that exceptions do incur some run-time penalty if triggered, so they should not be used just to inform about something that may routinely occur, e.g., in a function that searches for a substring and does not find it. Sometimes propagating a \texttt{NAN} is a cheaper alternative, used, for instance, in action finders if the energy is positive (does not correspond to bound motion).

\paragraph{Diagnostic output}  is a separate issue from error handling, and is handled by a dedicated printout routine \ttt{utils::msg} that may be used with different levels of verbosity and write the messages to console or a log file. Its behaviour is controlled at runtime by the environment variables \texttt{LOGLEVEL} (ranging from 0 to 3; default 0 means print only necessary messages, 1 adds some non-critical warnings, 2 prints ordinary debugging information, 3 dumps even more debugging information on screen and to text files) and \texttt{LOGFILE} (if set, redirects output to the given file, otherwise it is printed to \texttt{stderr}). The library's default handler may be reassigned to a user-provided function.

\paragraph{Parallelization}  in \Agama is using the \texttt{OpenMP} model, which is nearly transparent for the developer and user. Only a few operations that are supposed to occur in a serial context have internal loops parallelized: this includes the construction of \ttt{Multipole} and \ttt{CylSpline} potentials from density profiles or from \ttt{ParticleArray}s, initialization of interpolation tables in \ttt{ActionFinderAxisymFudge}, and \hyperref[sec:Sampling]{sampling} from a multidimensional probability density. The former operation appears, for instance, in the context self-consistent modelling (Section~\ref{sec:SCM}), when the evaluation of density at each point is a costly operation involving multidimensional integration of distribution function over velocities and thousands of calls to an action finder, and the values of density at different points are collected in parallel.
Other typical operations, such as computation of potential or action for many points simultaneously, should be paralellized in the caller code itself. Almost all classes and functions provided by the library can be used from multiple threads simultaneously, because they operate with read-only or thread-local data (exceptions from this rule are linear and quadratic optimization routines, which are not thread-safe, but hardly would need to be called in parallel); we do not have any mutex locks in the library routines.
For instance, in the \Python interface, a single call to the potential or action finder may provide an array of points to work with, and the loop is internally parallelized in the \Cpp extension module. 

An important thing to keep in mind is that an exception that may occur in a parallel section should be handled in the same section, otherwise the program immediately aborts. Thus in such loops it is customary to provide a general handler that stores the error text, and then re-throws an exception when the loop is finished.
Also, the \Python interface provides a way to supply a user-defined \Python callback function to some of the routines implemented in \Cpp, but the standard \Python interpreter has a global locking mechanism preventing its simultaneous usage from multiple threads. Therefore, when such callback functions are used with \Cpp routines, this temporarily disables \texttt{OpenMP} parallelization. However, these user-defined functions are invoked with a vector of input points that may be processed in a single call (e.g. using \texttt{numpy} array operations), instead of one point at a time, thus reducing the cost of transferring the control flow between \Cpp and \Python.

%%%%%%%%%%%%%%%%%%%%%%%%%%%%%%%%%
\subsection[Mathematical methods]
{Mathematical methods\protect\footnote{
In this chapter, we use $\boldsymbol{boldface}$ for column-vectors and \textsf{Sans-serif} font for matrices, while keeping the ordinary $cursive$ script for writing element-wise expressions.}}

%%%%%%%%%%%%%%
\subsubsection{Basis-set approximation of functions}  \label{sec:MathBasisSetDetails}

We often need to represent approximately an arbitrary function of one variable, defined on a finite or infinite interval $[a,b]$.
This is conventionally achieved by defining the inner product of two functions $f(x), g(x)$:
\begin{align}  \label{eq:InnerProduct}
\langle f, g \rangle \equiv \int_a^b f(x)\, g(x)\, \d x \,,
\end{align}
and introducing a complete set of basis functions $B_i(x)$, so that any sufficiently well-behaved function $f(x)$ can be approximated by a weighted sum with a finite number of terms $M$ to any desired accuracy:
\begin{align}  \label{eq:BasisApproximation}
\tilde f^{(M)}(x) = \sum_{j=1}^M A_j\,B_j(x) \,.
\end{align}
The coefficients of this expansion (amplitudes) $A_j$ are determined from the requirement that the inner product $\mathcal P_i$ of the original function $f$ and each of the basis elements $B_i$ is the same as the inner product of the approximated function $\tilde f^{(M)}$ and the same basis element (Galerkin projection):
\begin{subequations}  \label{eq:BasisExpansionCoefs}
\begin{align}
\mathcal P_i \{f\} &\equiv \langle f, B_i \rangle = \\
\mathcal P_i \{\tilde f^{(M)}\} &\equiv \langle \tilde f^{(M)}, B_i \rangle =
\int_a^b \tilde f^{(M)}(x)\, B_i(x)\, \d x = \sum_{j=1}^M G_{ij}\, A_j  \,,
\end{align}
\end{subequations}
where $\mathsf G$ is the Gram matrix of inner products of basis functions:
\begin{align}  \label{eq:BasisExpansionMatrix}
G_{ij} &\equiv \langle B_i, B_j \rangle  =  \int_a^b B_i(x)\,B_j(x)\,\d x \,.
\end{align}

Classical basis sets are usually orthonormal, i.e., $G_{ij} = \delta_{ij}$, and addition of each subsequent term does not change existing expansion coefficients (e.g., Fourier series, orthogonal polynomials). Of course, it is possible to construct an orthogonal set by employing the Gram--Schmidt procedure for any sequence of independent basis functions. However, it is not always necessary, as long as we can solve efficiently the linear system $\mathsf G\,\boldsymbol A = \boldsymbol{\mathcal P}$ (\ref{eq:BasisExpansionCoefs}) to find $A_j$ from $\mathcal P_i$ (this is the case for the B-spline basis set discussed in the Section~\ref{sec:MathBSplineDetails}).

The computation of projection integrals $\langle f, B_i\rangle$ ideally should be performed with a method that gives an exact result if the function $f$ is itself a basis element (or, consequently, a weighted sum of these elements, such as $\tilde f$); equivalently, the Gram matrix (\ref{eq:BasisExpansionMatrix}) needs to be computed exactly. This implies the use of the trapezoidal rule with equidistant points for the Fourier basis, the Gauss--Legendre rule for a basis of Legendre polynomials (spherical harmonics), the Gauss--Hermite rule for the eponymous basis set  (Section~\ref{sec:MathGaussHermiteDetails}), or again the Gauss--Legendre rule \textit{separately on each grid segment} for a B-spline basis (Section~\ref{sec:MathBSplineDetails}).

We now discuss several commonly encountered tasks and the techniques for solving them with the aid of basis sets.

\paragraph{Estimation of the probability distribution function $f(x)$ from discrete samples} $\{x_n\}_{n=1}^{N}$ drawn from this function can be formulated in the framework of basis functions as follows. The discrete realization of the probability density is $\hat f(x) \equiv \frac{1}{N} \sum_{n=1}^{N} \delta(x-x_n)$, and we identify it with the smooth approximation $\tilde f$ expressed in terms of a weighted sum of basis functions (\ref{eq:BasisApproximation}). According to (\ref{eq:BasisExpansionCoefs}), the amplitudes $\boldsymbol A$ of this expansion satisfy the linear equation system
\begin{align}
\sum_{j=1}^M G_{ij}\, A_j = \mathcal{P}_i \{\hat f\} = \frac{1}{N} \sum_{n=1}^{N} B_i(x_n) \;.
\end{align}
This expression is trivially generalized to the case of unequal-weight samples.\\
The drawback of this formulation is that the approximated density $\tilde f$ is not guaranteed to be non-negative, unlike the original function $f$. An alternative approach, discussed in Section~\ref{sec:MathSplineDensityDetails}, is to represent the \textit{logarithm} of $f$ as a basis-set approximation, ensuring the non-negativity property; however, it is a non-linear operation, unlike the simpler approach introduced above. Of course, there exist various other methods for estimating the density from discrete samples, for instance, using the kernel density estimation (KDE). However, the latter is fundamentally a smoothing operation, producing the estimate of the convolution of $f$ with the kernel, rather than of the original function $f$. [illustration?]

\paragraph{Linear operators} acting on the function $f$ can be represented as linear transformations of the vector of amplitudes of $\tilde f$: $A_k' = T_{kj} A_j$. The matrix $\mathsf T$ may correspond to an exact representation of the transformed function $\tilde f'(x) \equiv T\{\tilde f\}(x)$, possibly in terms of a different basis set $B_k'$, or to its approximation constructed in the same way as for the original function $f$ (by Galerkin projection of the transformed function $\tilde f$ onto the basis).

An example of the first kind is the differentiation or integration of $\tilde f$, represented by its expansion in terms of a similarly transformed set of basis functions. For instance, in the case of Fourier or Chebyshev series, the transformed basis functions can be exactly represented by the same or a related basis set (increasing the degree $M$ in the case of integration). For a $M$-th degree B-spline basis set described in the next section, the integrals or derivatives of basis functions are also B-splines of degree $M+1$ or $M-1$, correspondingly, defined by the same grid. This feature is used in the finite-element analysis to represent differential equations in a discretized form, and solve the equivalent linear algebra equations.

An example of the second kind is a convolution operation $\tilde f \ast K \equiv \int_a^b \tilde f(y)\, K(x-y)\, \d y$. If $\tilde f$ is represented by the vector of amplitudes $\boldsymbol A$, and the convolved function $\breve f \equiv \tilde f \ast K$ is represented by the vector of amplitudes $\boldsymbol{\breve A}$, the relation between these two vectors is found by applying the projection operator $\mathcal P_i$ to $\breve f$:
\begin{align}
\mathcal P_i \{ \breve f \} = \sum_j K_{ij} A_j, \qquad
K_{ij} \equiv \int_a^b \d x \int_a^b \d y\; B_i(x)\, B_j(y)\, K(x-y) \,.
\end{align}
Hence, according to the general rule, $\boldsymbol{\breve A} = \mathsf G^{-1} \, \mathsf K\, \boldsymbol A$.

\paragraph{Change of basis:}  \label{sec:MathBasisSetChange}
if one needs to express the function $\tilde f(x) = \sum_j A_j\, B_j(x)$ in terms of a different basis set as $\sum_m \mathcal A_m\, \mathcal B_m(x)$, the amplitudes of this expansion are given by $\boldsymbol{\mathcal A} = \mathcal G^{-1}\, \mathsf H\, \boldsymbol A$, where $H_{mj} \equiv \langle \mathcal B_m, B_j \rangle$, and $\mathcal G$ is the Gram matrix of the basis set $\mathcal B_m$. This is, in general, a lossy operation (one may call it a \textit{reinterpolation} onto the new basis), in a sense that constructing an approximation for a function $f$ directly in terms of the new basis is not equivalent to the approximation for $f$ in terms of the old basis and then a further approximation of this $\tilde f$ in terms of the new basis. Depending on the properties of both the function and the basis, the extra error may be negligible or not.

%%%%%%%%%%%%%%
\subsubsection{B-splines}  \label{sec:MathBSplineDetails}

The B-spline set of basis function is defined by a grid of points (knots) on the interval $[a,b]$: $a=k_1<k_2<\dots<k_K=b$. Each basis function is a piecewise polynomial of degree $N\ge 0$ that is non-zero on at most $N+1$ consecutive segments of the grid (or fewer at the edges of the grid). Specifically, it is a polynomial inside each grid segment, and its $N-1$-th derivative is continuous at each knot (except the endpoints $a,b$, but including all interior knots). The total number of basis functions is $M=K+N-1$. These functions are defined through the following recursion (de Boor's algorithm):
\begin{subequations}
\begin{align}
B^{[0]}_j(x) &\equiv \left\{ \begin{array}{ll} 1 &\mbox{if }k_j \le x \le k_{j+1} \\ 0 & \mbox{otherwise} \end{array} \right., \\
B^{[N]}_j(x) &\equiv B^{[N-1]}_j(x)\,\frac{x-k_j}{k_{j+N}-k_j} +
B^{[N-1]}_{j+1}(x)\,\frac{k_{j+N+1}-x}{k_{j+N+1}-k_{j+1}} \;.
\end{align}
\end{subequations}

B-splines have the following convenient properties:
\begin{itemize}
\item At any grid segment, at most $N+1$ basis functions are nonzero. This makes the computation of interpolant (\ref{eq:BasisApproximation}) very efficient -- the grid segment enclosing the point $x$ is located in $\mathcal{O}(\log M)$ operations, and the computation of all $N$ possibly non-zero basis functions takes $\mathcal{O}(N^2)$ operations (with $N$ typically ranging from 0 to 3), instead of $\mathcal{O}(M)$ as for traditional basis sets.
\item The basis is not orthogonal, but the matrix $\mathsf G$ (\ref{eq:BasisExpansionMatrix}) is block-diagonal with bandwidth $N$, thus the coefficients of decomposition $A_j$ are obtained from $\mathcal P_i$ in $\mathcal{O}(N^2M)$ operations.
\item Although the number and degree of basis functions must be fixed in advance before computing any decompositions, we may use the freedom to put the knots at the most suitable locations to improve the accuracy of approximation.
\item The basis functions are non-negative, thus to ensure that $f(x) \ge 0$, it is sufficient to have $A_i\ge 0$. 
\item The sum of all basis functions is always 1 at any $x$.
\end{itemize}

\begin{figure}
\begin{center}
\includegraphics[width=16cm]{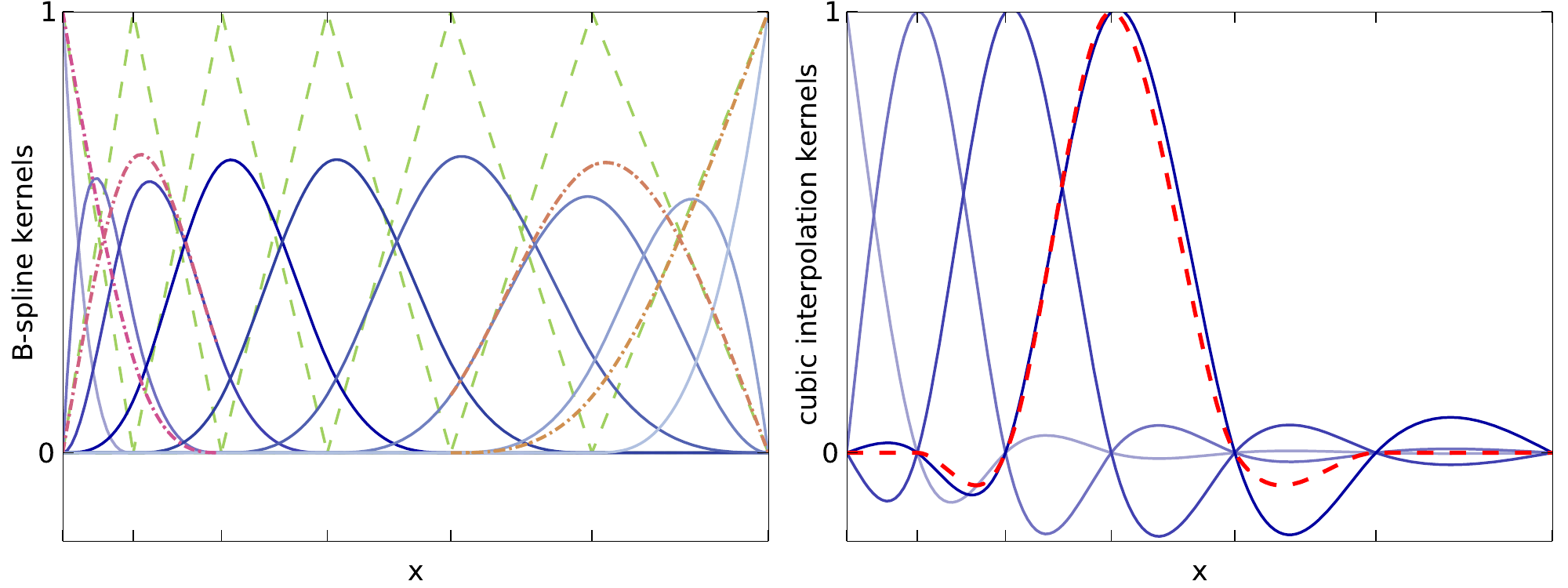}
\end{center}
\caption{Various interpolation kernels. To obtain the value of interpolated function, one sums up the values of all interpolating kernels with appropriate weight coefficients. \protect\\
\textit{Left panel:} B-splines of degree $N=1$ (dashed) and $N=3$ (solid lines), defined by the nodes of a non-uniform grid (marked by $x$-axis ticks). The former are piecewise-linear and non-zero on at most two segments, and the latter are piecewise-cubic, with two continuous derivatives, and nonzero on at most four consecutive segments. The sum of all B-spline functions at any point $x$ is unity, and they are always non-negative; thus the cubic kernels never reach unity (except the endpoint ones), because at any point more than one of them is positive.
The total number of functions is $K+N-1$, where $K$ is the number of grid nodes.
Dash-dotted lines show modified cubic B-spline basis functions that have natural boundary conditions (zero second derivative at endpoints); they are obtained by replacing three leftmost $N=3$ B-splines with two new functions (essentially distributing the middle one between the other two), and similarly for the three rightmost B-splines.\protect\\
\textit{Right panel:} interpolation kernels of a natural cubic spline defined by the same grid (solid lines). This type of spline is constructed from the array of function values at grid nodes, thus each kernel reaches unity at the corresponding grid point, and the number of kernels is $K$ (not all of them are shown). Consequently, they must attain negative values for their sum to be unity at any $x$. Moreover, since the weights of kernels are computed by solving a global linear system of equations, each kernel spans the entire grid, although its amplitude rapidly decreases away from its central node. For comparison, a cubic interpolating Catmull--Rom kernel (dashed curve) is nonzero only on four adjacent grid segments, although it also attains negative values on the two outermost segments. } \label{fig:SplineKernels}
\end{figure}

The case $N=0$ corresponds to a histogram (piecewise-constant function), $N=1$ -- to a piecewise-linear interpolator, and $N=3$ -- to a cubic spline with clamped boundary conditions (it is defined by $K$ nodes but has $K+2$ independent components, unlike the more familiar natural cubic spline, in which the extra two degrees of freedom are used to make the second derivative zero at endpoints). It is possible to construct an equivalent representation of a natural cubic spline in terms of a modified $N=3$ B-spline basis set, in which the first three basis functions $B_0, B_1, B_2$ are replaced with two linear combinations that have zero second derivative: $\tilde B_1 \equiv B_0 + \frac{x_2-x_0}{x_1+x_2-2x_0}\,B_1$, $\tilde B_2 \equiv B_2 + \frac{x_1-x_0}{x_1+x_2-2x_0}\,B_1$, and similarly for the last three basis functions, see Figure~\ref{fig:SplineKernels} (left panel).

B-splines are well suited for constructing the approximating function with a relatively small number of terms from a possibly large array of points (essentially replacing the integral in (\ref{eq:BasisExpansionCoefs}) by a discrete sum, see Sections~\ref{sec:MathSplineApproxDetails} and \ref{sec:MathSplineDensityDetails}). %They are also used in the finite-element analysis (Section~\ref{sec:MathFiniteElementDetails}. 
On the other hand, if one needs to construct an interpolating function passing through the given set of points (the standard interpolation problem), B-splines are less convenient, and the evaluation of the interpolant is also less efficient than for the ``classical'' textbook splines discussed in the next section.

%%%%%%%%%%%%%%
%%%%%%%%%%%%

%%%%%%%%%%%%%%
\subsubsection{Spline interpolation}  \label{sec:MathSplineDetails}

The task of interpolation in 1, 2 and 3 dimensions is performed by several kinds of splines: linear interpolators (not particularly exciting), cubic splines and quintic splines (the latter -- only in 1d and 2d). The degree of spline (1, 3 or 5) refers to the degree of the piecewise polynomial at each grid segment (in more than one dimension, along each axis). However, the continuity conditions at grid nodes may be different from B-splines (in which the function has $N-1$ continuous derivatives at all interior nodes).

Let us first consider the 1d case with $K\ge 2$ grid points ($K-1$ segments).

Interpolation by piecewise-cubic polynomials requires 4 coefficients for each segment, which are conveniently taken to be the values and first derivatives of the function at two adjacent nodes: $f(x_i), f(x_{i+1}), f'(x_i), f'(x_{i+1})$ -- this is called the Hermite interpolation. Thus the total number of coefficients is $2K$; the function and its derivative is continuous across segments, but the second derivative may change discontinuously.

What if we are only given the function values $f(x_i)$, but not the derivatives? One may come up with a plausible approximation for derivatives at each node, and then use the Hermite interpolation on each segment -- this will yield a continuously differentiable curve no matter what the values of $f'(x_i)$ are. Of course, we want it not only to be smooth, but also to approximate the true function accurately, and this requires a judicious assignment of derivatives. One possibility is to use finite-differences to estimate $f'(x_i)$ as $[f(x_{i+1})-f(x_{i-1})]/[x_{i+1}-x_{i-1}]$, with a suitable modification for boundary points, or a generalization for unequally-spaced grids. This is called the Catmull--Rom spline, and is frequently used in resampling (especially in more than one dimension), due to its locality: the value of interpolated function on each interval $x_i\le x\le x_{i+1}$ depends on four nearby values -- $f(x_{i-1}), f(x_i), f(x_{i+1}), f(x_{i+2})$.
Another possibility is the familiar cubic spline, in which the first derivatives are computed from the requirement that the \textit{second} derivatives are continuous at each interior node (i.e. $K-1$ points). This results in a tridiagonal linear equation system, relating $f'(x_i)$ to $f'(x_{i-1})$ and $f'(x_{i+1})$%
\footnote{Usually this is expressed as a relation between second derivatives at grid nodes, and the spline function is defined in terms of $f(x_i)$ and $f''(x_i), i=1..K$. Since all cubic splines are a subset of Hermite piecewise-cubic polynomial interpolators, we may equivalently parametrize them by the values and \textit{first} derivatives at grid nodes, and this automatically ensures that the second derivative is continuous because the first derivative was computed from this condition. Furthermore, the computed derivatives may subsequently be modified by the regularization filter to improve the behaviour around sharp discontinuities (see below). }
Two additional boundary conditions are required to close the equation system; most commonly, these are $f''(x_1) = f''(x_K) = 0$ (so-called natural cubic splines), but alternatively, one may specify the first derivatives at the endpoints (clamped cubic spline). Thus the derivatives, and consequently the interpolated curve, depend on the values of $f$ at all grid nodes, not just the adjacent ones; since $f'(x_i)$ are expressed as a linear function of $f(x_1)\dots f(x_K)$, we may consider the result of interpolation as a smoothing kernel defined by the grid nodes and linearly depending on all input values $f(x_i)$ (see Figure~\ref{fig:SplineKernels}, right panel, for a comparison of Catmull--Rom and natural cubic spline interpolation kernels).

Thus natural cubic splines are defined by $K$ function values at grid points and are a subset of all cubic splines with $K$ grid points (parametrized by $K+2$ numbers); the latter, in turn, are a subset of a wider class of piecewise-cubic Hermite interpolators (which are fully specified by $2K$ coefficients). The set of all cubic splines is also equivalent to B-splines of degree 3 over the same grid. The evaluation of cubic splines and optionally their derivatives is more efficient than the B-splines, and the array of B-spline amplitudes may be directly used to initialize an equivalent clamped cubic spline.

If, on the other hand, one can independently compute both $f(x_i)$ and $f'(x_i)$ at all grid nodes, and use these $2K$ numbers to construct a piecewise-cubic Hermite interpolator, this should generally improve the accuracy of approximation (even though will decrease its \textit{smoothness} compared to the case of cubic splines). In practice, within the class of piecewise-cubic polynomials the improvement is not dramatic. However, one may instead go to higher order and use piecewise-quintic polynomials, specified by 6 coefficients on each segment, which are again conveniently taken to be the values and first two derivatives of the function at two adjacent nodes -- a natural extension of cubic Hermite interpolation. The resulting curve will be twice continuously differentiable for any choice of $f''(x_i)$, but the accuracy of approximation will be good only if these second derivatives are assigned carefully. In close analogy to cubic splines, one may compute them from the requirement that the 3rd derivative is continuous at all interior grid nodes, augmented with two boundary conditions; the natural choice for them is $f''''(x_1)=f''''(x_K)=0$, because in a degenerate case $K=2$ they simply lead to a cubic Hermite interpolator.
\begin{subequations}
\begin{align}
\frac{1}{3} f'''_i
&= 20 \frac{f_{i+1}-f_i}{(x_{i+1}-x_i)^3} - \frac{12f'_i+8f'_{i+1}}{(x_{i+1}-x_i)^2} - \frac{3f''_i-f''_{i+1}}{x_{i+1}-x_i} = \\
&= 20 \frac{f_i-f_{i-1}}{(x_i-x_{i-1})^3} - \frac{12f'_i+8f'_{i-1}}{(x_i-x_{i-1})^2} + \frac{3f''_i-f''_{i-1}}{x_i-x_{i-1}}\quad \mbox{for }2\le i \le K-1, \nonumber \\
0 &= 30\frac{f_2-f_1}{(x_2-x_1)^3} - \frac{16f'_1+14f'_2}{(x_2-x_1)^2} - \frac{3f''_1-2f''_2}{x_2-x_1} \quad\mbox{for }i=1,\mbox{ and similarly for }i=K.
\end{align}
\end{subequations}

This tridiagonal system results in a quintic spline, which provides 5th degree piecewise-polynomial interpolation with three continuous derivatives. It is \textit{not} equivalent to a 5th degree B-spline -- the latter would have 4 continuous derivatives and is fully specified by $K+4$ coefficients, whereas a quintic spline is specified by $2K+2$ numbers (the values and first derivatives at all nodes, plus two endpoint conditions). The accuracy of approximation with a quintic spline is far better than anything achievable with a cubic interpolation -- but only in the case when one may compute the function derivatives independently and accurately enough (i.e., using a cubic spline to find the derivatives results in a quintic spline which is almost equivalent to a cubic one in terms of accuracy).

\begin{figure}[t]
\begin{center}
\includegraphics[width=15cm]{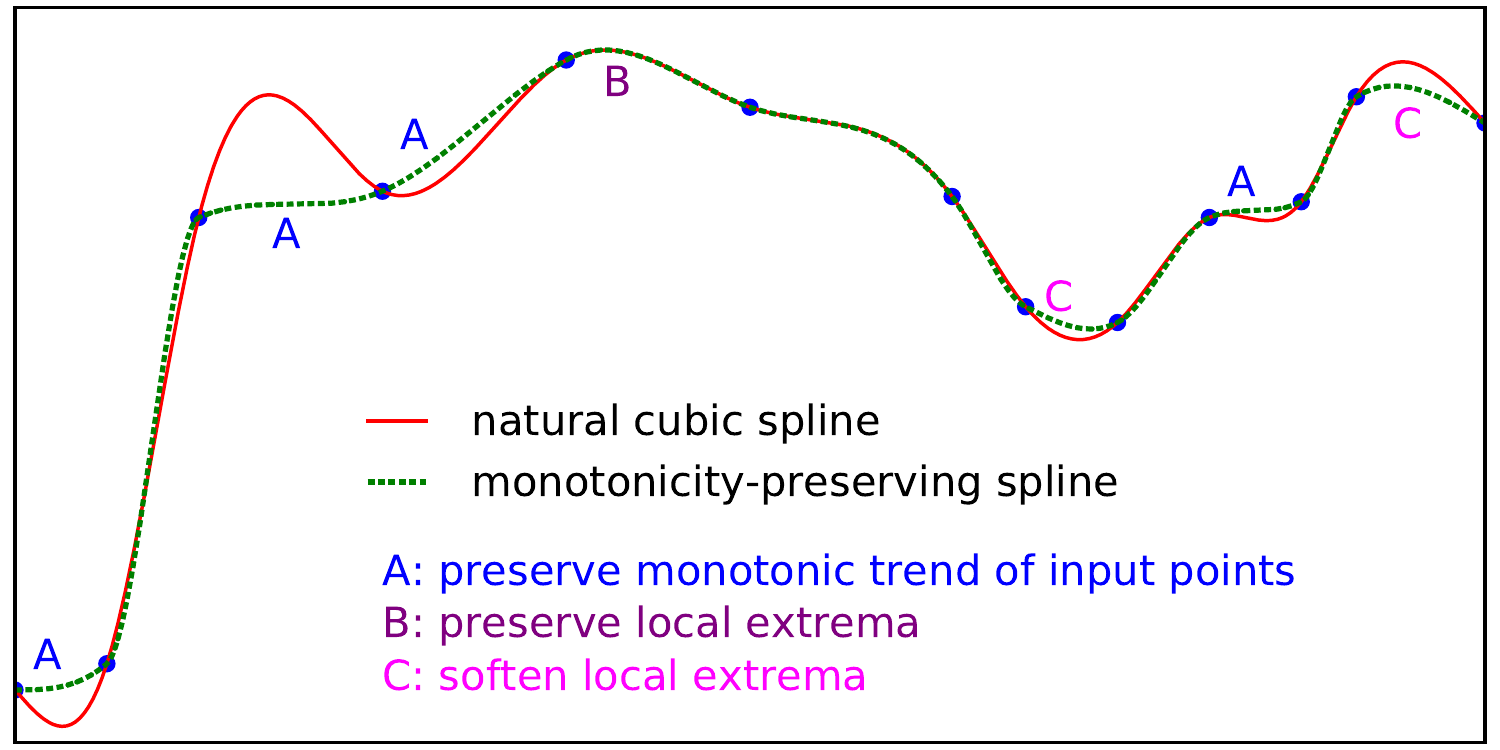}
\end{center}
\caption{Monotonicity-preserving spline (dotted line) compared to the natural cubic spline (solid line). This demonstrates several aspects of regularization filter: it preserves a monotonic trend of input points, avoiding spurious bumps (left part of the plot), does nothing if the spline is smooth enough (center), and preserves the information about local minima/maxima but may soften them somewhat (right). } \label{fig:SplineMonotonic}
\end{figure}

A common problem with high-order interpolation arises when there are sharp discontinuities in the input data, which lead to overshooting and nasty oscillations in the interpolated curve. To cope with these cases, there is a possibility of applying a regularizing (monotonicity-preserving) filter for a natural cubic spline \cite{Hyman}. As usual, the first derivatives at each node are computed from the standard tridiagonal system under the requirement that the second derivative is continuous. Then the filter examines the behaviour of the cubic interpolation polynomial at each segment and determines whether it is monotonic or not. In the latter case, and if the input data values were monotonic, it adjusts the first derivative, so that the interpolated curve also becomes monotonic. This does not preclude it from having a local maximum or minimum on a segment adjacent to a local extremum of input data points, but sometimes may also soften the amplitude of this maximum. Figure~\ref{fig:SplineMonotonic} illustrates various aspects of the regularization filter. The downside of this filter is that it converts a twice continuously differentiable curve into a generic piecewise-cubic Hermite interpolator with only one continuous derivative, but it typically does so only in the cases when the more smooth curve is actually a worse approximation of the function. One should also keep in mind that this procedure breaks the linearity of spline construction (i.e. a sum of two filtered splines may not be equal to a filtered spline of a summed input points). Thus the regularization filter does not apply by default, but is useful in particular for log-scaled interpolators ($\ln f(\ln x)$ is represented as a cubic spline), where an occasional very small positive value of an input point results in a large jump in the log-scaled curve, which then overshoots and oscillates on nearby segments, aggravated by the inverse exponential scaling.

%%%%%%%%%%%%%%
%%%%%%%%%%%%

Let's move to the multidimensional case, where the interpolation is provided on a separable grid with piecewise-polynomial functions of degree $N$ (per dimension) on each cell: $\sum_{p=0}^{N}\sum_{q=0}^N\dots C_{pq\dots} x_1^p\,x_2^q\dots$
A straightforward way of performing a multidimensional interpolation is to employ a sequence of 1d interpolation operations for each dimension separately. The 1d interpolation function is defined by $N+1$ coefficients -- the values and certain derivatives of the polynomial at two adjacent nodes in the $d$-th coordinate. These coefficients are obtained from $(N-1)$-dimensional interpolation along other coordinates in each of these two nodes.

To illustrate this, consider the case of piecewise-cubic Hermite interpolation, specified by the value and the first derivative in each dimension. For a 2d point $\{x,y\}$, we first locate the indices of grid nodes $\{i,j\}$ in each dimension that enclose the point: $x_i\le x \le x_{i+1},\; y_j\le y \le y_{j+1}$. Then we perform four 1d interpolation operations in $x$ to find $f(x, y_j),\; f(x, y_{j+1}),\; f'_y(x, y_j),\; f'_y(x, y_{j+1})$, using 16 coefficients at 4 corners of the cell (i.e., $f(x_i,y_j),\; f(x_{i+1},y_j),\; f'_x(x_i,y_j),\; f'_x(x_{i+1},y_j)$ for the first value, etc.). Finally the last interpolation in $y$ produces the required value. The same result would be obtained, had we reversed the order of coordinates. 
If we also want partial derivatives in both dimensions, then we need to compute higher derivatives on the first stage (i.e., from the first four coefficients we get not only $f(x, y_j)$, but also $f'_x(x, y_j)$ and $f''_{xx}(x, y_j)$), and then use extra 1d interpolation in $y$ per each output derivative on the second stage (e.g., $f''_{xx}(x, y)$ is obtained from $f''_{xx}(x, y_j),\; f''_{xx}(x, y_{j+1}),\; f'''_{xxy}(x, y_j),\; f'''_{xxy}(x, y_{j+1})$, and the latter one is computed on the first stage from $f'_y(x_i, y_{j+1}),\; f'_y(x_{i+1}, y_{j+1}),\; f''_{xy}(x_i, y_{j+1}),\; f''_{xy}(x_{i+1}, y_{j+1})$). 

Thus we see that for a cubic Hermite interpolation we need the values and derivatives in each direction at all grid nodes: $f(x_i, y_j),\; f'_x(x_i, y_j),\; f'_y(x_i, y_j),\; f''_{xy}(x_i, y_j)$. How do we find them, given only the function values at grid nodes? It turns out that the concept of cubic splines can naturally be extended to the multidimensional case, using the following multi-stage procedure. At the beginning, the first derivatives $f'_x, f'_y$ in each dimension are found from the condition that the corresponding \textit{second} derivatives $f''_{xx}, f''_{yy}$ are continuous at grid nodes. Then the mixed second derivative $f''_{xy}(x_i, y_j)$ is computed by establishing a cubic spline for $f'_y(x, y_j)$ at each $j$ and taking its derivative in $x$. The elegance of this approach is in its symmetry: the same value for $f''_{xy}$ is also produced by constructing a cubic spline for $f'_x(x_i, y)$ at each $x_i$ and taking the derivative in $y$. To understand why, recall that the natural cubic spline for $g(x)$ is a linear function of the input values $g_i\equiv g(x_i)$ defined by the array of grid nodes $\{x_i\}$. Equivalently, it is described by the matrix $\mathsf{X}$ that transforms the array of input values to the array of spline derivatives at grid nodes: $g'_i\equiv g'(x_i) = \sum_{k=1}^{K_x} X_{ik}\,g_k$. The complete 2d matrix $\mathsf{f}'_x$ of $x$-derivatives $f'_{x;\,ij} \equiv f'_x(x_i, y_j)$ is then given by $\mathsf{f}'_x = \mathsf{X\,f}$, where the elements of matrix $\mathsf{f}$ are $f_{ij} \equiv f(x_i, y_j)$. On the other hand, the interpolation in the $y$ direction is provided by the matrix $\mathsf{Y}$ such that for any array of values $h_j\equiv h(y_j)$, the $y$-derivatives are given by $h'_j\equiv h'(y_j) = \sum_{k=1}^{K_y} Y_{jk}\,g_k$. The complete 2d matrix $\mathsf{f}'_y$ of $y$-derivatives $f'_{y;\,ij} \equiv f'_y(x_i, y_j)$ is given by $\mathsf{f}'_y = (\mathsf{Y}\,\mathsf{f}^T)^T = \mathsf{f\,Y}^T$. Now if we compute the mixed second derivatives $\mathsf{f}''_{xy}$ by constructing the $y$-spline from $\mathsf{f}'_x$, this results in $\mathsf{f}''_{xy} = \mathsf{f}'_x\,\mathsf{Y}^T = (\mathsf{X\,f})\,\mathsf{Y}^T$, whereas computing them from the $\mathsf{f}'_y$ results in $\mathsf{X\,f}'_y = \mathsf{X}\,(\mathsf{f\,Y}^T)$ -- which are identical matrices.

The same scheme works in three dimensions: now we need eight coefficients at each grid node $\{x_i,y_j,z_k\}$ -- $f, f'_x, f'_y, f'_z, f''_{xy}, f''_{xz}, f''_{yz}, f'''_{xyz}$, which are all found in three steps, using just the values of $f$ and continuity conditions on higher derivatives. The evaluation of $f(x,y,z)$ also proceeds in three stages: first a total of 64 coefficients (8 numbers at 8 cell corners) are fed into 16 Hermite cubic interpolation operations in $x$, then the resulting 16 numbers are used in 4 interpolations in $y$, and finally one interpolation in $z$. As before, the outcome does not depend on the order. This should also work in higher dimensions, although would clearly become more clumsy.

We now consider a more complicated case of 2d quintic interpolation. Similarly to the cubic Hermite case, we may interpolate with a piecewise-quintic polynomial in each direction using the following 9 quantities stored at each node: $f, f'_x, f''_{xx}, f'_y, f''_{xy}, f'''_{xxy}, f''_{yy}, f'''_{xyy}, f''''_{xxyy}$, first performing 6 interpolations in $y$ to compute $f$, $f'_x$, $f''_{xx}$ at two nodes of the $x$-grid, and then the final interpolation in $x$, or vice versa.
By a similar argument, if we are provided with four matrices $\mathsf{f}, \mathsf{f}'_x, \mathsf{f}'_y, \mathsf{f}''_{xy}$, we can construct 1d quintic splines in each direction and use them to initialize the remaining five derivatives at each node from the conditions of continuity of still higher derivatives; again this will not result in a conflicting assignment of the mixed fourth derivative $f''''_{xxyy}$. Unfortunately, in practice we often have only three matrices to work with -- the function and its two partial derivatives at each node, but no mixed second derivative. We were not able to come up with an equally elegant method in this case, although the following kludge seems to deliver satisfactory results. We first construct the 1d quintic splines for $f(x, y_j)$ and $f(x_i, y)$ at each node, and use them to compute the second derivatives ($f''_{xx}$ and $f''_{yy}$, correspondingly). To compute the mixed derivatives, we construct natural cubic splines in $y$ from the values of $f'_x$ and $f''_{xx}$, and similarly in $x$ from the values of $f'_y$ and $f''_{yy}$, and then differentiate them. The resulting values do not agree with each other, so we take either the average of the two estimates, or the one that is expected to be more accurate: for instance, if $i=1$, we would compute $f''_{xy,ij}$ by differentiating the spline for $f'_x(x_i,y)$ by $y$, because the alternative variant (differentiating the spline for $f'_y(x,y_j)$ by $x$) does not provide a good estimate at the endpoint of its $x$-grid. In this way, all remaining derivatives are assigned.

%%%%%%%%%%%%%%
\subsubsection{Penalized spline regression}  \label{sec:MathSplineApproxDetails}

Suppose we have $N_\mathrm{data}$ points $\{\bx, \by\}$ with weights $\boldsymbol{w}$, and we need to find a smooth function $y=f(x)$ that approximates the data in the weighted least-square sense, but does not fluctuate too much -- in other words, minimize the functional
\begin{align}  \label{eq:ObjectiveSplineFit}
\mathcal{Q} &\equiv \sum_{i=1}^{N_\mathrm{data}} w_i\;\left[y_i - f(x_i)\right]^2 + \lambda \int \left[f''(x)\right]^2 \,\d x .
\end{align}
Here $\lambda\ge 0$ is the smoothing parameter that controls the tradeoff between approximation error and wiggliness of the function \cite{GreenSilverman}; its choice is discussed below.

We represent $f(x)$ as a sum of B-spline basis functions of degree $N$, $B^{[N]}_k(x)$, defined by the grid of $N_\mathrm{grid}$ knots, with adjustable amplitudes $A_k$:
\begin{align}  \label{eq:FncSplineFit}
f(x) &\equiv \sum_{k=1}^{N_\mathrm{basis}} A_k\,B_k(x).
\end{align}
We use the basis set of modified cubic ($N=3$) B-splines with natural boundary conditions, so that the number of basis functions $N_\mathrm{basis}$ is equal to the number of grid knots. Note that the value of interpolated function $f(x)$ at a grid point $x_k$ is \textit{not} equal to the amplitude of the corresponding basis function $A_k$, but rather to a linear combination of three adjacent amplitudes $A_{k-1},A_k,A_{k+1}$, see Figure~\ref{fig:SplineKernels}, left panel; however, for convenience we represent the result in terms of the array of function values $f(x_k)$, which may be used to initialize a natural cubic spline in the usual way.

Let the matrix $\mathsf{B}$ with $N_\mathrm{data}$ rows and $N_\mathrm{basis}$ columns contain the values of basis functions at data points: $B_{ik} = B_k(x_i)$. Due to the locality of B-splines, this matrix is sparse -- each row contains at most $N+1$ nonzero consecutive elements. The grid in $x$ does not need to encompass all data points -- the function $f(x)$ is linearly extrapolated outside the grid boundaries.
Define the ``roughness matrix'' $\mathsf{R}$ containing the integrals of products of second derivatives of basis functions: $R_{kl} \equiv \int B_k''(x) \, B_l''(x)\, \d x$. This matrix is also sparse (band-diagonal) and symmetric.
The functional $\mathcal{Q}$ to be minimized (\ref{eq:ObjectiveSplineFit}) may be rewritten as
\begin{align}  \label{eq:ObjectiveSplineFit2}
\mathcal{Q}\equiv (\by - \mathsf{B} \bA)^{T}\, \mathsf{W}\, (\by - \mathsf{B} \bA) +
\lambda\, \bA^T\,\mathsf{R}\, \bA, \qquad \mathsf{W}\equiv \mathrm{diag}(\boldsymbol{w}),
\end{align}
and its minimum is obtained by solving the normal equations $\d\mathcal{Q}/\d\bA=0$ for the amplitudes $\bA$:
\begin{align}  \label{eq:SplineFitSol}
\left(\mathsf{B}^T\mathsf{W\,B} + \lambda \mathsf{R} \right) \bA &= \mathsf{B}^T\mathsf{W}\,\by .
\end{align}
Note that the size of this linear system is only $N_\mathrm{basis}\times N_\mathrm{basis}$, possibly much smaller than the number of data points $N_\mathrm{data}$. If one needs to solve the system for several values of $\lambda$ and/or different vectors $\by$ (with the same coordinates $\bx$ and weights $\boldsymbol{w}$), there is an efficient algorithm for this \cite{RuppertWandCarroll}:
\begin{enumerate}
\item Compute the Cholesky decomposition of the matrix $\mathsf{B}^T\mathsf{W\,B}$, representing it as $\mathsf{L}\mathsf{L}^T$, where $\mathsf{L}$ is a lower triangular matrix with size $N_\mathrm{basis}$. To avoid problems when $\mathsf{B}^T\mathsf{W\,B}$ is singular (which occurs when some grid segments contain no points), we add a small multiple of $\mathsf{R}$ before computing the decomposition.\\
Then compute the singular-value decomposition of the symmetric positive definite matrix $\mathsf{L}^{-1} \mathsf{R} \mathsf{L}^{-T}$, representing it as $\mathsf{U}\, \mathrm{diag}(\boldsymbol{S})\, \mathsf{U}^T$, where $\mathsf{U}$ is a square orthogonal matrix (i.e., $\mathsf{U}\mathsf{U}^T=\mathsf{I}$) with size $N_\mathrm{basis}$, and $\boldsymbol{S}$ is the vector of singular values.\\
Now the matrix in the l.h.s.\ of (\ref{eq:SplineFitSol}) can be written as\\
$\mathsf{B}^T\mathsf{W\,B}+\lambda \mathsf{R} = \mathsf{L} \mathsf{L}^T + \mathsf{L} \mathsf{L}^{-1} \mathsf{R} \mathsf{L}^{-T} \mathsf{L}^T = \mathsf{L} \mathsf{U} \mathsf{U}^T \mathsf{L}^T + \mathsf{L} \mathsf{U}\, \mathrm{diag}(\boldsymbol{S})\, \mathsf{U}^T \mathsf{L}^T$.\\
Finally, compute a matrix $\mathsf{M}\equiv \mathsf{L}^{-T}\mathsf{U}$.
\item For any vector of $\by$ values, pre-compute $\boldsymbol{p}\equiv \mathsf{B}^T\mathsf{W}\,\by$ and $\boldsymbol{q}\equiv \mathsf{M}^T\boldsymbol{p}$ (vectors of length $N_\mathrm{basis}$), and the weighted sum of squared $y$-values $V\equiv \by^T \mathsf{W}\,\by = \sum_{i=1}^{N_\mathrm{data}} w_i y_i^2$.
\item Now for any choice of $\lambda$, the solution is given by
\begin{align}
\bA = \mathsf{M}\, [\mathsf{I}+\lambda\,\mathrm{diag}(\boldsymbol{S})]^{-1}\,\boldsymbol{q},
\end{align}
i.e., involves only a multiplication of a vector by inverse elements of a diagonal matrix and a single general matrix-vector multiplication.
The residual sum of squares -- the first term in (\ref{eq:ObjectiveSplineFit2}) -- is given by
\begin{align}
\mathrm{RSS} \equiv (\by - \mathsf{B}\bA)^T\,\mathsf{W}\,(\by - \mathsf{B}\bA)^T =
V - 2\bA^T\boldsymbol{p} + |\mathsf{L}^T\bA|^2.
\end{align}
\end{enumerate}

In case of non-zero smoothing, the effective number of free parameters is lower than the number of basis functions, and is given by the number of equivalent degrees of freedom:
\begin{align}
\mathrm{EDF} \equiv \mathrm{tr}[\mathsf{I} + \lambda\,\mathrm{diag}(\boldsymbol{S})]^{-1} =
\sum_{k=1}^{N_\mathrm{basis}} \frac{1}{1+\lambda S_k} \;.
\end{align}
It varies from $N_\mathrm{basis}$ for $\lambda=0$ to 2 for $\lambda\to\infty$ (which corresponds to a two-parameter linear least-square fit). The amount of smoothing thus may be specified by EDF, which has a more direct interpretation than $\lambda$. The optimal choice of smoothing parameter is given by minimization of the (corrected) Akaike information criterion:
\begin{align}
\mathrm{AIC_C} \equiv \ln(\mathrm{RSS}) + \frac{2\,(\mathrm{EDF}+1)}{N_\mathrm{data}-\mathrm{EDF}-2}.
\end{align}

\begin{figure}[t]
\begin{center}
\includegraphics[width=12cm]{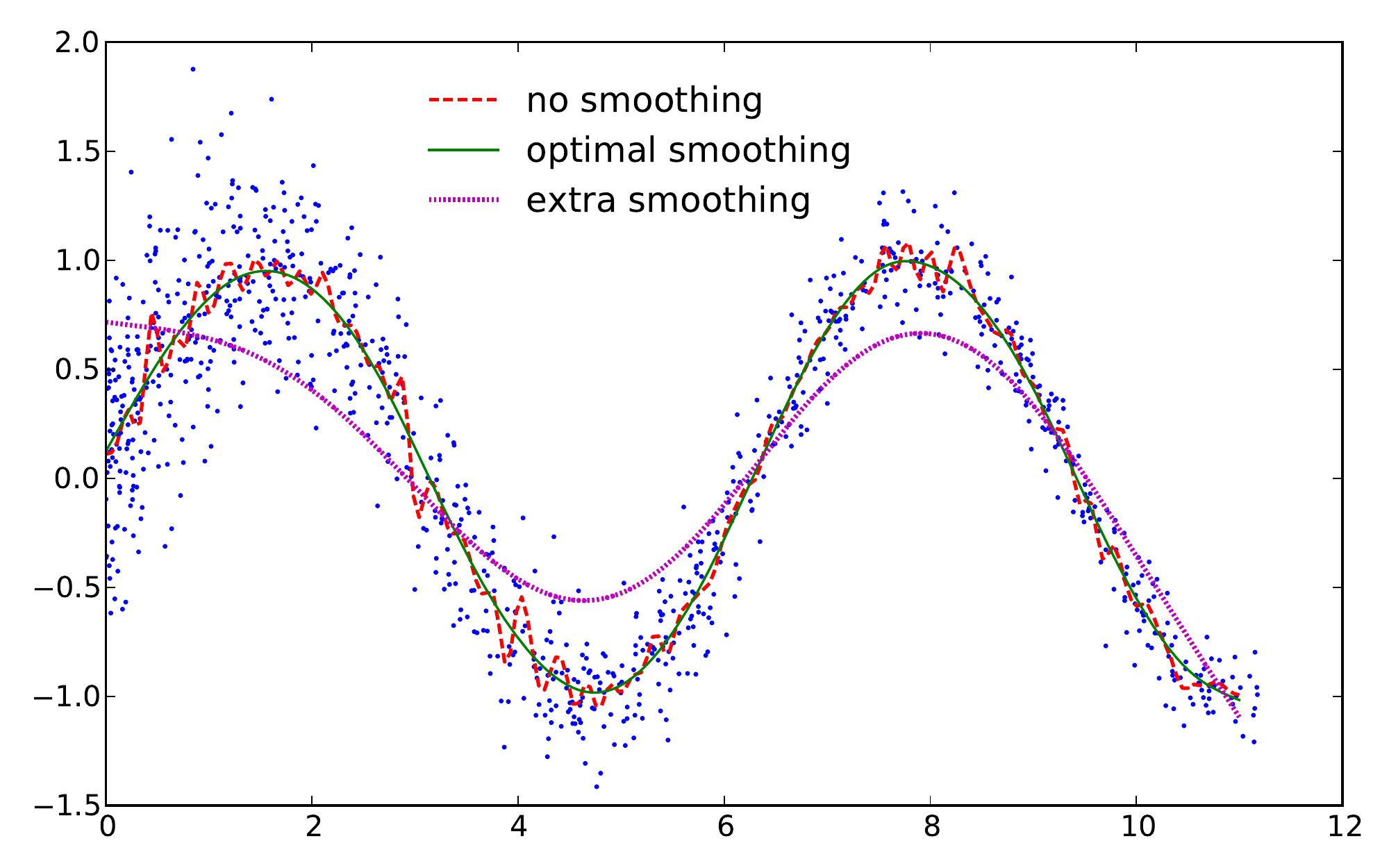}
\end{center}
\caption{Penalized spline fit to noisy data. $N_\mathrm{data}=1000$ points follow a sine curve with a random noise in $y$-coordinate. A non-penalized spline fit with 100 nodes (red dashed line) is overfitting the noise, whereas an optimally-smoothed spline with the same number of points (solid green line) recovers the original trend very well; of course, we could use a far smaller number of nodes ($\lesssim 10$) and still get a decent fit, but the optimal amount of smoothing prevents overfitting even if the node spacing is too dense. Finally, dotted magenta line illustrates the effect of oversmoothing.
} \label{fig:SplineFit}
\end{figure}

Often one may wish to apply a somewhat stronger smoothing than the one given by minimizing AIC, for instance, by allowing it to be larger than the minimum value by a specified amount $\Delta\mathrm{AIC}\sim \mathcal{O}(1)$. In both cases, the corresponding value of $\lambda$ is  obtained by standard one-dimensional minimization or root-finding routines. Figure~\ref{fig:SplineFit} illustrates the method.

%%%%%%%%%%%%%%
%%%%%%%%%%%%

%%%%%%%%%%%%%%
\subsubsection{Penalized spline density estimate}  \label{sec:MathSplineDensityDetails}

Let $P(x)>0$ be a density function defined on the entire real axis, a semi-infinite interval $[x_\mathrm{min},+\infty)$ or $(-\infty,x_\mathrm{max}]$, or a finite interval $[x_\mathrm{min},x_\mathrm{max}]$.
Let $\{x_i, w_i\}$ be an array of $N_\mathrm{data}$ samples drawn from this distribution, where $x_i$ are their coordinates, and $w_i\ge 0$ are weights. We follow the convention that $\int P(x)\,\d x$ over its domain is equal to $M\equiv \sum_i w_i$ (not necessarily unity).

We estimate $P(x)$ using a B-spline approximation to $\ln P$ constructed for a grid of $N_\mathrm{grid}$ nodes $\{X_k\}$ \cite{OSullivan1988}. We implemented two variants of basis set: linear B-splines and modified cubic B-splines with natural boundary conditions; in both cases the number of basis function $N_\mathrm{basis}$ is equal to the number of grid nodes. The estimated log-density is thus
\begin{align}
\ln P(x;\,\bA) = \sum_{k=1}^{N_\mathrm{basis}}  A_k\, B_k(x) - \ln G_0 + \ln M \equiv Q(x; \bA) - \ln G_0(\bA) + \ln M ,
\end{align}
where $A_k$ are the amplitudes -- free parameters that are adjusted during the fit,\\
$B_k(x)$  are  basis functions,\\
$Q(x; \bA) \equiv \sum_k  A_k\, B_k(x)$  is the weighted sum of basis function, \\
and $G_0(\bA) \equiv \int \exp[Q(x; \bA)]\, \d x$  is the normalization constant determined from the condition that $\int P(x)\,\d x = M$.
There is a gauge freedom in the choice of amplitudes $A_k$: if we add a constant to $Q(x; \bA)$, it would not have any effect on $\ln\mathcal{L}$ because this shift will be counterbalanced by $G_0$. We eliminate this freedom by fixing the amplitude of the last basis function to zero ($A_{N_\mathrm{basis}}=0$), thus retaining $N_\mathrm{ampl}\equiv N_\mathrm{basis}-1$ free parameters.

As in the case of penalized spline regression, we first compute the matrix $\mathsf{B}$ of weighted basis-function values at each input point: $B_{ik} \equiv w_i\,B_k(x_i)$. This matrix is large ($N_\mathrm{data}$ rows, $N_\mathrm{ampl}$ columns) but sparse, and is further transformed into two vectors of length $N_\mathrm{ampl}$ and a square matrix of the same size:
$\boldsymbol{V} \equiv \mathsf{B}^T\boldsymbol{1}_{N_\mathrm{data}}$ (i.e., $V_k = \sum_i w_i\,B_k(x_i)$), $\boldsymbol{W} \equiv \mathsf{B}^T\boldsymbol{w}$, $\mathsf{C} \equiv \mathsf{B}^T\,\mathsf{B}$. In the remaining steps of the procedure, only vectors and matrices of size $N_\mathrm{ampl}$ rather than $N_\mathrm{data}$ are involved, which allows to deal efficiently even with very large arrays of samples described by a moderate number of parameters (such as fitting the density profile of an \Nbody model with a few dozen grid points in radius).

The total penalized likelihood of the model given the vector of amplitudes $\bA$ is
\begin{align}
\ln\mathcal{L}\; &\equiv\; \ln\mathcal{L}_\mathrm{data} - \lambda \mathcal{R}(\bA) \;\equiv\;
\sum_{i=1}^{N_\mathrm{data}}  w_i  \ln P(x_i;\,\bA) - \lambda \int \left[\ln P''(x_i)\right]^2 \,\d x \nonumber \\
&= \sum_{i=1}^{N_\mathrm{data}} w_i \left(\sum_{k=1}^{N_\mathrm{ampl}} A_k B_k(x_i) - \ln G_0(\bA) + \ln M \right) - \lambda \sum_{k=1}^{N_\mathrm{ampl}} \sum_{l=1}^{N_\mathrm{ampl}} A_k\,A_l\,R_{kl}  \nonumber\\
&= \big[ \boldsymbol{V}^T\bA - M\, \ln G_0(\bA)  + M\, \ln M \big] - \lambda\,\bA^T\,\mathsf{R}\,\bA ,
\label{eq:MaxLikelihoodDensity}
\end{align}
where $\lambda \mathcal{R}$ is the roughness penalty term, and the matrix $R_{kl}\equiv \int B_k''(x)\,B_l''(x)\,\d x$ is also pre-computed at the beginning of the procedure%
\footnote{It may be advantageous to use third derivatives here \cite{Silverman1982}, in which case the solution in the limit of infinite smoothing ($\mathcal{R}\to 0$) corresponds to a Gaussian density profile; however, for our choice of \textit{natural} cubic splines it is unattainable, since a quadratic function (logarithm of a Gaussian) has non-zero second derivatives at endpoints, and hence cannot be represented by a spline with natural boundary conditions.}.
The smoothing parameter $\lambda$ controls the tradeoff between the likelihood of the data and the wiggliness of the estimated density; its choice is discussed below.

Unlike the penalized spline regression problem, in which the amplitudes are obtained from a linear equation, the problem of penalized spline density estimation is nonlinear because of the normalization factor $G_0(\bA)$. The amplitudes $\bA$ that minimize $-\!\ln\mathcal{L}$ (\ref{eq:MaxLikelihoodDensity}) are found by solving the system of equations $\D \ln\mathcal{L}/\D A_k=0$ iteratively, using a multidimensional Newton method with explicit expressions for the gradient and hessian:
\begin{subequations}
\begin{align}
-\frac{\D \ln\mathcal{L}}{\D A_k} &= -V_k + M \frac{\D \ln G_0}{\D A_k} + 2\lambda \sum_l R_{kl}\,A_l \;, \label{eq:lnLgrad} \\
-\frac{\d^2 \ln\mathcal{L}}{\D A_k\, \D A_l} &= M \frac{\D ^2 \ln G_0}{\D A_k\, \D A_l} + 2\lambda R_{kl} \,\equiv H_{kl}\;, \\
\mbox{where }G_0 &\equiv \int \exp[Q(x; \bA)]\,\d x \;, \quad
Q(x; \bA) \equiv \sum_k  A_k\, B_k(x) \;, \nonumber \\
\frac{\D \ln G_0}{\D A_k} &= \frac{\int B_k(x)\,\exp[Q(x; \bA)]\,\d x}{G_0} \;, \nonumber\\
\frac{\D ^2 \ln G_0}{\D A_k\, \D A_l} &= \frac{\int B_k(x)\,B_l(x)\,\exp[Q(x; \bA)]\,\d x}{G_0} - \frac{\D \ln G_0}{\D A_k}\; \frac{\D \ln G_0}{\D A_l} \;. \nonumber
\end{align}
\end{subequations}

The choice of smoothing parameter $\lambda$ may be done by cross-validation: for each sample $i$, we compute its likelihood using best-fit parameters $\bA^{(i)}$ calculated for all samples except this one, and then sum these values over all samples. 
\begin{align}  \label{eq:CrossValidation}
\ln\mathcal{L}_\mathrm{CV}(\lambda) &\equiv \sum_{i=1}^{N_\mathrm{data}} w_i \, \ln P(x_i;\, \bA^{(i)}) =
\sum_{i=1}^{N_\mathrm{data}} w_i \left(\sum_{k=1}^{N_\mathrm{basis}} A_k^{(i)} B_k(x_i) - \ln G_0(\bA^{(i)}) \right) + M\ln M.
\end{align}

Of course, it would be prohibitively expensive to compute the best-fit amplitudes $\bA^{(i)}$ separately for each omitted point; instead, we express them as small perturbations of $\bA$, by demanding that the l.h.s. of (\ref{eq:lnLgrad}) is zero for each $i$ at the corresponding $\bA^{(i)}$:
\begin{align*}
0 &= -V_k + w_i\,B_k(x_i) + M \frac{\D \ln G_0}{\D A_k} + M \frac{\D ^2 \ln G_0}{\D A_k\,\D A_l}(A_l^{(i)}-A_l) + 2\lambda \sum_l R_{kl}\,A_l^{(i)} \;, \\
\delta A^{(i)}_l &\equiv A^{(i)}_l - A_l = - \left[ M \frac{\D ^2 \ln G_0}{\D A_k\,\D A_l} + 2\lambda R_{kl} \right]^{-1} w_i\,B_k(x_i) \;,\mbox{ or }\;
\delta \mathsf{A} = -\mathsf{H}^{-1} \mathsf{B}^T .
\end{align*}

Here the gradient and hessian of $G_0$ are taken at the overall best-fit amplitudes $\bA$ for the entire sample, computed for the given value of $\lambda$. The matrix $\delta \mathsf{A}$ with $N_\mathrm{ampl}$ rows and $N_\mathrm{data}$ columns needs not be computed explicitly each time. Finally, the cross-validation score (\ref{eq:CrossValidation}) is expressed as
\begin{align}  \label{eq:CrossValidation2}
\ln\mathcal{L}_\mathrm{CV}(\lambda) = \ln\mathcal{L}_\mathrm{data}
- \mathrm{tr}(\mathsf{H}^{-1}\mathsf{C}) + \frac{\d\ln G_0(\bA)}{d\bA}\, \mathsf{H}^{-1}\, \boldsymbol{W} .
\end{align}

Here $\ln\mathcal{L}_\mathrm{data}$ is the expression in brackets in (\ref{eq:MaxLikelihoodDensity}). The optimal value of $\lambda>0$ that maximizes the cross-validation score is found by a simple one-dimensional search. We first assign a reasonable initial guess for amplitudes (approximating the density as a Gaussian with the mean and dispersion computed from input samples). At each step, the multidimensional nonlinear root-finder routine is invoked to find best-fit amplitudes $\bA$ for the given $\lambda$, starting from the current initial guess. If it was successful and $\ln\mathcal{L}_\mathrm{CV}(\lambda)$ is higher than the current best estimate, the initial guess is replaced with the best-fit amplitudes: this not only speeds up the root-finder, but also improves the convergence. The range of $\lambda$ is progressively narrowed until the maximum has been located with sufficient accuracy, at which point we return the last successful array of $\bA$. 

%%%%%%%%%%%%%%
\begin{figure}[t]
\begin{center}
\includegraphics[width=12cm]{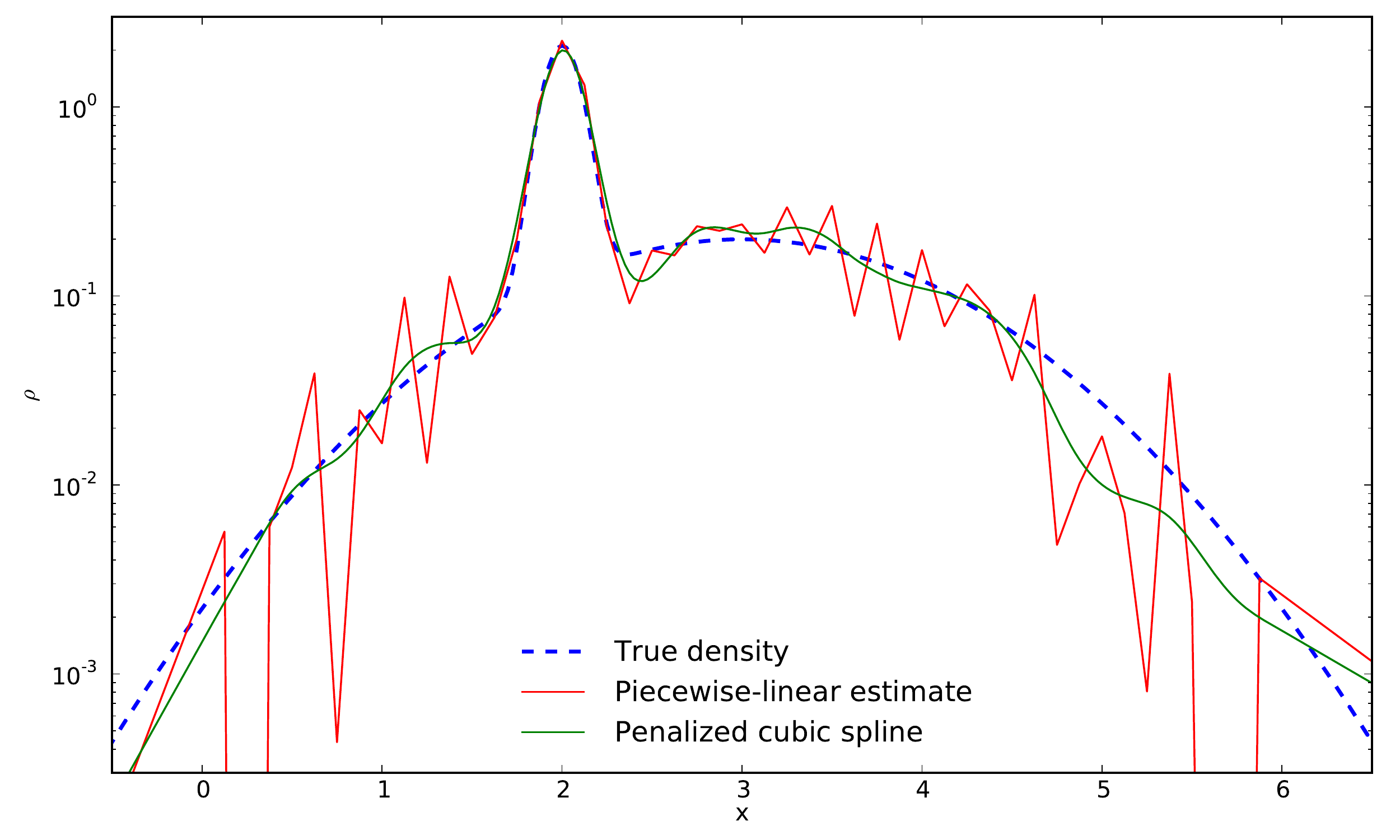}
\end{center}
\caption{Penalized spline estimate of density from sample points. Here $N_\mathrm{data}=1000$ were drawn from the original density profile (shown in dashed blue) described by a sum of two gaussians, with dispersions equal to 0.1 and 1. We reconstruct the logarithm of density using a linear ($N=1$) and cubic ($N=3$) B-splines with 50 nodes uniformly placed on the interval $[0..6]$, so that it is linearly extrapolated beyond the extent of this grid. The linear B-spline estimate (shown in red) is rather wiggly, because the grid spacing is intentionally made too fine for the given number of samples -- some elements do not contain any samples. The non-penalized cubic B-spline (not shown) is very close to the linear one, and also close to a standard Gaussian kernel density estimate with the same bandwidth as the grid spacing (also not shown). By contrast, the penalized cubic B-spline with the smoothing parameter determined automatically in order to maximize the cross-validation likelihood (shown in blue) is much closer to the true density.
} \label{fig:SplineLogDensity}
\end{figure}
%%%%%%%%%%%%

Figure~\ref{fig:SplineLogDensity} illustrates the application of linear (non-smoothed) and cubic spline with optimal smoothing to a test problem. In this case the grid spacing was deliberately too dense for the given number of samples, so that some grid segments do not contain any samples, but nevertheless the penalized density estimate comes out rather close to the true one.
This regime is not very stable, though, and for normal operation the grid should be assigned in such a way that each segment contains at least a few samples -- this ensures that even the un-smoothed estimate is mathematically well defined.

Maximization of cross-validation score is considered to be the ``optimal'' smoothing; however, in some cases the inferred $\ln P(x)$ may still be too wiggly. An alternative approach is to estimate the expected scatter in $\ln\mathcal{L}_\mathrm{data}$ for a sample of finite size $N_\mathrm{data}$, and allow the likelihood score to be worse than the best-fit score by an amount comparable to this expected scatter. In the case of uniform-weight samples and zero smoothing, the mean and dispersion in $\ln\mathcal{L}$ are
\begin{align}
\big\langle\ln\mathcal{L}\big\rangle &=
  \int P(x)\,\ln P(x)\, \d x = M\,\left[\frac{G_1}{G_0} + \ln M - \ln G_0\right] ,\\
\Big\langle\big(\ln\mathcal{L}-\langle\ln\mathcal{L}\rangle\big)^2\Big\rangle &=
  N_\mathrm{data}^{-1} \left( M\!\int\! P(x)\,[\ln P(x)]^2\, \d x - \big\langle\ln\mathcal{L}\big\rangle^2 \right) =
  M^2 N_\mathrm{data}^{-1} \left[\frac{G_2}{G_0} - \left(\frac{G_1}{G_0}\right)^{\!2}\right]\! , \nonumber\\
\mbox{where }\; G_n &\equiv \int \big[Q(x)\big]^n\, \exp\big[Q(x)\big]\,\d x\;.  \nonumber
\end{align}

We first determine the best-fit $\bA$ for the optimal value of $\lambda_\mathrm{opt}$ and compute the expected r.m.s. scatter $\delta\ln\mathcal{L}\equiv \sqrt{\langle(\ln\mathcal{L}-\langle\ln\mathcal{L}\rangle)^2\rangle}$ from the above equation; then we search for $\lambda$ such that $\ln\mathcal{L}_\mathrm{data}(\lambda) = \ln\mathcal{L}_\mathrm{data}(\lambda_\mathrm{opt}) - \kappa\, \delta\ln\mathcal{L}$, where $\kappa\sim 1$ is a tunable parameter. The resulting density is less fluctuating, but the asymptotic behaviour near or beyond grid boundaries, where the number of samples is low, may be somewhat biased as a result of more aggressive smoothing.

%%%%%%%%%%%%%%
\subsubsection{Gauss--Hermite series}  \label{sec:MathGaussHermiteDetails}

Gauss--Hermite (GH) expansion is another type of basis set, useful for representing velocity distribution functions, which are typically not too dissimilar to a Gaussian profile. Unlike the B-spline basis set, which is specified by the number and location of grid nodes and the degree of polynomials, the GH basis is specified by the parameters of the zero-order function (a Gaussian with the given amplitude $\Xi$, mean value $\mu$, and width $\sigma$) and the order of expansion $M$. The total number of basis functions is $M+1$, and they are defined as
\begin{subequations}
\begin{align}  \label{eq:GaussHermiteBasis}
\mathcal B_m(x) \equiv \frac{\Xi}{\sqrt{2\pi}\;\sigma} \;
\exp\left[ -\frac{1}{2} \left( \frac{x-\mu}{\sigma} \right)^2 \right] \;
\mathcal H_m \left( \frac{x-\mu}{\sigma} \right) \,,
\end{align}
where $\mathcal{H}_m(y)$ are the (astrophysical) Hermite polynomials \cite{Gerhard1993, vdMarelFranx1993}, defined by a recurrence relation
\begin{align}  \label{eq:HermitePoly}
\mathcal H_0 = 1 ,\quad 
\mathcal H_1(y) = \sqrt{2}\,y ,\quad
\mathcal H_{m+1}(y) = 
\Big[ \sqrt{2}\,y\, \mathcal H_m(y) - \sqrt{m}\, \mathcal H_{m-1}(y) \Big] \Big/ \sqrt{m+1} \,.
\end{align}
\end{subequations}
They differ in normalization from the physicist's definition of Hermite polynomials $H(y)$: 
$\mathcal H_m(y) = H_m(y) / \sqrt{2^m\,m!}$. The GH basis functions are orthogonal on the real axis, with the Gram matrix being
\begin{align}
\mathcal G_{mn} \equiv \int_{-\infty}^{\infty} \mathcal B_m(x)\, \mathcal B_n(x)\, \d x =
\frac{\Xi^2}{2\sqrt{\pi}\,\sigma}\, \delta_{mn} \,.
\end{align}

According to the general definition given in Section~\ref{sec:MathBasisSetDetails}, any function $f(x)$ can be approximated as a sum of GH basis functions multiplied by amplitudes $\boldsymbol{h}$ (also called GH coefficients):
\begin{align}  \label{eq:GHcoefs}
\tilde f(x) = \sum_{m=0}^{M} h_m\, \mathcal B_m(x), \quad
h_m = \frac{2\sqrt{\pi}\,\sigma}{\Xi^2} \int_{-\infty}^{\infty} f(x)\, \mathcal B_m(x)\, \d x \,.
\end{align}

For the fixed parameters $\Xi, \mu, \sigma, M$, this expansion is defined in a unambiguous way. However, there is some redundancy (most obviously in the overall amplitude factor $\Xi$, which merely rescales the amplitudes $\boldsymbol h$). If one has freedom to adjust $\Xi$, $\mu$ and $\sigma$, it makes sense to define them in such a way that the first three coefficients are $h_0=1, h_1=h_2=0$, which is always possible (more on this later). In this case, if one limits the order of expansion $M$ to 2 (or, in fact, to 0), the approximated function $\tilde f(x)$ is actually the best-fit Gaussian for the original function $f(x)$. Addition of higher-order coefficients $h_m (m\ge 3)$ allows one to represent the deviations from this Gaussian, but does not change the first three coefficients, thanks to the orthogonality of the basis set.

On the other hand, if one needs to compare two functions represented by the GH series, this can only be done if the parameters of the basis set $\Xi, \mu, \sigma$ are the same. For instance, when the observational constraints on the velocity distribution come in the form of GH coefficients $h_m$ (with the given $\Xi, \mu, \sigma$ and with the default values of $h_0=1, h_1=h_2=0$), the same function in the model must be represented in the same basis, even though the coefficients $h_{0,1,2}$ do not have the default values.

In practice, it may be convenient to represent the velocity distribution of the model in terms of a B-spline expansion with amplitudes $A_j$, and then convert it into the GH coefficients $h_m$. This is, of course, a linear operation $\boldsymbol h = \mathcal G^{-1}\, \mathsf C \,\boldsymbol A$ (for the given parameters of the two basis sets), with the matrix $C_{mj} = \langle \mathcal B_m, B_j \rangle$.

It is important to keep in mind that $\Xi, \mu, \sigma$ are \textit{not} identical to the overall normalization, mean and dispersion of the function $\tilde f$. These are given by an amplitude-weighted sum over the corresponding values for \text{all} basis functions; for instance, the overall normalization is
\begin{subequations}
\begin{align}  \label{eq:GHnormalization}
\int_{-\infty}^{\infty} \tilde f(x)\, \d x =
\sum_{m=0}^M  h_m\, \int_{-\infty}^{\infty} \mathcal B_m(x)\, \d x =
\Xi\!\!\sum_{\mbox{$m$ even}}\!  h_m\, \frac{\sqrt{m!}}{m!!} \,,
\end{align}
and the second moment times the above quantity is
\begin{align}  \label{eq:GHdispersion}
\int_{-\infty}^{\infty} \tilde f(x)\, x^2\, \d x =
\Xi\!\!\sum_{\mbox{$m$ even}}\!  h_m\, \frac{(2m+1)\,\sqrt{m!}}{m!!} \,.
\end{align}
\end{subequations}

As mentioned above, the GH expansion is defined by the amplitude $\Xi$, center $\mu$ and width $\sigma$ of the zero-order function (a pure Gaussian). Given a sufficiently large number of terms $M$, it can approximate any reasonably smooth function $f(x)$ to a desired accuracy; however, the speed of convergence naturally depends on the choice of $\mu$ and $\sigma$. There are several possible approaches for setting these parameters.

\begin{enumerate}
\item van der Marel \& Franx \cite{vdMarelFranx1993} choose the parameters $\mu, \sigma$ so that the lowest-order term approximates the function as closely as possible (this implies that only $h_0$ is nonzero, and hence the GH expansion is a pure Gaussian -- note that the best-fit Gaussian is \textit{not} a Gaussian with the width equal to the dispersion of the function!), then construct a full GH expansion with these parameters, computing the coefficients $h_m$ according to (\ref{eq:GHcoefs}). Thanks to the orthogonality of basis functions, the addition of subsequent terms does not change the previous expansion coefficients, hence in this (and \textit{only in this}) case $h_1=h_2=0$.
\item The same authors find it ``more convenient in practice'' to fit a truncated GH series $\mathscr L(x) \equiv \mathcal B_0(x) + \sum_{m=3}^M h_m\,\mathcal B_m(x)$ to the function $f(x)$ with $\Xi, \mu, \sigma, h_3 \dots h_M$ as free parameters adjusted during the fit to minimize the rms deviation between the function and its approximation, while still fixing $h_0=1, h_1=h_2=0$.
In this case, all parameters in the fit depend on the order of expansion $M$; in other words, this is the best approximation \textit{at the given order}, not the approximation in which the lowest-order function is chosen to be the best-fit one. Naturally, this results in a better overall fit, but note that this is \textit{not} a true GH expansion: if we compute the coefficients $h_{0,1,2}$ for the original function $f(x)$ with the parameters $\mu,\sigma$ having the best-fit values as described above, they will not have the values $1,0,0$ as implied during the fit, while the values of the remaining coefficients will be the same. Consequently, the actual GH expansion $\tilde f(x)$ constructed with these parameters will be somewhat less accurate than the fitted function $\mathscr L(x)$, but typically still more accurate than the GH expansion constructed around the best-fit Gaussian (as in the first approach). 
Since the fitted function $\mathscr L(x)$ has zero $h_1,h_2$, the best-fit values of $\mu,\sigma$ (and consequently all GH coefficients) will converge to the ones produced by the first method as $M \to \infty$, because only for this choice of $\sigma$ the first two GH moments vanish.
\item Gerhard \cite{Gerhard1993} independently introduced the GH parametrization of velocity profiles. In his approach, the scale parameter $\sigma$ is not necessarily fixed to any particular value, hence $h_2$ (called $s_2$ in that paper) is not zero (his eq.~3.10). To avoid ambiguity, he suggests to use the true dispersion of the original function%
\footnote{When fitting the noisy data, true dispersion is unknown a priori, so Gerhard suggests a two-step procedure: first perform a GH fit with the scale parameter $\sigma$ set to some reasonable value (e.g. the width of the main peak of the function), then compute the true dispersion from this GH series, and re-fit another GH expansion with the scale set to the true dispersion. The total dispersion determined from the GH fit is less sensitive to the poorly measured wings of the profile. van der Marel \& Franx caution that this procedure could underestimate of the dispersion if the truncated GH expansion has significant negative wings, and suggest to take $\mathrm{max}\big(\tilde f(x), 0\big)$ when computing the dispersion according to Equation~\ref{eq:GHdispersion}.}
$f(x)$ as the scale parameter $\sigma$, but notices that sometimes a smaller value of $\sigma$ could be preferred, e.g., when the function has a sharp and narrow peak. In common with the previous choice, all coefficients $h_m$ are, in general, nonzero if $\sigma$ is not equal to the width of the best-fit Gaussian.
\item However, if one dispenses with the constraint that $h_1=h_2=0$, then a \textit{still} better approximation (for a given order $M$) could be achieved by fitting an unrestricted GH series (\ref{eq:GHcoefs}), simultaneously optimizing $\mu,\sigma$ and all $h_{m\ge 0}$.
\end{enumerate}

%%%%%%%%%%%%%%
\begin{figure}
\begin{center}
\includegraphics[width=16.5cm]{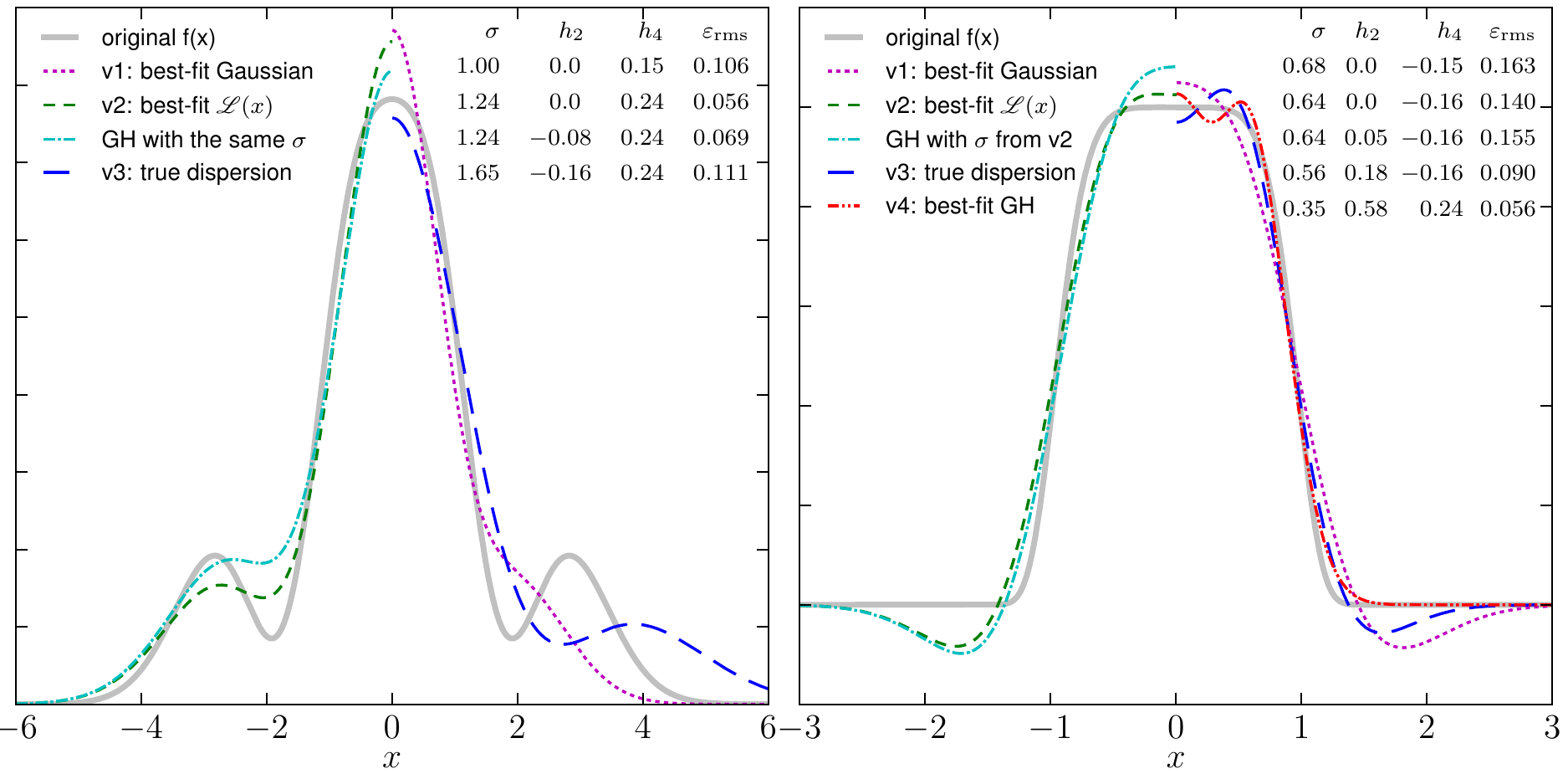}
\end{center}
\caption{Gauss--Hermite expansions with $M=4$ and different choices of $\sigma$. \protect\\
Left panel: $f(x) = \frac{1}{\sqrt{2\pi}} \exp\big(-\frac{1}{2} x^2 \big) \;\big[1 + 0.15\, \mathcal H_4(x) + 0.2\, \mathcal H_6(x) \big]$ (same as in Appendix A of \cite{Joseph2001}) is actually a pure GH expansion with 6 terms, but we fit it with only 4 terms. \protect\\
Right panel: $f(x) = \exp(-x^6)$ is a slightly smoothed top-hat function. \protect\\
Solid gray line shows the original function, and other lines show various approximations (each one is plotted for either positive or negative $x$ only, to reduce congestion): \protect\\
Purple dotted line: GH expansion with the scale $\sigma$ equal to that of the best-fit Gaussian (variant 1 in the list) has $h_2=0$ by construction, and recovers the true value of $h_4$ on the left panel, but is not the best fit by itself.\protect\\
Dashed green line: best-fit function $\mathscr L(x)$ which looks like a GH expansion with $h_2=0$ and adjustable $\sigma,h_4$ (variant 2 in the list). It is not the \textit{true} GH expansion of the function $f(x)$ with this scale $\sigma$, however: the latter is shown by dot-dashed cyan line, and has non-zero $h_2$.\protect\\
Long-dashed blue line: GH expansion with the scale $\sigma$ equal to the true dispersion of the original function (variant 3 in the list) also has non-zero $h_2$.\protect\\
Dash-double-dotted red line: the absolute best-fit GH expansion with $\sigma, h_2$ and $h_4$ all being free parameters, which has the smallest deviation from the original function (not shown on the left panel because it is very similar to the variant 1).
} \label{fig:GHmoments}
\end{figure}
%%%%%%%%%%%%

Among these approaches, the first one (taking $\mu$ and $\sigma$ to be the parameters of the best-fit Gaussian) is the only one that gives the same values for all $h_m$ coefficients regardless of the order of expansion $M$ (because its parameters are fixed and do not depend on $M$), and has $h_1=h_2=0$ by construction. This makes it attractive from the conceptual point of view, because it can be ``gracefully degraded'' (truncated at a lower order while still producing a reasonable fit). In other approaches, $\mu$, $\sigma$ and all other coefficients depend on the choice of $M$, and because of two additional free parameters ($h_1,h_2$), the approximation is generally more accurate for a fixed $M$ than in the first approach. Figure~\ref{fig:GHmoments} illustrates the effect of different choices of $\sigma$ on the approximation accuracy of the resulting GH series truncated at $M=4$. The two examples shown in that figure are rather extreme, and in practice the difference is much smaller (and as mentioned before, vanishes for sufficiently high $M$, regardless of the choice of $\sigma$).
We stress again that if the original function $f(x)$ is significantly non-Gaussian, one cannot interpret $\sigma$ from the best-fit approximation as the ``width'' of the function, because the higher-order terms cannot be neglected. Moreover, there is no uniquely defined set of ``true'' GH parameters -- different choices of $\sigma$ will lead to different GH expansions, and it is not clear a priori which one will converge faster as the number of terms $M$ increases (for the function shown on the right panel of the above figure, the optimal $\sigma$ appears to be substantially smaller than either the width of the best-fit Gaussian or the true dispersion).

The above discussion assumed that we know perfectly the ``true'' original function. In practice, the GH expansion is often used to parametrize the velocity distribution functions extracted from or fitted to the spectra, which are often both noisy and undersampled.
\cite{CappellariEmsellem2004} find that the first approach in the above list (fitting the data by a Gaussian with free parameters $\mu$ and $\sigma$ first and then using it to construct a GH expansion) becomes biased towards a pure Gaussian when the data is undersampled, even in the limit of zero noise, while the second approach (fitting a function $\mathscr L(x)$ with $\mu,\sigma,h_{m\ge 3}$) produces large and correlated uncertainties when the signal-to-noise ratio (SNR) is low. They design a hybrid method (the \textsc{pPXF} code) which uses $\mathscr L(x)$ at high SNR, but degenerates into fitting a pure Gaussian at low SNR. As explained above, the values of $\sigma$ (and hence all GH coefficients) produced by the second approach will tend to those of the first approach as $M$ increases (in the limit of well-sampled and noiseless data).
Alternatively, one may use an entirely different method for extracting the velocity distribution from the spectrum, and only then fit a GH expansion to the resulting function. \cite{Joseph2001} use a nonparametric maximum penalized likelihood method to determine $f(x)$ and then follow the first approach (determine $\sigma$ from the best-fit Gaussian) to construct a GH expansion $\tilde f(x)$.

%%%%%%%%%%%%%%
\subsubsection{Sampling}  \label{sec:MathSamplingDetails}

The sampling routine performs the following task: given an $D$-dimensional function $f(\bx)\ge 0$ defined in a rectangular domain (without loss of generality may take it to be a unit hypercube $[0..1]^D$), generate an array of $N$ points $\bx_i$ such that their density in the vicinity of any point $\bx$ is proportional to $f(\bx)$. In other words, $f$ is interpreted as a probability distribution function and is sampled with equal-weight samples; in fact the integral $\int f(\bx)\, \d ^D x = A$ needs not be unity, and is itself estimated by the routine.

We employ an adaptive rejection sampling method.
The entire domain $\mathcal{V}$ is partitioned hierarchically into smaller hyperrectangles (cells) $\mathcal{V}_c$, an envelope function $\bar f_c$ (constant in each cell) is constructed, and a rejection sampling is used to draw output points from each cell. This requires several (up to 10--20) iterations, during each one this partitioning is further refined in regions with where the function values are largest, and more trial points are drawn. The cells are organized into a tree structure, where each non-leaf cell is split into two equal-volume child cells along some dimension. Each leaf cell keeps the list of trial points $\bx_{c,k}\in\mathcal{V}_c$ that belong to this cell; when a cell is split in two halves, these points are distributed among the child cells.
The procedure, illustrated in Figure~\ref{fig:Sampling}, can be summarized as follows:

%%%%%%%%%%%%%%
\begin{figure}
\begin{center}
\includegraphics[width=15cm]{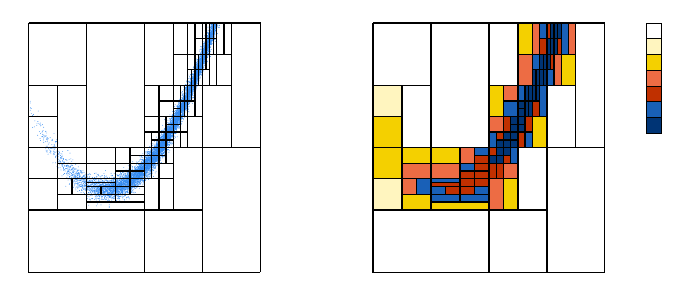}
\end{center}
\caption{Illustration of the adaptive rejection sampling algorithm in the domain $[-1..2]^2$, recursively partitioned into $\sim 100$ rectangular cells. Left panel shows $10^4$ points sampled from $f(x,y) = \exp[-R(x,y)]$, where $R(x,y) \equiv (1-x)^2-100(y-x^2)^2$ is the Rosenbrock function. In the right panel cells are \href{https://www.google.com/search?q=mondrian&tbm=isch}{colored} according to the density of trial points (successive shades are $2\times$ denser).
} \label{fig:Sampling}
\end{figure}
%%%%%%%%%%%%

\begin{enumerate}  \setcounter{enumi}{-1} \setlength{\parskip}{2pt} \setlength{\itemsep}{2pt}
\item We start with only one cell $\mathcal{V}_1$ covering the entire domain, and sprinkle $M_1$ sampling points $\bx_k$ uniformly in this cell. At this stage, the estimate of the integral is just
\begin{align}
A = \frac{\mathrm{vol}(\mathcal{V}_1)}{M_1} \sum_{k=1}^{M_1} f(\bx_k).
\end{align}
\item \label{step:computeIntegral}
At each iteration, we first loop over all leaf cells $\mathcal{V}_c$ in the tree (those that are not split into child cells), and consider all $M_c$ trial points $\bx_{c,k}$ belonging to $\mathcal{V}_c$. The integral $A$ is estimated as \begin{align}
A = \sum_{c=1}^{N_\mathrm{cell}} \frac{\mathrm{vol}(\mathcal{V}_c)}{M_c} \sum_{k=1}^{M_c} f(\bx_{c,k}).
\end{align}
\item \label{step:refinement} We then perform another scan over all leaf cells and check whether they contain enough trial points for the rejection sampling procedure. A trial point $\bx_{c,k}$ can be selected as one of $N$ output samples with probability $f(\bx_{c,k}) / \bar f_c$, where $\bar f_c \equiv \frac{A}{N} \frac{M_c}{\mathrm{vol}(\mathcal{V}_c)}$ is the "envelope" function, constant across the cell; clearly we need $\bar f_c$ to be larger than the maximum value of $f(\bx\in\mathcal{V}_c)$. If this condition is not satisfied, the cell is scheduled for refinement.
\item For each such cell that needs to be refined, we have two choices: either to add more trial points into the entire cell (thus increasing $M_c$ and hence pushing up the envelope function), or first split the cell in two halves and consider both of them in turn. There is a lower limit $M_\mathrm{min}$ on the number of trial points in a cell, and splitting is possible only if child cells would contain at least that many points. We examine all possible dimensions along which the cell could be split, and choose the one with the lowest entropy (i.e. the function varies more significantly along this dimension). The two new child cells inherit the parent cell's trial points, are added to the end of the list and will be scanned in turn during the same pass (at least one of them will need to be further refined). If the cell contains less than $2M_\mathrm{min}$ trial points and hence cannot be split, we double the number of trial points $M_c$ in it (add the same number of new points uniformly distributed in the cell), so that it could be split (if necessary) on the next iteration.
\item If we have added new trial points into any cell, we repeat the procedure from step~\ref{step:computeIntegral}. 
\item Otherwise the iterative refinement is complete, and we draw $N$ output points from the trial points, with the probability described in step~\ref{step:refinement}, and finish.
\end{enumerate}

%%%%%%%%%%%%%%%%%%%%%%%%%%%%%%%%%%%%%%%%%%%%%%%%%%%%%
\subsection{Coordinates}  \label{sec:CoordinateDetails}

%%%%%%%%%%%%%%
\begin{figure}
\begin{center}
\includegraphics[width=6cm]{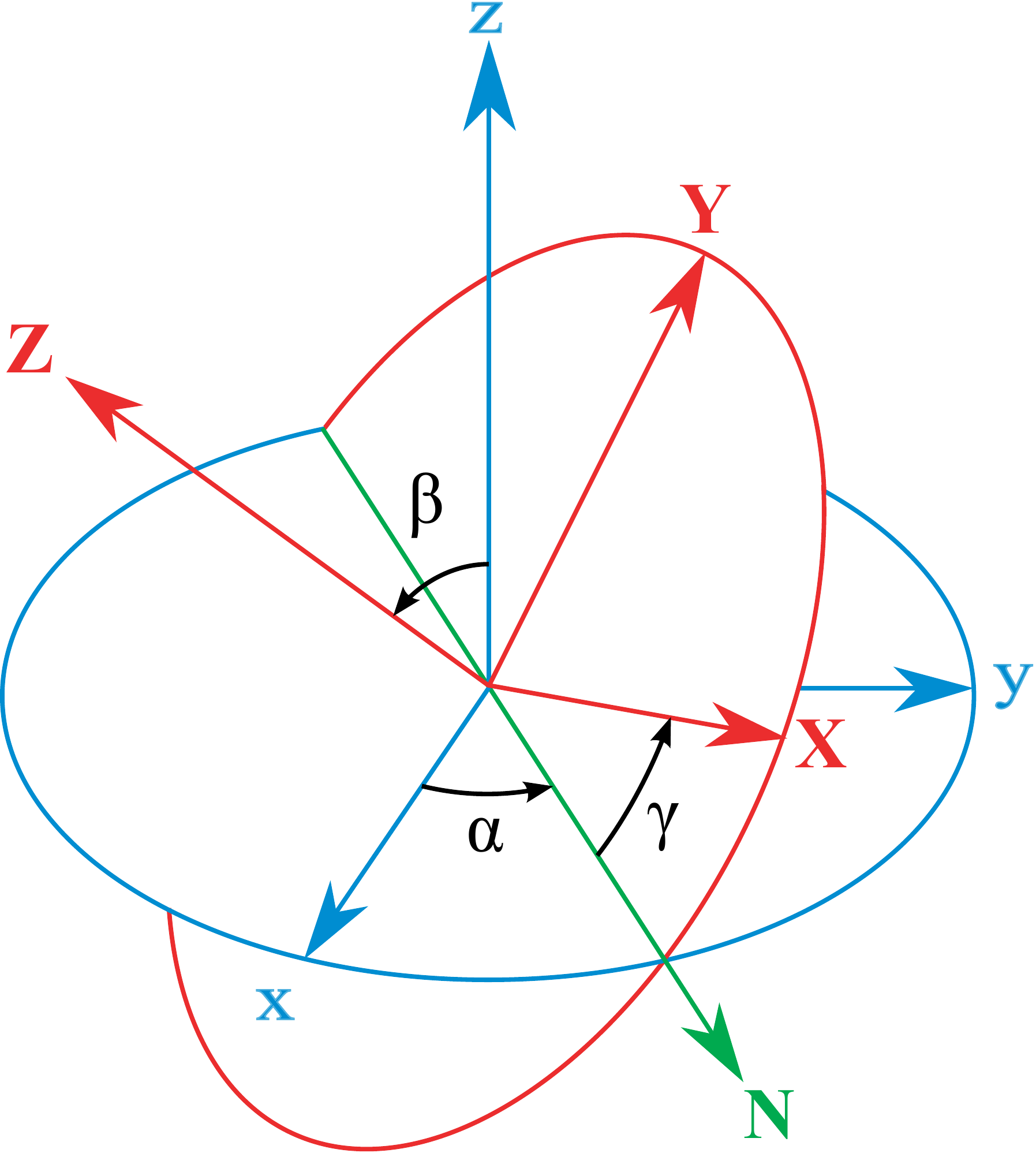} \hspace{1cm}
\includegraphics{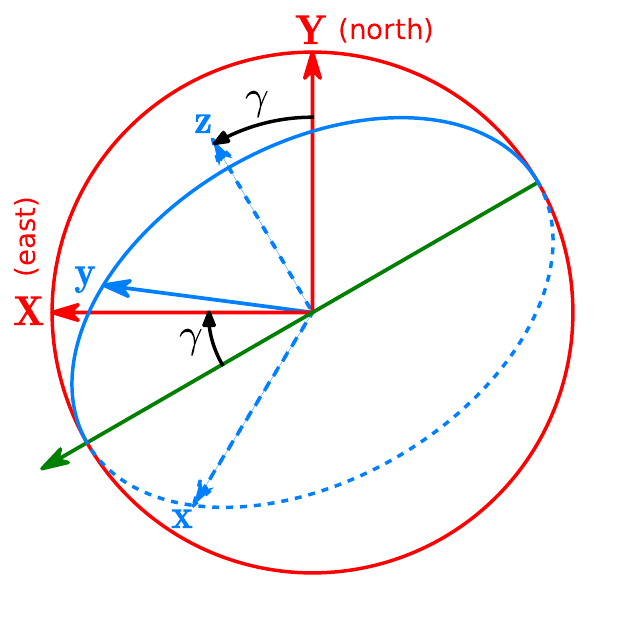}
\end{center}
\caption{\textit{Left panel:} specification of rotation of the cartesian coordinate system in terms of Euler angles $\alpha,\beta,\gamma$.\protect\\
Let the original reference frame be specified by the right-handed triplet of basis vectors (axes) $\bx,\by,\bz$, and the rotated frame -- by basis vectors $\bX,\bY,\bZ$.
The first rotation of the $xy$ plane (blue) by angle $\alpha$ about the $\bz$ axis creates an intermediate basis $\bx',\by',\bz'$, where the axis $\bx'$ points along the line of nodes of the overall transformation (denoted by the green arrow $N$).
The second rotation by angle $\beta$ about the $\bx'$ axis (line of nodes) tilts the $x'y'$ plane by angle $\beta$, creating a second intermediate basis $\bx'',\by'',\bz''$; the angle between basis vectors $\bz''$ and $\bz$ is also $\beta$.
The third rotation of the $x''y''$ plane (red) by angle $\gamma$ about the $\bz''$ axis does not change the direction of that axis, hence the final axis $\bZ$ is the same as $\bz''$. \protect\\
\textit{Right panel:} the same two reference frames as in the left panel, viewed in the image plane $XY$. The $\bZ$ axis points perpendicular to the plane away from the observer; the $\bY$ axis points up / north, and the $\bX$ axis -- left / east. The equatorial plane $xy$ of the intrinsic coordinate system is shown by blue ellipse (dashed when it is behind the image plane). Its intersection with the image plane is the line of nodes, marked by the green arrow. The angle of clockwise rotation of the $\bX$ axis w.r.t.\ the line of nodes is $\gamma$ (the last of the three rotations). The position angle (PA) in the $XY$ plane is measured from the $\bY$ axis counter-clockwise (towards the $\bX$ axis), hence the PA of the projection of $\bz$ axis onto the image plane is also $\gamma$, and the PA of the line of nodes is $\gamma+\pi/2$.
%\vspace{2cm}
} \label{fig:EulerAngles}
\end{figure}
%%%%%%%%%%%%

The coordinate transformation subsystem in \Agama consists of several concepts.
First, we define several commonly used coordinate systems (or \textit{representations} in the terminology of \texttt{astropy}): Cartesian, Cylindrical, Spherical, and Prolate Spheroidal. The same point in 3d space can be represented in any of these coordinate systems (having a common center and orientation). There are routines for transforming coordinates, velocities, gradients and hessians of scalar functions between these representations.

Second, we define rotations of coordinate basis.
Consider a right-handed Cartesian reference frame defined by basis vectors (axes) $\bx,\by,\bz$, and a rotated frame with the same origin defined by basis vectors $\bX,\bY,\bZ$. The relative orientation of two frames can be specified by Euler angles (see Figure~\ref{fig:EulerAngles}).
The same point has coordinates $x,y,z$ in the original frame and $X,Y,Z$ in the rotated frame.
The transformation between these coordinates (passive rotation) is given by the following orthogonal rotation matrix $\mathsf R$, produced by the routine \texttt{makeRotationMatrix}:
\begin{equation}
\left( \begin{array}{c} X \\ \,Y \\ Z \end{array} \right) = \mathsf R 
\left( \begin{array}{c} x \\   y \\ z \end{array} \right) , \qquad
\mathsf R \equiv \left( \begin{array}{ccc}  \phantom{-}
 c_\alpha c_\gamma - s_\alpha c_\beta s_\gamma\;\; & \phantom{-}
 s_\alpha c_\gamma + c_\alpha c_\beta s_\gamma\;\; &
 s_\beta  s_\gamma \\
-c_\alpha s_\gamma - s_\alpha c_\beta c_\gamma\;\; &
-s_\alpha s_\gamma + c_\alpha c_\beta c_\gamma\;\; &
 s_\beta  c_\gamma \\
 s_\alpha s_\beta &
-c_\alpha s_\beta &
 c_\beta 
\end{array} \right) ,
\end{equation}
where we used $c_\circ, s_\circ$ as shortcuts for $\cos\circ, \sin\circ$.
The inverse transformation is described by the same angles, but with negative signs and in reverse order: $-\gamma,-\beta,-\alpha$, and its rotation matrix is $\mathsf R^{-1} = \mathsf R^T$. 
The rotation matrix is invariant under the simultaneous substitution $\beta \to -\beta, \alpha \to \alpha+\pi, \gamma \to \gamma+\pi$; hence it is convenient to restrict the range of $\beta$ to $[0,\pi]$ and the other two angles -- to $(-\pi,\pi]$.\\
\makebox[-2mm]{}\begin{tabular}{m{9.5cm} r}
In the case of a triplanar symmetry (invariance under reflection about any of the three principal planes, or equivalently under change of sign of any coordinate), one may simultaneously change the sign of any two coordinates while preserving the right-handedness of the coordinate system, and this will not change any physical property of the system. These simultaneous sign changes and associated transformations of Euler angles are listed in the table on the right.
&
\begin{tabular}{c c c r r r}
x & y & z & & \makebox[0pt][r]{angles} \\
\raisebox{-8pt}[0pt][0pt]{$+$} & \raisebox{-8pt}[0pt][0pt]{$+$} & \raisebox{-8pt}[0pt][0pt]{$+$}
&   $\alpha$ & $\beta$ & $\gamma$ \\
&&& $\pi+\alpha$ & $-\beta$ & $\pi+\gamma$ \\
\raisebox{-8pt}[0pt][0pt]{$+$} & \raisebox{-8pt}[0pt][0pt]{$-$} & \raisebox{-8pt}[0pt][0pt]{$-$}
&   $\pi-\alpha$ & $\pi-\beta$ & $\pi+\gamma$ \\
&&& $-\alpha$ & $\pi+\beta$ & $\gamma$ \\
\raisebox{-8pt}[0pt][0pt]{$-$} & \raisebox{-8pt}[0pt][0pt]{$+$} & \raisebox{-8pt}[0pt][0pt]{$-$}
&   $-\alpha$ & $\pi-\beta$ & $\pi+\gamma$ \\
&&& $\pi-\alpha$ & $\pi+\beta$ & $\gamma$ \\
\raisebox{-8pt}[0pt][0pt]{$-$} & \raisebox{-8pt}[0pt][0pt]{$-$} & \raisebox{-8pt}[0pt][0pt]{$+$}
&   $\pi+\alpha$ & $\beta$ & $\gamma$ \\
&&& $\alpha$ & $-\beta$ & $\pi+\gamma$
\end{tabular}
\end{tabular}

This kind of transformation is used to convert between the intrinsic coordinates of the stellar system (original frame $\bx\by\bz$) and the observed coordinates (rotated frame $\bX\bY\bZ$), where the $\bZ$ axis points along the line of sight away from the observer, the $\bY$ axis points upward / north in the image plane, and the $\bX$ axis -- leftward (!) / east in the image plane.
Note that this creates a rather awkward situation that the $\bX$ axis is directed opposite to the usual rightward orientation in conventional two-dimensional plots: this results from the unfortunate fact that the observer sits ``inside'' the celestial sphere. This transformation is often referred to as projection onto the sky or image plane $XY$. As another example of wrecked but venerable astronomical convention, position angles (PA) in the $XY$ plane are measured from the $\bY$ (north) direction towards $\bX$ (east), i.e., counter-clockwise in this configuration; hence the PA of $\bX$ is $\pi/2$.
In some studies, the orientation of the rotated frame is given by the spherical polar angles $\theta,\phi$ of the line of sight $\bZ$ in the intrinsic frame; in this convention, $\alpha=\phi+\pi/2$, $\beta=\theta$.

If the shape of an object in the intrinsic reference frame is a triaxial ellipsoid with the major axis $a\bx$, intermediate axis $b\by$ and minor axis $c\bz$, then the projection along the major axis corresponds to $\{\alpha,\beta\} = \{\pm\pi/2, \pi/2\}$, along the intermediate axis -- to $\{0, \pi/2\}$ or $\pi,\pi/2$, and along the minor axis -- to $\beta=0$ and any $\alpha$. In a general case, $\beta$ is the inclination angle (0 is ``face-on'' orientation and $\pi/2$ is ``edge-on''), and $\gamma$ is the PA of the projection of the intrinsic minor axis $\bz$ onto the sky plane. 
In the case of an axisymmetric system, the first rotation ($\alpha$) does not change anything in its properties, hence we may set $\alpha=0$. Then the intrinsic major axis $\bx$ coincides with the line of nodes, and its PA is $\gamma+\pi/2$.
In a general case, the projection of a triaxial ellipsoid is also an ellipse in the image plane, but its major and minor axes $A,B$ and orientation (PA $\eta$) are related to the intrinsic shape $a,b,c$ and viewing angles $\alpha,\beta,\gamma$ in a rather complicated way. In particular, the major axis of the ellipse may not be aligned with any of the three projected principal axes. There are routines \texttt{getProjectedEllipse}, \texttt{getIntrinsicShape} and \texttt{getViewingAngles} for computing one of these triplets from the other two, but note that the deprojection (determination of either the intrinsic shape or the viewing angles) is not always possible. For an axisymmetric system, there are two possible choices of viewing angles, depending on which side of the system is nearer, and for a triaxial system, there are four possible combinations of viewing angles. %, illustrated in Figure~\ref{fig:TriaxialProjection}.

In case that the distance to the stellar system is not overwhelmingly larger than its characteristic size (or equivalently, when it occupies a non-negligible area on the sky), one needs to employ more sophisticated transformations between the cartesian intrinsic coordinates within the stellar system and the spherical coordinates on the sky (no longer ``sky plane''). An extreme case is our Galaxy, where the observer is located well within the stellar system, and it occupies the entire sky. Also in this case, we may need to consider shifts, not only rotations of the coordinate systems. The projection transformations will be extended to these cases in the future.

%%%%%%%%%%%%%%%%%%%%%%%%%%%%%%%%%%%%%%%%%%%%%%%%%%%%%
\subsection{Potentials}  \label{sec:PotentialDetails}

%%%%%%%%%%%
%\subsubsection{Analytic potentials}  \label{sec:PotentialAnalyticDetails}

%%%%%%%%%%%%%%
\subsubsection{Multipole expansion}  \label{sec:PotentialMultipoleDetails}

The potential in the multipole expansion approach is represented as a sum of individual spherical-harmonic terms with coefficients being arbitrary functions of radius:
\begin{align}
\Phi(r,\theta,\phi) &= \sum_{l=0}^{l_\mathrm{max}}\sum_{m=-m_0(l)}^{m_0(l)}
\Phi_{l,m}(r)\: \sqrt{4\pi} \tilde P_l^m(\cos\theta)\:\trig m\phi, \\
\trig m\phi &\equiv \left\{\begin{array}{rcl} 
  1 &,& m=0 \\
  \sqrt{2}\,\cos  m \phi &,& m > 0 \\
  \sqrt{2}\,\sin |m|\phi &,& m < 0 
\end{array}\right.   \nonumber
\end{align}
Here $\tilde P_l^m(x) \equiv \sqrt{\frac{2l+1}{4\pi}\frac{(l-|m|)!}{(l+|m|)!}} \;P_l^{|m|}(x)$ 
are normalized associated Legendre polynomials, $l_\mathrm{max}$ is the order of expansion in meridional angle $\theta$, and $m_0(l) = \mathrm{min}(l, m_\mathrm{max})$, where $m_\mathrm{max} \le l_\mathrm{max}$ is the order of expansion in azimuthal angle $\phi$ (they do not need to coincide, e.g., if the model is considerably flattened but only weakly triaxial, then $m_\mathrm{max}=2$ may be sufficient, while $l_\mathrm{max}$ may be set to 8 or 10). The normalization is chosen so that for a spherically-symmetric potential, $\Phi_{0,0}(r)=\Phi(r)$, and that for each $l$, the sum of squared coefficients over all $m$ is invariant under rotations of coordinate system.

In the \ttt{Multipole} class, individual terms are approximated as suitably scaled functions in log-scaled radius. The logarithmic transformation of radius is intended to attain high dynamic range with a moderate number (a few dozen) of grid points, while the transformation of amplitude of each term increases the accuracy of interpolation. The main $l=0$ term is log-scaled when possible (i.e., if it is everywhere negative, then the actual function to be interpolated is $\ln\big[1/\Phi(0)-1/\Phi_{0,0}(\ln r)\big]$, where $\Phi(0)$ is the value of potential at origin, which may be finite or $-\infty$). The other terms are normalized to the value of the $l=0$ term (i.e., the spline is constructed for $[\Phi_{l,m}/\Phi_{0,0}](\ln r)$). Each term is interpolated as a quintic spline, defined by the values and first derivatives at nodes of a radial grid; usually this grid would be linear in log-radius, with a constant ratio between consecutive radii $f\equiv r_{k+1}/r_k$.

If the minimum/maximum grid radii are not provided, they are assigned automatically using the following approach. First we locate the radius at which the logarithmic curvature of the spherically-symmetric part of the density profile ($d^2\ln(\rho_{0,0})/d(\ln r)^2$), weighted by the mass at the given radius ($\propto r^2\rho$), reaches the maximum. For most finite-mass models, this would be the near the half-mass radius, but even for models with infinite mass (such as NFW) this criterion still estimates the "radius of interest" quite well. This fiducial radius $r_\star$ is taken as the center of the logarithmic grid, a suitable grid spacing factor $f$ is assigned, and the grid is extended both ways from this radius: $r_\mathrm{max/min} = r_\star\: f^{\pm N_R/2}$. As $N_R$ gets larger, both the dynamical range $D\equiv r_\mathrm{max}/r_\mathrm{min}$ is increased, and the resolution gets better (nodes are spaced more densely); e.g., for $N_R=20$, these are $D\sim 10^6$ and $f\sim 2$. If the input density drops to zero beyond some radius, the upper extent of the grid is moved inward to this radius, Likewise, if the density is declining towards small $r$, the potential has a very flat core, so that the inner grid point is shifted up to a "safe" radius $r_\mathrm{min}$ at which the potential is sufficiently different from $\Phi(0)$ (at least in the last few digits), to prevent the loss of precision of floating-point numbers. This automatic algorithm gives reasonable results in vast majority of cases, but if necessary, the min/max radii may be provided by the user.

To compute the potential and its derivatives at a given point, one needs to sum the contributions of each harmonic term. For systems with certain symmetries, many of these terms are identically zero, and this is taken into account thereby reducing the amount of computation. By convention, negative $m$ correspond to sine terms and positive -- to cosine; if a triaxial model is aligned with the principal axes, all sine terms must be zero; symmetry w.r.t. reflection about one of the principal planes also zeroes down some terms, and axisymmetry retains only $m=0$ terms; Table~\ref{tab:Symmetry} lists the most common cases. All possible combinations of symmetries are encoded in the \ttt{coords::SymmetryType} class, and each one corresponds to a certain combination of non-trivial spherical-harmonic terms (\ttt{math::SphHarmIndices}), as described in \texttt{math_sphharm.h}. For instance, a model of a disk galaxy with two spiral arms is symmetric w.r.t. $z$-reflection (change of sign of $z$ coordinate) and $xy$-reflection (change of sign of both $x$ and $y$ simultaneously), and this retains only terms with even $l$ and even $m$ (both positive and negative).

At each nontrivial $m$, we may need to compute up to $l_\mathrm{max}-|m|$ 1d interpolating splines in $r$ multiplied by Legendre polynomials in $\cos\theta$. This may be replaced with a single evaluation of a 2d interpolation spline in $\ln r,\theta$ plane (in fact a suitably scaled analog of $\theta$ is used to avoid singularities along $z$ axis), which was pre-computed during potential initialization -- this is more efficient for $l_\mathrm{max}>2$. In this variant, the main log-scaled term is the 2d spline for the $m=0$ component, while the other azimuthal harmonics are normalized to the value of the main term.

Extrapolation to small and large radii (beyond the extent of the grid) is performed using the assumption of a power-law behaviour of individual multipole components: $\Phi_{l,m}(r) = U_{l,m}\, r^{s_{l,m}} + W_{l,m}\, r^{v}$, where $v\equiv l$ or $-1-l$ for the inward or outward extrapolation, correspondingly. The term with $r^v$ represents the ``principal'' component with a zero Laplacian, while $r^s$ corresponds to a power-law density profile $\rho_{l,m}\propto r^{s-2}$, and is typically much smaller in magnitude. This allows to describe very accurately the asymptotic behaviour of potential beyond the extent of the grid, if the coefficients $U,W$ and $s$ can be determined reliably. In order to do so, we use the value and derivative of each harmonic coefficient at the first or the last grid node, plus its value at the adjacent node, to obtain a system of 3 equations for these variables. Thus the value and derivative of each term are continuous at the boundaries. 

A \ttt{Multipole} potential may be constructed either from an existing potential object (in which case it simply computes a spherical-harmonic transform of the original potential at radial grid nodes), or from a density profile (thereby solving the Poisson equation):
\begin{align}
\Phi_{l,m}(r) &= -\frac{4\pi}{2l+1} \left[ r^{-l-1} \int_0^r \rho_{l,m}(r')\,{r'}^{\,l+2}\,\d r' + r^l\int_r^\infty \rho_{l,m}(r')\,{r'}^{\,1-l}\,\d r' \right],  \label{eq:SphHarmPoisson} \\
\rho_{l,m}(r) &\equiv \frac{1}{\sqrt{4\pi}} \int_{-1}^1 \d \cos\theta\, \tilde P_l^m(\cos\theta) \int_0^{2\pi}\d \phi\:\trig m\phi\:\rho(r,\theta,\phi) .  \label{eq:SphHarmDensity}
\end{align}

A separate class \ttt{DensitySphericalHarmonic} serves to approximate any density profile with its spherical-harmonic expansion, with coefficients being cubic splines in $\ln r$. 
Similarly to the Multipole potential class, we extrapolate the profile to small or large radii using power-law asymptotes, with slopes deduces from the values of the $l=0$ coefficient at two inner- or outermost grid points. This class is mainly used in self-consistent modelling (Section~\ref{sec:SCM}) to provide a computationally cheap way of evaluating the density at any point in space, once it is initialized by computing the costly integrals over distribution function at a small number of points (grid nodes in radius and nodes of Gauss--Legendre quadrature rule in $\cos\theta$). This interpolated density is then used to construct the Multipole potential: the solution of Poisson equation requires integration of harmonic terms in radius using a more densely spaced internal grid, and the values of these terms are easily evaluated from the density interpolator. Note that this process involves two forward and one reverse spherical-harmonic transformation (first time during the construction of density interpolator, then the reverse transformation to obtain the interpolated values at the required spatial points, and then again in the Multipole potential). However, since the spherical-harmonic transformation is invertible (reproduces the source density at this special set of points to machine precision), this double work does not add to error, and incurs negligible overhead.

\ttt{DensitySphericalHarmonic} may also be constructed from an array of particles, and then used to create the \ttt{Multipole} potential in a usual way. To do so, we first compute the spherical-harmonic expansion coefficients at each particle's radius:
\begin{align*}
\rho_{l,m;i} &\equiv m_i\,\sqrt{4\pi}\,\tilde P_l^m(\cos\theta_i)\,\trig m\phi_i .
\end{align*}
Then the $l=0$ coefficients (which contain just particle masses) are used to determine the spherically-symmetric part of the density profile. We use penalized spline log-density fit (Section~\ref{sec:MathSplineDensityDetails}) to estimate the logarithm of an auxiliary quantity $P(\ln r) \equiv dM(<r)/d\ln r$ from the array of point masses and log-radii; the actual density is $\rho_{0,0}(r) = P(\ln r) / (4\pi\,r^3)$. Finally, we create smoothing splines (Section~\ref{sec:MathSplineApproxDetails}) for all non-trivial $\rho_{l,m}(\ln r)$ terms.
This temporary density model is used to construct the \ttt{Multipole} potential from an \Nbody model -- even though the Poisson equation (\ref{eq:SphHarmPoisson},\ref{eq:SphHarmDensity}) can be solved directly by summing over particles (the approach used in \cite{Vasiliev2013}), this results in a noisier and less accurate potential than the intermediate smoothed density can provide.

%%%%%%%%%%%%%%
\subsubsection{CylSpline expansion}  \label{sec:PotentialCylSplineDetails}

The \ttt{CylSpline} potential is represented as a sum of azimuthal Fourier harmonics in $\phi$, with coefficients of each term intepolated on a 2d grid in $R,z$ plane with suitable scaling.
Namely, both $R$ and $z$ coordinates are transformed to $\tilde R \equiv \ln(1+R/R_0), \tilde z \equiv \ln(1+z/R_0)$, where $R_0$ is a characteristic radius. The amplitudes of each interpolated term are also transformed in the same way as for the \ttt{Multipole} potential (for the same purpose -- improving the accuracy of interpolation), namely, the main $m=0$ term uses log-scaling of its amplitude, and the remaining ones are normalized to the value of the main term. We use either 2d quintic splines or 2d cubic splines to construct the interpolator, depending on whether the partial derivatives of potential by $R$ and $z$ are available. Normally, if the potential is constructed from a smooth density profile or from a known potential, it is advantageous to use 5th order interpolation to improve accuracy, even though this increases the computational cost of construction (but not of evaluation of the potential). On the other hand, in the case of a potential constructed from an array of particles, estimates of derivatives are too noisy and in fact deteriorate the quality of approximation.

\begin{figure}
\includegraphics[width=16cm]{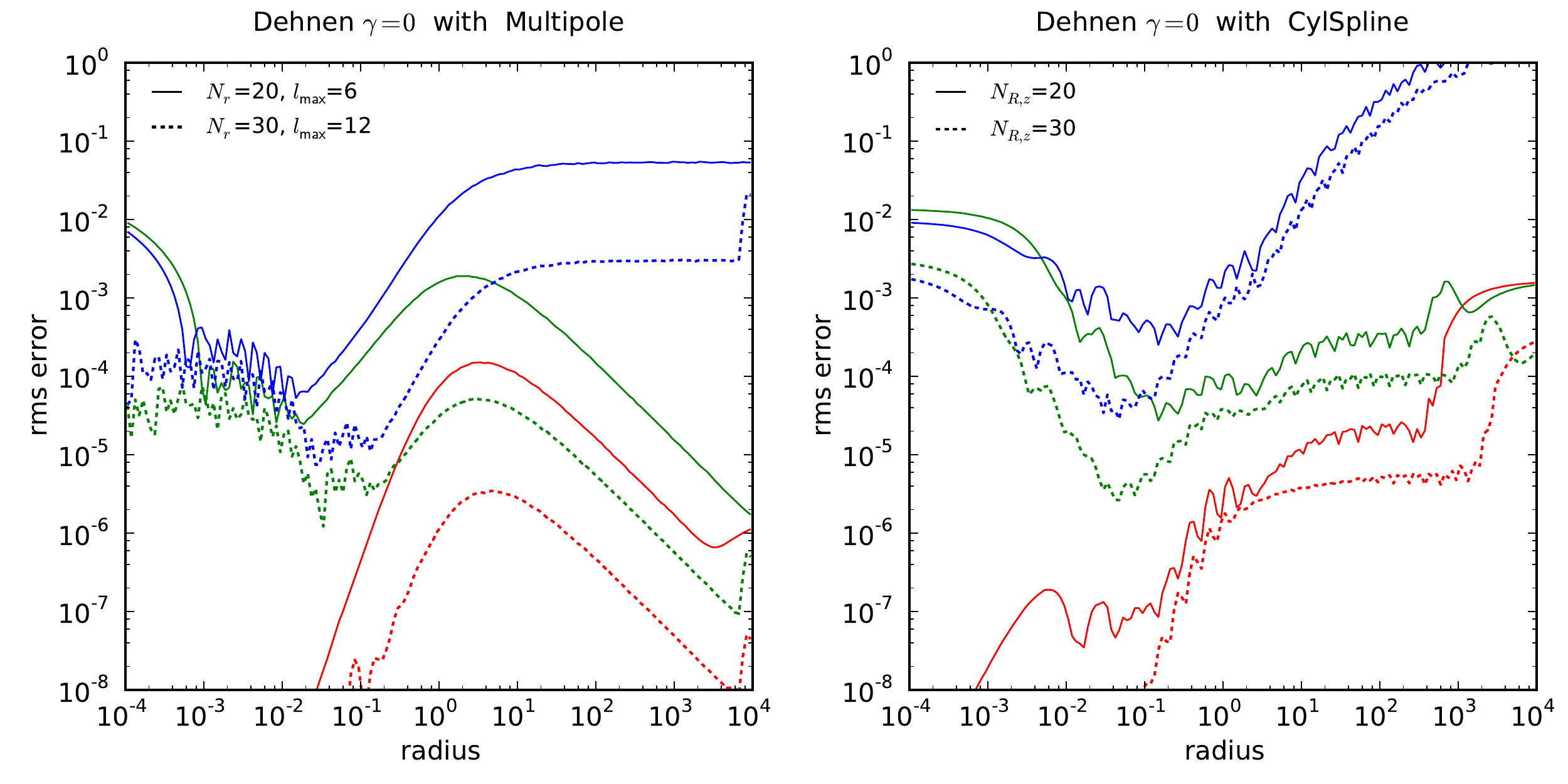}
\includegraphics[width=16cm]{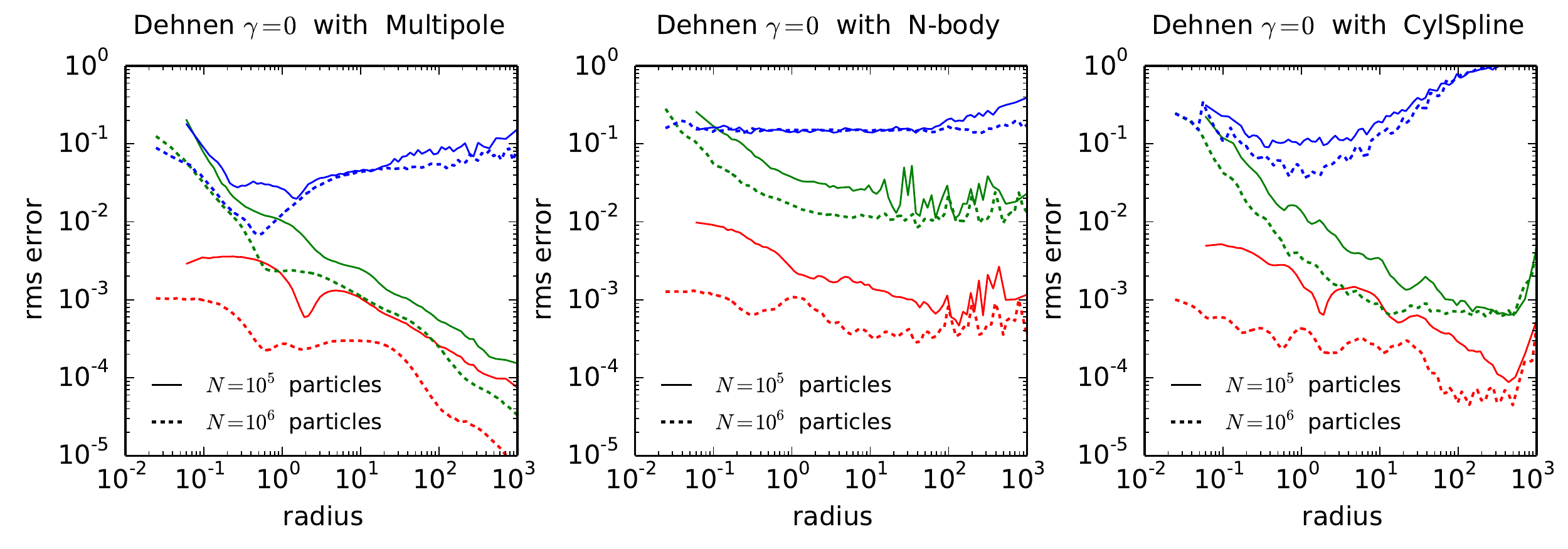}
\caption{Accuracy of potential approximations in the case of initialization from a smooth density profile (top panels) and from an array of $N$ particles (bottom panels). In both cases we compare the potential (red), force (green) and density (blue) computed using the potential expansions (left: Multipole, right: CylSpline) with the ``exact'' values for a triaxial $\gamma=0$ Dehnen profile ($x:y:z=1:0.8:0.5$), obtained by numerical integration, and plot the relative errors as functions of radius. In the top panels we vary the order of spherical-harmonic expansion and the number of grid nodes. Both potential approximations deliver fairly high accuracy, which increases with resolution. In the bottom panels we additionally show these quantities computed with a conventional \Nbody approach (direct-summation and SPH density estimate). Here the error is dominated by noise in computing the potential from discrete samples, and not by the approximation accuracy (it is almost independent of the grid parameters, but decreases with $N$). Notably, both smooth potential approximations are closer to the true potential than the \Nbody estimate.
}  \label{fig:PotentialAccuracy1}
\end{figure}

\begin{figure}
\includegraphics[width=16cm]{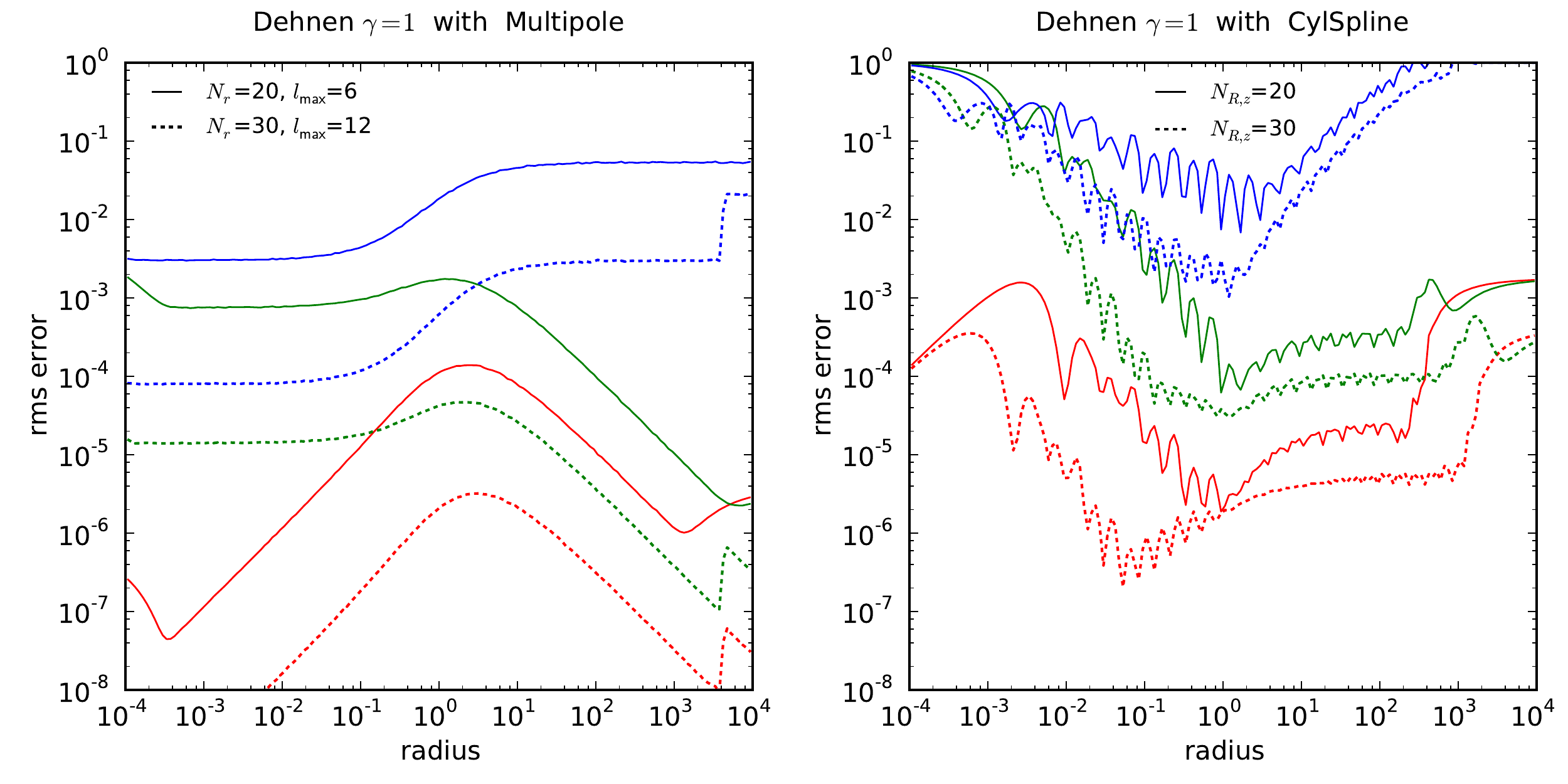}
\includegraphics[width=16cm]{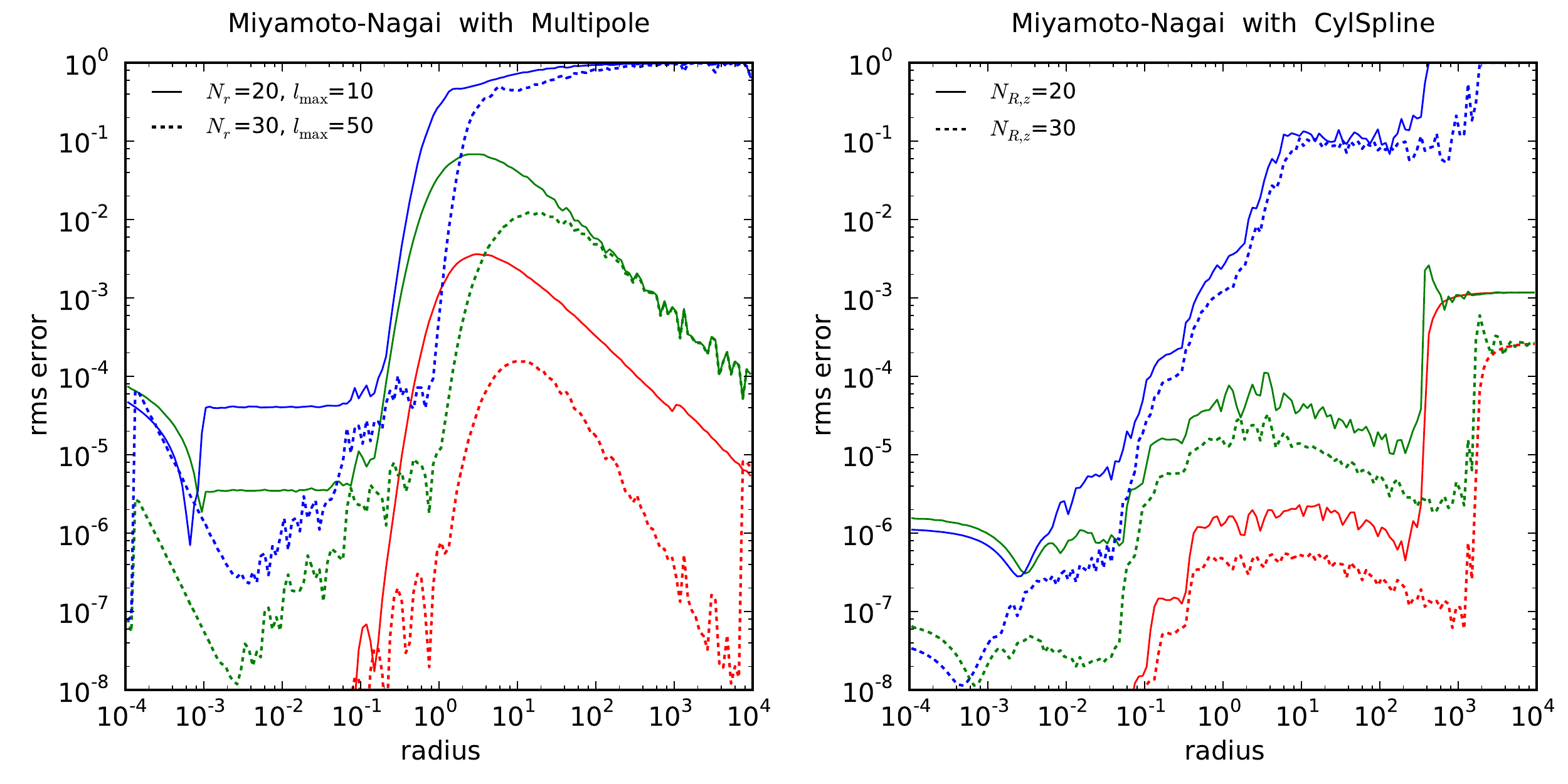}
\caption{Accuracy of potential approximations for different types of density profiles. As in the previous figure, we plot the relative errors in potential (red), force (green) and density (blue) for Multipole (left) and CylSpline (right) potential expansions. Top panels are for a triaxial $\gamma=1$ Dehnen model, and bottom -- for a Miyamoto--Nagai disk. In the former case, CylSpline cannot efficiently deal with cuspy density profiles, while Multipole is able to deliver accurate results even for a $\gamma=2$ cusp without any difficulty. On the other hand, in the latter case the strongly flattened density model is poorly represented by the spherical-harmonic expansion even with $l_\mathrm{max}=50$, whereas CylSpline performs well.
}  \label{fig:PotentialAccuracy2}
\end{figure}

Unlike the \ttt{Multipole} potential, which can handle a power-law asymptotic behaviour of density both at small and large radii, \ttt{CylSpline} is more restricted -- since the grid covers the origin, it can only represent a model with finite density at $r=0$. Extrapolation to large radii (beyond the extent of the rectangular grid in $R,z$) is performed using a similar approach to \ttt{Multipole}, but keeping only the principal spherical-harmonic terms $W r^{-l-1}$ with zero Laplacian, i.e., corresponds to a zero density outside the grid. The coefficients for a few low-order multipoles (currently $l_\mathrm{max}=8$) are determined from a least-square fit to the values of potential at the outer boundary of the grid; thus the potential values inside and outside the boundary are not exactly the same, but still are quite close -- the relative error in potential and force in the extrapolated regime is typically $\lesssim 10^{-3}$ (see Figures~\ref{fig:PotentialAccuracy1},~\ref{fig:PotentialAccuracy2}).

Since the grid spacing is near-uniform at small and near-exponential at large $R,z$, the dynamical range of \ttt{CylSpline} is also very broad. If the values of first/last grid nodes are not specified, they are determined automatically using the same approach as for \ttt{Multipole}. Typically, $20-25$ grid nodes are enough to span a range from 0 to $\gtrsim 10^3\,r_\mathrm{half-mass}$.
The automatic procedure is somewhat less optimal than in case of Multipole, so it may be advisable to set up the grid manually, with two considerations in mind. First, the inner grid point should be comparable with the smallest scale of variation of the density profile (e.g., in the case of a thin disk, $z_\mathrm{min} \simeq{}$\texttt{scaleHeight}), but not much smaller, because the relative difference between the potential at adjacent grid points should exceed the accuracy of its computation ($\sim10^{-6}$), or else the high-order interpolation scheme greatly amplifies the errors in its derivatives. Second, the outer edge of the grid should be far enough from the region where the density is concentrated, so that the extrapolation outside the grid using only a few spherical-harmonic terms is accurate enough. The potential and its derivatives are discontinuous at the grid boundary, and it's important to keep this discontinuity at a negligible level.

The main advantage of \ttt{CylSpline} is in its ability to efficiently represent even very flattened density profiles, which are not suitable for \ttt{Multipole} expansion. When \ttt{CylSpline} approximation is constructed from another potential, this boils down to taking the Fourier transform in $\phi$ of potential and forces of the original potential at the nodes of 2d grid in $R,z$ plane. When it is constructed from a density profile, this involves the solution of Poisson equation in cylindrical coordinates, which is performed in two steps. First, a Fourier transform of the source model is created (if it was neither axisymmetric nor a \ttt{DensityCylGrid} class, see \hyperref[sec:DensityCylGrid]{below}). Next, for each $m$-th harmonic $\rho_m$, the potential is computed at each node $R,z$ of the 2d grid using the following approach \cite{CohlTohline1999}:
\begin{align}
\Phi_m(R,z) &= -\int_{-\infty}^{+\infty} \d z' \int_0^{\infty} \d R' \,2\pi R'\,\rho_m(R',z')  
  \,\Xi_m(R,z,R',z')\;,  \label{eq:PoissonCylindric} \\
\Xi_m &\equiv \int_0^\infty \d k\, J_m(kR)\, J_m(kR')\, \exp(-k|z-z'|) \;,\quad
  \mbox{which evaluates to} \\
\Xi_m&= \frac{1}{\pi\sqrt{RR'}}\, Q_{m-1/2}\left( \frac{R^2+R'^2+(z-z')^2}{2RR'} \right) \quad
  \mbox{if }R>0,R'>0,  \nonumber \\
\Xi_m&= \frac{1}{\sqrt{R^2+R'^2+(z-z')^2}}\quad
  \mbox{if }R=0\mbox{ or }R'=0\mbox{, and }m=0\mbox{, otherwise 0}. \nonumber
\end{align}
Here $Q$ is the Legendre function of the second kind, which is computed using a hand-crafted Pad\'e approximation for $m\le 12$ or Gauss' hypergeometric function otherwise (more expensive).
%The two-dimensional integral in (\ref{eq:PoissonCylindric}) is calculated numerically with the \texttt{integrateNdim} routine (Section~\ref{sec:MathDetails}) to an accuracy of $10^{-6}$.
For an array of particles, 
\begin{align}  \label{eq:PoissonCylindricParticles}
\Phi_m(R,z) = -\sum_k m_k\,\Xi_m(R,z,R_k,z_k)\,\trig m\phi_k .
\end{align}

%%%%%%%%%%%%%%

%%%%%%%%%%%%

The computation of \ttt{CylSpline} coefficients is much more expensive than that of \ttt{Multipole}, because at each of $\mathcal{O}(N_R\times N_z \times m_\mathrm{max})$ nodes we need to evaluate a 2d integral in (\ref{eq:PoissonCylindric}) or a sum over all particles in (\ref{eq:PoissonCylindricParticles}). On a typical workstation, this may take from from a few seconds to a few minutes, depending on the resolution and the number of CPU cores. Nevertheless, this is a one-time cost; once the coefficients are calculated, the evaluation of both \ttt{Multipole} and \ttt{CylSpline} potentials is very fast -- the cost depends very weakly on the number of grid nodes, and is proportional to the number of azimuthal-harmonic terms ($m_\mathrm{max}$, but not $l_\mathrm{max}$ in the case of Multipole).
Symmetries of the model are taken into account in the choice of non-trivial azimuthal Fourier terms (in the case of axisymmetry, only $m=0$ term is retained; for triaxial models only even $m\ge 0$ are used, etc.); and for models with $z$-reflection symmetry, coefficients are computed and stored only for the $z\ge 0$ half-space.

Figure~\ref{fig:PotentialAccuracy1} demonstrates that the accuracy of both approximations is fairly good (relative error in force $\lesssim 10^{-3}$) with default settings ($N_R=25, l_\mathrm{max}=m_\mathrm{max}=6$) and improves with resolution. For the case of initialization from an array of particles, discreteness noise is the main limiting factor.
Figure~\ref{fig:PotentialAccuracy2} illustrates that each of the two potential expansions has its weak points: \ttt{Multipole} is not suitable for strongly flattened systems and \ttt{CylSpline} performs poorly in systems with density cusps; but for most density profiles at least one of them should deliver a good accuracy.

\phantomsection\label{sec:DensityCylGrid} A separate class \ttt{DensityCylGrid} serves the same task as \ttt{DensitySphericalHarmonic}: provides an interpolated density model that is initialized from the values of source density at nodes of a 2d grid in $R,z$ plane (for an axisymmetric model) or 3d grid in $R,z,\phi$ (in general). The density is represented as a Fourier expansion in $\phi$, with each term being a 2d cubic spline in $\tilde R, \tilde z$ coordinates (scaled in the same way as in \ttt{CylSpline}). Interpolated density is zero outside the grid. This class serves as a counterpart to \ttt{DensitySphericalHarmonic} in the context of DF-based self-consistent models for disk-like components: the values of density at grid nodes are computed by (expensive) integration of DF over velocities, and density in the entire space, necessary for computing the potential, is given by the interpolator.

%%%%%%%%%%%%%%%%%%%%%%%%%%%%%%%%%%%%%%%%%%%%%%%%%%%%%%%%%%%%%%%%%%%%
\subsection{Action/angle transformation}  \label{sec:ActionsDetails}

\subsubsection{St\"ackel approximation}  \label{sec:ActionsStaeckelDetails}

A prolate spheroidal coordinate system is characterized by a single parameter $\Delta$ -- the distance between the origin and any of the two focal points (located on the $z$ axis). The coordinate lines are ellipses and hyperbolae defined by the focal points. How exactly the coordinate values along these lines are chosen is a matter of preference: various studies use different definitions, and we suggest yet another one for the reasons to be explained shortly. 
\begin{itemize}
\item  The triplet $\lambda,\nu,\phi$ and their canonically conjugate momenta $p_\lambda, p_\nu, p_\phi$ is used in \cite{deZeeuw1985,Sanders2012,SandersBinney2016}. 
The transformation between cylindrical and prolate spheroidal coordinates is given by
\begin{subequations}
\begin{align}
R^2 &= (\lambda-\Delta^2)(\Delta^2-\nu)/\Delta^2,\quad z^2= \lambda\nu/\Delta^2, \\
\lambda,\nu &= \textstyle\frac12 (R^2+z^2+\Delta^2) \pm \frac12\sqrt{(R^2+z^2-\Delta^2)^2+4R^2\Delta^2}
\quad\mbox{(+ for $\lambda$, -- for $\nu$)}, \\
p_\lambda &= \frac{R\,v_R}{2(\lambda-\Delta^2)} + \frac{z\,v_z}{2\lambda}, \quad
p_\nu      = \frac{R\,v_R}{2(\nu    -\Delta^2)} + \frac{z\,v_z}{2\nu},     \quad
p_\phi     = R v_\phi.
\end{align}
The allowed range of variables is $0\le \nu \le \Delta^2 \le \lambda$ \footnote{
The above papers introduce different parameters of the coordinate system: $\alpha\equiv -a^2, \gamma\equiv -c^2$, such that $\Delta^2=a^2-b^2$, and the range of variables is $c^2 \le \nu \le a^2 \le \lambda$, but we may always set $c=0$. A slightly different convention is used in \cite{Bienayme2015}: $\lambda$ in that paper corresponds to $\lambda-\Delta^2$ here, and similarly $\nu$. Moreover, in some papers $\Delta$ stands for the squared focal distance. }.
An arbitrary separable axisymmetric St\"ackel potential is given by
\begin{align}  \label{eq:PotentialStaeckel}
\Phi(\lambda, \nu) = - \frac{f_\lambda(\lambda) - f_\nu(\nu)}{\lambda-\nu}.
\end{align}
\end{subequations}
The advantage of this choice is a near-symmetry between $\lambda$ and $\nu$ and the fact that they occupy distinct ranges, which makes possible to use a single function $f$ of one variable in place of both $f_\lambda$ and $f_\nu$.
The disadvantages are that the coordinates only describe the half-space $z\ge 0$ (this may be amended by extending the range of $\nu$ to $-\Delta^2\le \nu \le \Delta^2$ with the convention that $\nu<0$ corresponds to $z<0$), that the transformation is quadratic (not linear) at small $R$ or $z$, and that it does not apply in the spherical limit ($\Delta=0$).

\item  The triplet $u, v, \phi$ with the corresponding momenta is used in \cite{Binney2012, BinneyTremaine}. The transformation is given by
\begin{subequations}
\begin{align}
R &= \Delta\,\sinh u\,\sin v, \quad z = \Delta\,\cosh u\,\cos v,\qquad u\ge 0,\; 0\le v\le \pi,
\end{align}
thus $\lambda = \Delta^2\,\cosh^2 u,\; \nu = \Delta^2\,\cos^2 v$, and the expression for a generic axisymmetric St\"ackel potential is
\begin{align}
\Phi(u,v) = \frac{ U(u) - V(v) }{ \sinh^2 u + \sin^2 v }\;,\qquad
U(u)\equiv -\frac{f_\lambda[u(\lambda)]}{\Delta^2},\;V(v)\equiv -\frac{f_\nu[v(\nu)]}{\Delta^2}.
\end{align}
\end{subequations}
This form is advantageous because it covers the entire space, and $v$ tends to the spherical polar angle $\theta$ at large radii, however $u$ does not have an equally straightforward asymptotic meaning, and there is still no valid limit $\Delta\to 0$.

\item  Instead of $u$, one may use the quantity $\varpi \equiv \Delta\,\sinh u = \sqrt{\lambda-\Delta^2}$, which exactly equals the cylindrical radius $R$ whenever $z=0$ (equivalently $v=\pi/2$), regardless of $\Delta$. At large distances ($\gg \Delta$), $\varpi$ and $v$ tend to the spherical radius $r$ and the polar angle $\theta$, respectively; thus in the limit $\Delta\to 0$ these are just the spherical coordinates. The transformation is thus
\begin{subequations}
\begin{align}
R &= \varpi\,\sin v, \quad z = \sqrt{\varpi^2+\Delta^2}\,\cos v, \qquad \varpi\ge 0,\; 0\le v \le \pi,
\end{align}
and the expression for the potential is
\begin{align}
\Phi(\varpi,v) = -\frac{ f_\varpi(\varpi) - f_v(v) }{ \varpi^2 + \Delta^2\,\sin^2v }\;,\qquad
f_\varpi(\varpi)\equiv f_\lambda[\varpi(\lambda)],\; f_v(v)\equiv f_\nu[v(\nu)].
\end{align}
\end{subequations}
\end{itemize}

In the subsequent discussion, we will use the coordinates $\lambda$ and $\nu$ for consistency with the previous work, even though internally the calculations are performed in terms of $\varpi$ and $v$.

Since the two functions $f_\lambda,f_\nu$ may be shifted by an arbitrary constant simultaneously, we assume that $f_\nu(0)=0$. The continuity of potential at focal points requires that $f_\nu(\Delta^2)=f_\lambda(\Delta^2)$. Thus to obtain the values of these functions at an arbitrary point $\{\lambda,\nu\}$, we compute the potential at this point and at $\{\lambda,0\}$ (in the equatorial plane), then take $f_\lambda(\lambda) = -\Phi(\lambda,0)\,\lambda,\; f_\nu(\nu) = \Phi(\lambda,\nu)\,(\lambda-\nu) - \Phi(\lambda,0)$.

The third integral is also introduced in different forms across various studies.
Here we adopt the definition used by \cite{Sanders2012} (their eq.~3), given by
\begin{subequations}
\begin{align}
%I_3 &\equiv \lambda \left( E - \Phi(\lambda,0) - \frac{L_z^2}{2(\lambda-\Delta^2)} \right) -
%\frac{\dot\lambda^2 \,(\lambda-\nu)^2}{8(\lambda-\Delta^2)\lambda}.
I_3 &= f_\tau(\tau) + \left( E - \frac{L_z^2}{2(\tau-\Delta^2)} - 2(\tau-\Delta^2)\,p_\tau^2 \right) \tau \qquad\mbox{($\tau$ is either $\lambda$ or $\nu$)}  \label{eq:I3} \\
&= f_\tau(\tau) + \Phi(\lambda,\nu)\,\tau +
\textstyle \frac12 \big( L^2 - L_z^2 + v_z^2\Delta^2 \big) \nonumber \\
&= [\Phi(\lambda,\nu) - \Phi(\lambda,0)]\,\lambda +
\textstyle \frac12 \big( z^2 v_\phi^2 + (R v_z - z V_R)^2 + v_z^2\Delta^2 \big) . \label{eq:I3init}
\end{align}
\end{subequations}
A slightly different expression is used in \cite{Bienayme2015}: their eq.~1 introduces $I_s$ which equals $-I_3/\Delta^2$ (they denote the focal distance as $z_0$). 
The quantity introduced in \cite{BinneyTremaine} (eq.~3.248), and also used in \cite{Binney2012} (where it was also called $I_3$), is equivalent to $(I_3 + L_z^2/2) / \Delta^2 - E$.

The actions $J_\tau$, where $\tau=\{\lambda,\nu\}$, are computed as
\begin{align}  \label{eq:ActionsStaeckel}
J_\tau = \frac{1}{\pi} \int_{\tau_\mathrm{min}}^{\tau_\mathrm{max}} p_\tau\,\d \tau
\qquad\mbox{($\tau$ is either $\lambda$ or $\nu$)},
\end{align}
where the canonical momentum $p_\tau(\lambda,\nu;E,L_z,I_3)$ is expressed from (\ref{eq:I3}), and the limits of integration $\tau_\mathrm{min,max}$ are defined by the condition $p_\tau^2=0$. In the spherical limit, $J_\lambda = J_r$, $J_\nu = J_z = L-L_z$.

The essence of the St\"ackel approximation is to pretend that the potential is of the St\"ackel form and use the above expressions to compute the actions, substituting the actual potential where needed. The procedure is the following:
\begin{enumerate}
\item Choose the focal distance $\Delta$, presumably in such a way as to maximize the resemblance of the potential to a separable form (\ref{eq:PotentialStaeckel}). This defines the transformation between $\{R,z\}$ and $\{\lambda,\nu\}$.
\item Compute the potential at two points, $\{\lambda,\nu\}$ and $\{\lambda,0\}$, and assign the three integrals of motion $E, L_z$ and $I_3$ (the latter from \ref{eq:I3init}).
\item Find the integration limits $\{\lambda,\nu\}_\mathrm{min,max}$ (assuming that the orbits looks like a rectangle in the $\{\lambda,\nu\}$ plane, which is of course only an approximation if the potential is not separable). In doing so, we solve for $p_\tau^2=0$ in (\ref{eq:I3}), where $\tau$ is either $\lambda$ or $\nu$, and the other coordinate is kept at its initial value.
This step costs $\sim 30$ potential evaluations to find the three roots (the fourth one is always $\nu_\mathrm{min}=0$).
\item Compute the actions from (\ref{eq:ActionsStaeckel}), again integrating along each of the two coordinates $\{\lambda,\nu\}$ while keeping the other one fixed at its initial value.
We use a fixed-order Gauss--Legendre integration (with ten points) in a suitably scaled coordinate (to neutralize the singular behaviour of $p_\tau(\tau)$ near endpoints), hence this step costs 20 potential evaluations.
\item If the frequencies and angles are needed, follow the procedure described in the Appendix A of  \cite{Sanders2012}. This involves computation of six additional integrals for frequencies, and further six -- for the angles (note that they are carried along the same paths as the integrals for the actions, so one could store and re-use the values of potential, but this is not yet implemented).
\end{enumerate}

The accuracy of this approximation may be judged by computing numerically an orbit in the given potential, determining the actions at each point along the orbit, and estimating their variation. If actions returned by the St\"ackel approximations were true integrals, their variation would be zero. In practice, the variation depends crucially on the choice of the only free parameter -- the focal distance $\Delta$. Clearly, the best choice may depend on the two classical integrals of motion ($E$ and $L_z$). There are two alternative approaches to assigning $\Delta$:
\begin{itemize}
\item If $\Phi(\lambda,\nu)$ is a St\"ackel potential, then from (\ref{eq:PotentialStaeckel}) it follows that 
\begin{subequations}
\begin{align}
\frac{\D ^2 [ (\lambda-\nu) \Phi(\lambda,\nu)] }{\D \lambda\,\D \nu} = 0\;,
\quad\mbox{or, in the cylindrical coordinates,} \\
3 z\, \frac{\D \Phi}{\D R} - 3 R\, \frac{\D \Phi}{\D z} +
R\,z\,\left(\frac{\D ^2\Phi}{\D R^2} - \frac{\D ^2\Phi}{\D z^2}\right) +
(z^2 - R^2 - \Delta^2)\,\frac{\D ^2\Phi}{\D R\,\D z} = 0 .
\end{align}
\end{subequations}
Thus one may seek the value of $\Delta$ that minimizes the deviation of the above quantity from zero in the region occupied by the orbit.
\item A shell orbit (the one with $J_r=0$) in a St\"ackel potential has $\lambda=\mathrm{const}$; thus one may find such an orbit for each $E$ and $L_z$, %and compute the best-fit value of $\Delta$ that minimizes the variation of $\lambda$ along this orbit.
and assign $\Delta$ from the condition that $p_\lambda^2(R,z=0)$ has a maximum value 0 reached at $R=R_\mathrm{shell}$ (in other words, the range of oscillation in $\lambda$ shrinks to zero).
\end{itemize}
In either case, to make the action computation procedure efficient, we need to pre-compute a suitable table of $\Delta(E,L_z)$ and calculate the suitable value of $\Delta$ by 2d interpolation as the first step of the above procedure.
We found that the second approach generally leads to a somewhat better action conservation accuracy.

A further significant speedup may be achieved by pre-computing a 3d interpolation table for the actions as functions of three integrals of motion -- $E, L_z$ and $I_3$. In this way, we only follow the first two steps of the procedure, avoiding the costly part of finding the integration limits and performing the integration itself. The table may be constructed by following the entire procedure for a family of orbits started at $R=R_\mathrm{shell}(E,L_z),\; z=0$ and velocity directed at various angles in the meridional plane. From (\ref{eq:I3init}) it follows that in this case, $I_3 = (R^2+\Delta^2)\,v_z^2$, and since $v_z^2 = 2[E-\Phi(R,0)] - L_z^2/R^2 - v_R^2$, the maximum value of $I_3$ at the given $E$ and $L_z$ is
\begin{align}
I_3^{\mathrm{(max)}}(E, L_z) = \big( R_\mathrm{shell}^2 + \Delta^2 \big)
\big( 2[E-\Phi(R_\mathrm{shell},0)] - L_z^2/R_\mathrm{shell}^2 \big),
\end{align}
where both $R_\mathrm{shell}$ and $\Delta$ are also functions of $E$ and $L_z$.
The interpolation table is constructed in terms of scaled variables: $E$, $L_z / L_\mathrm{circ}(E)$, $I_3 / I_3^{\mathrm{(max)}}(E, L_z)$, and the values to be interpolated are $J_{r,z} / (L_\mathrm{circ}-L_z)$. The cost of construction of this 3d table is comparable to the cost of pre-computing the 2d table for $\Delta(E,L_z)$, and both take only a few CPU seconds (and are trivially parallelized).

However, since $I_3$ is only an approximate integral (which furthermore also depends on $\Delta$), this introduces an additional error in the approximation, which cannot be reduces by making the interpolation grid finer. The error comes from the fact that the accuracy of conservation of $I_3$ along the orbit is typically worse than the accuracy of conservation of $J_r,J_z$ computed at each point numerically. It turns out that the variation of $I_3$ from one point to another is largely balanced by performing the integration along the lines of constant $\lambda$ and $\nu$ that pass through the given point, as depicted in the left panel of the same figure. By contrast, in performing the interpolation from a pre-computed table, we essentially always follow the integration paths passing through $\{R_\mathrm{shell}(E,L_z),0\}$, but for a ``wrong'' $I_3$.

Nevertheless, the overall accuracy of the St\"ackel approximation is reasonably good (for low-eccentricity orbits it is typically much better than shown in the above example). For disk orbits, the relative errors are typically better than 1\%, while for halo orbits they may reach 10\% -- this happens mostly at resonances, when the orbit is not at all well described by a box in any prolate spheroidal coordinate system. The error in the interpolated approximation is a factor of 1.5--3 worse, but still tolerable in many contexts (e.g., construction of equilibrium models), and it leads to a $\sim 10\times$ speedup in action computation, for a very moderate overhead in construction of interpolation tables. In terms of accuracy and speed, the interpolated St\"ackel approximation is thus similar to the use of un-adorned $I_3$ as the approximate third integral of motion, as advocated in \cite{Bienayme2015}; however, actions have clear conceptual advantages.

%%%%%%%%%%%%%%%%%%%%%%%%%%%%%%%%%%%%%%%%%%%%%%%%%%%%%%%%%%
\subsection{Distribution functions}  \label{sec:DFdetails}

\subsubsection{Spherical anisotropic DFs}  \label{sec:DFsphericalDetails}

This type of DF, represented by the class \ttt{df::QuasiSpherical} and its descendants, is constructed from a given combination of density and potential, under certain assumptions about the functional form of the DF. At the moment, only one specific subtype is implemented in the \ttt{df::QuasiSphericalCOM} class: the Cuddeford--Osipkov--Merritt model:
\begin{align}  \label{eq:DFanisotropic}
f(E,L) &= \hat f(Q) \; L^{-2\beta_0}, \qquad Q\equiv E + L^2 / (2 r_a^2) , \\[3mm]
\hat f(Q) &= \left\{  \begin{array}{ll}
\displaystyle \frac{2^{\beta_0}}{(2\pi)^{3/2}\, \Gamma(1-\beta_0)\; \Gamma(3/2-\beta)}\,
\int_Q^0 \frac{\d \hat\rho}{\d \Phi} \frac{\d \Phi}{(\Phi - Q)^{3/2-\beta_0}}, &
1/2<\beta_0<1, \\[7mm]
\displaystyle \frac{1}{2\pi^2}\, 
\left. \frac{\d \hat\rho}{\d \Phi} \right|_{\Phi=Q}, &
\beta_0=1/2, \\[7mm]
\displaystyle \frac{2^{\beta_0}}{(2\pi)^{3/2}\, \Gamma(1-\beta_0)\; \Gamma(1/2-\beta)}\,
\int_Q^0 \frac{\d ^2\hat\rho}{\d \Phi^2} \frac{\d \Phi}{(\Phi - Q)^{1/2-\beta_0}}, &
-1/2<\beta_0<1/2, \\[7mm]
\displaystyle \frac{1}{2\pi^2}\, 
\left. \frac{\d ^2\hat\rho}{\d \Phi^2} \right|_{\Phi=Q} , &
\beta_0=-1/2,
\end{array} \right. \nonumber \\[3mm]
\hat\rho(\Phi) &\equiv \left. \rho(r) \,r^{2\beta_0}\, \big[1 + (r/r_a)^2\big]^{1-\beta_0} \right|_{r=r(\Phi)}.  \nonumber
\end{align}

Here $\hat\rho$ is the augmented density, expressed as a function of potential and then differentiated once or twice.
We use a finite-difference estimate for the radial derivatives of the original density $\rho(r)$, but this becomes inaccurate at small $r$ when both $\rho$ and $\Phi$ tend to finite limiting values. To cope with this issue, we fit a Taylor series expansion for $\rho(\Phi)$ as $\Phi \to \Phi(r=0)$, and use it at small radii where it can be differentiated analytically (only if the series produce a reasonable approximation of the density). We limit the range of the anisotropy coefficient to $\beta_0 \ge -1/2$, since lower values would need a third or even higher derivative of density, becoming too challenging to compute accurately.
The energy-dependent part of the DF $\hat f(Q)$ is represented by a log-scaled cubic spline in the scaled energy coordinate $\scE$ defined below. Negative values of $f(Q)$ are replaced by zeros.

%%%%%%%%%%%%%%
\subsubsection{Spherical isotropic DFs and the phase-volume formalism}  \label{sec:DFsphericalIsotropicDetails}

In the isotropic case ($\beta_0=0$, $r_a=\infty$), the DF is a function of $E$ alone, but it can also be expressed in terms of an action-like variable $h$.

The correspondence between energy $E$ and phase volume $h$ in the given potential is provided by the class \ttt{PhaseVolume}. Phase volume $h(E)$ and its derivative (density of states) $g(E)\equiv dh(E)/dE$ are defined as
\begin{subequations}
\begin{align}
h(E) &= \frac{16\pi^2}3 \int_0^{r_\mathrm{max}(E)} r^2\, v^3(E,r)\, \d r =
  8\pi^3\int_0^{L^2_\mathrm{circ}(E)} J_r(E,L)\,\d L^2 =
  \int_{\Phi(0)}^E g(E')\, \d E' ,\\
g(E) &= 16\pi^2 \int_0^{r_\mathrm{max}(E)} r^2\, v(E,r)\, \d r =
  4\pi^2\, \int_0^{L_\mathrm{circ}^2(E)} T_\mathrm{rad}(E,L)\,\d L^2,
\end{align}
\end{subequations}
where  $v = \sqrt{2(E-\Phi(r))}$ is the velocity,  $L_\mathrm{circ}(E)$ is the angular momentum of a circular orbit with energy $E$,  and  $T_\mathrm{rad}(E,L) \equiv 2 \int_{r_-}^{r_+} \d r/v_r \equiv 2 \int_{r_-}^{r_+} \d r/\sqrt{v^2-L^2/r^2} = 2\pi\,\D J_r/\D E$ is the radial period (its dependence on $L$ at a fixed $E$ is usually weak).
In other words, phase volume is literally the volume of phase space enclosed by the energy hypersurface.

The bi-directional correspondence between $E$ and $h$ is given by two 1d quintic splines (with derivative at each node given by $g$) in scaled coordinates. Namely, we use $\ln h$ as one coordinate, and the scaled energy $\scE \equiv \ln[1/\Phi(0) - 1/E]$ as the other one (both when the potential has a finite value $\Phi(0)$ at origin, or when it tends to $-\infty$). The purpose of this scaling is twofold. First, in the case of a finite $\Phi(0)$, any quantity that depends on $E$ directly is poorly resolved as $E\to \Phi(0)$ because of finite floating-point precision: e.g., if $\Phi(0)=-1$, and $E=-1+10^{-8}$ (corresponding to the radius as large as $10^{-4}$ in a constant-density core), we only have half of the mantissa available to represent the variation of $E$. By performing this scaling, we ``unfold'' the range of $\scE$ down to $-\infty$ with full precision. Second, this scaling converts a power-law asymptotic behaviour of $h(\Phi(r))$ at small and large radii into a linear dependence between $\ln h$ and $\scE$, suitable for extrapolation. Namely, as $E \to 0$ and $\Phi \propto -1/r$ (which is true for any finite-mass model in which the density drops faster than $r^{-3}$ at large radii), $h(E)\propto (-E)^{-3/2}$ and $g(E)\propto (-E)^{-5/2}$. At small radii, if the density behaves as $\rho \propto r^{-\gamma}$ and the corresponding potential -- as $\Phi \propto r^{2-\gamma}$, then $h(E) \propto [E-\Phi(0)]^{(12-3\gamma)/(4-2\gamma)}$ in the case $\gamma<2$ (when $\Phi(0)$ is finite), or $h(E) \propto (-E)^{(12-3\gamma)/(4-2\gamma)}$ if $2\le \gamma \le 3$ (including the case of the Kepler potential, $\gamma=3$); in both cases, $g(h) \propto h^{(8-\gamma)/(12-3\gamma)}$.
The interpolation in scaled coordinates typically attains a level of accuracy better than $10^{-9}$ over the range of $h$ covered by the spline, and $\sim 10^{-5}$ in the extrapolated regime (if the potential indeed has a power-law asymptotic behaviour).

Any non-negative function $f(h)$ may serve as a spherical isotropic DF. One possible representation is provided by the \ttt{math::LogLogSpline} class -- an interpolating spline in doubly-logarithmically scaled coordinates (i.e., $\ln f (\ln h)$ is a cubic spline and is extrapolated linearly to small and large $h$). Such DFs are constructed, e.g., by routines \ttt{createSphericalIsotropicDF} and \ttt{fitSphericalIsotropicDF} defined in \texttt{df_spherical.h}.

The main application of these DFs is for simulating the effect of two-body relaxation, used in the Monte Carlo code \textsc{Raga} \cite{Vasiliev2015}. We assume that test stars are moving in the background of fields stars with distribution function $f(h)$, and both test and field stars have the same mass. There are two possible descriptions of relaxation phenomena: either locally, as a perturbation to the velocity $v\equiv \sqrt{2(E-\Phi(r))}$ of the test star at the given position $r$, or, in the orbit-averaged approach, as a perturbation to the star's energy $E$ averaged over its radial motion.
In both cases, the rate of change of the given quantity per unit time is denoted by $\langle \dots \rangle$.

The local (position-dependent) drift and diffusion coefficients in velocity are given by
\begin{subequations}
\begin{align}
v\dvpar &= \textstyle -2\Gamma\, J_{1/2} \;,\\
\dvsqpar&= \textstyle \;\frac{2}{3}\Gamma\, \left(I_0 + J_{3/2}\right) , \\
\dvsqper&= \textstyle \;\frac{2}{3}\Gamma\, \left(2 I_0 + 3 J_{1/2} - J_{3/2}\right) , \quad\mbox{where} \\
I_0(E)     &\equiv \int_E^0 f(E')\, \d E' = \int_{h(E)}^\infty \frac{f(h')}{g(h')}\, \d h', \\
J_{n/2}(E,\Phi) &\equiv \int_{\Phi(r)}^E f(E') \left(\frac{E'-\Phi}{E-\Phi}\right)^{n/2} \d E' = 
 \int_{h(E)}^\infty \frac{f(h')}{g(h')}\, \left(\frac{E'(h')-\Phi}{E-\Phi}\right)^{n/2} \d h'.
\end{align}
\end{subequations}

Orbit-averaged energy drift and diffusion coefficients are given by
\begin{subequations}  \label{eq:DriftDiffusionE}
\begin{align}
\langle \Delta E   \rangle_\mathrm{av} &= \phantom{2} \Gamma\; \left[I_0 - K_g/g \right], \\
\langle \Delta E^2 \rangle_\mathrm{av} &= 2\Gamma\; \left[I_0\,h + K_h\right]/g, \\
K_g(E) &\equiv \int_{\Phi(0)}^E f(E')\,g(E')\,\d E' = \int_0^{h(E)} f(h')\,\d h', \\
K_h(E) &\equiv \int_{\Phi(0)}^E f(E')\,h(E')\,\d E' = \int_0^{h(E)} \frac{f(h')\,h'}{g(h')}\,\d h'.
\end{align}
\end{subequations}

In these expressions, $\Gamma  \equiv 16\pi^2 G^2 M_\mathrm{total} \times (N_\star^{-1}\ln\Lambda)$, where the term in brackets is the amplitude of relaxation term for the given number of stars $N_\star$ representing the stellar system ($\Lambda \sim N$ is the Coulomb logarithm). We note that $K_g(E)$ is the mass of stars with energies less than $E$ (and thus $M_\mathrm{total} = K_g(0)$), and $K_h(E)$ is their kinetic energy (up to a factor 3/2).

Of course, an efficient evaluation of diffusion coefficients again requires interpolation from pre-computed tables, which are provided by the class \ttt{SphericalIsotropicModelLocal}. From the above expressions it is clear that $I_0$, $K_g$ and $K_h$ can be very accurately approximated by quintic splines in $h$, log-scaled in both coordinates and linearly extrapolated (provided that $f(h)$ also has power-law asymptotic behaviour at large and small $h$). 
Moreover, $J_0(E,\Phi) = I_0(\Phi)-I_0(E)$, and $J_{n/2}(E,\Phi)\lesssim J_0$ thanks to the weighting factor (the ratio of velocities of field and test stars to the power of $n$). Indeed, for $E\to\Phi$, $J_{n/2} \to 1/(n+1)$. We interpolate the ratio $J_{n/2}/J_0$ as a function of $\ln h(\Phi)$ and $\ln h(E)-\ln h(\Phi)$ on a 2d grid covering a very broad range of $h$; the accuracy of this cubic spline interpolation is $\sim 10^{-4}..10^{-6}$, and it is extrapolated as a constant outside the definition region (while this is a good asymptotic approximation for large $h$, there is no easy way of delivering a reasonably correct extrapolation to small $h(\Phi)$ -- fortunately, the volume of this region is negligible in practice).

Another method for studying the evolution of stellar distribution driven by the two-body relaxation is the Fokker--Planck (FP) equation for $f(h,t)$ coupled with the 1d Poisson equation for $\Phi(r,t)$. The latter provides the potential corresponding to the density profile which is obtained by integrating the DF over velocity: $\rho(r,t)=\int f(h,t) \d ^3v$. This system of two PDEs -- parabolic for the DF and elliptic for the potential -- is solved using interleaved steps: first the orbit-averaged drift and diffusion coefficients entering the FP equation are obtained from (\ref{eq:DriftDiffusionE}), then the DF is evolved for some interval of time in a fixed potential, then the density is recomputed and the potential is updated through the Poisson equation.
Traditionally, the DF is expressed as a function of energy, but this has a disadvantage when it comes to solving the Poisson equation: as the potential changes, the DF should be kept fixed as a function of phase volume, not energy (e.g., \cite{Cohn1980}). Thus the formulation entirely in terms of $f(h)$ is preferrable, and does not introduce any additional complications. It is convenient to write down the FP equation in the flux-conservative form:
\begin{align}
\frac{\D f(h,t)}{\D (\Gamma t)} &= \frac{\D }{\D h} \left[ D_{hh} \frac{\D f(h,t)}{\D h} + D_h f(h,t) \right] , \label{eq:FokkerPlanck} \\
D_{hh} &= g(h)\,\big[ h\, I_0(h) + K_h(h) \big], \qquad D_h = K_g(h).  \label{eq:FokkerPlanckCoefs}
\end{align}
To achieve high dynamical range, $h$ is further replaced by $\ln h$, with a trivial modification of the above expressions.

The Fokker--Planck solver, dubbed \textsc{PhaseFlow} \cite{Vasiliev2017}, has several ingredients:
\begin{itemize}
\item The distribution function $f(h,t)$ for the evolving population of stars, represented by its values on a grid in $\ln h$. Values of $f(h)$ for an arbitrary argument are obtained from a cubic spline interpolator for $\ln f$ as a function of $\ln h$.
\item The potential $\Phi(r)$ corresponding to the density profile $\rho(r)$ of the evolving population, plus optionally an external component (e.g., a central point mass). The 1d Poisson equation is solved by the \ttt{Multipole} class (of course, a monopole is a particular case of a multipole).
\item The density $\rho(\Phi(r)) = \int_{\Phi}^0 f\big(h(E)\big)\,4\pi\sqrt{2(E-\Phi)}\,\d E$ is obtained from the DF \textit{in the given potential}. We recompute the density after the DF has been evolved in the Fokker--Planck step, using the potential extrapolated from its previous evolution to the current time. This extrapolation makes in unnecessary to iteratively improve the solution by substituting the self-consistent potential back to the r.h.s.\ of this equation.
\item The drift and diffusion coefficients $D_h, D_{hh}$ are computed using a dedicated class \ttt{SphericalIsotropicModel} which combines the DF $f(h)$ with a potential $\Phi(r)$ and a mapping $h\leftrightarrow E$ (\ttt{PhaseVolume}) constructed for the given $\Phi$.
\item The FP equation (\ref{eq:FokkerPlanck}) in the discretized form is solved with the Chang--Cooper scheme (e.g., \cite{ParkPetrosian1996}) or with a finite-element method, described in the appendix of \cite{Vasiliev2017}.
\end{itemize}

%%%%%%%%%%%%%%
\subsection{Schwarzschild modelling}  \label{sec:SchwarzschildDetails}

\begin{figure}
\includegraphics[width=16cm]{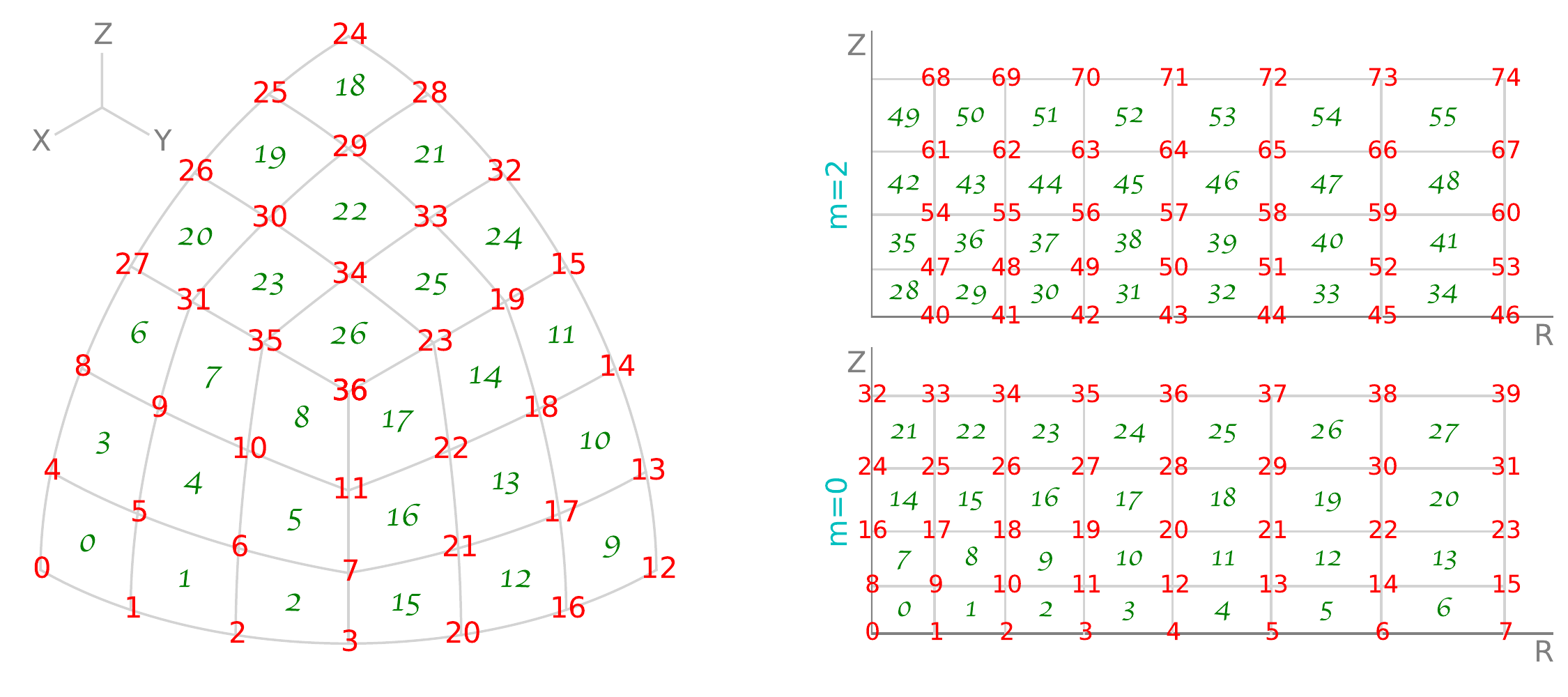}
\caption{Discretization schemes for the 3d density profile in Schwarzschild models.\protect\\
Left panel shows one radial shell in the ``classic'' grid scheme: one octant of a sphere is divided into three equal panes, and each pane -- into $K\times K$ cells, where $K$ is given by the parameter \ppp{stripsPerPane}. The volume discretization elements are 0th degree B-splines ($\sqcap$-shaped blocks) in the case of \ppp{type="DensityClassicTopHat"} (green indices at cell centers), or 1st degree B-splines ($\wedge$-shaped elements) in the case of \ppp{type="DensityClassicLinear"} (red indices at grid vertices). Indices further increase along the third dimension (radius), with radial shells that could be placed at arbitrary intervals (parameter \ppp{gridR}); for \ppp{DensityClassicLinear}, a single vertex at origin is added as the 0th element. In both cases, the grid may be further stretched along $Y$ and $Z$ axes by an amount controlled by \ppp{axisRatioY}, \ppp{axisRatioZ}; the radial grid then refers to the elliptical radius.\protect\\
Right panels show the meridional section ($R,Z$) of the grid in the ``cylindrical'' scheme; the grid is rectangular but arbitrarily spaced in both directions (parameters \ppp{gridR, gridz}), and each even azimuthal harmonic term $m \le {}$\ppp{mmax} has a separate set of indices. Again, the volume discretization elements are 0th degree B-splines in the case of \ppp{type="DensityCylindricalTopHat"} (green indices at cell centers) and 1st degree B-splines in the case of \ppp{type="DensityCylindricalLinear"} (red indices at grid vertices, excluding the $Z$ axis for $m>0$, where the coefficients of the Fourier expansion of density are always zero).
}  \label{fig:DensityGrid}
\end{figure}

The framework for constructing Schwarzschild orbit-superposition models is centered around the concept of \ttt{Target} -- an abstract interface for representing some features of the model in a discretized form. We denote the required values $U_n$ of these features as $N_\mathrm{cons}$ model constraints, and the contributions of $i$-th orbit as $u_{i,n}$. The goal of the modelling procedure is to reproduce the constraints by a weighted superposition of orbit contributions.
There are two categories of targets: density and kinematic.

\paragraph{Density targets} are used to produce a gravitationally self-consistent solution, in which the total density of the weighted superposition of orbits agrees with the Laplacian of the gravitational potential in which the orbits are integrated.
There are three discretization schemes, which differ in the geometry of the spatial grid: classic, cylindrical, and spherical-harmonic. The first two have two variants each, differing in the degree of B-spline basis set (0 -- top-hat, 1 -- linear), and the latter is always 1st degree. 
In the classic scheme, the volume is divided into spherical or concentric ellipsoidal shells, the surface of each shell -- into three panes, and each pane -- into $K\times K$ cells, as shown in Figure~\ref{fig:DensityGrid}, left panel. In \ppp{DensityClassicTopHat}, the volume of the model is divided into these 3d cells, and the mass of each cell (density integrated over the volume of the cell) serves as a single constraint. Equivalently, the volume discretization elements are non-overlapping quasi-rectangular blocks, and the basis functions have amplitude 1 in each cell and 0 elsewhere. In \ppp{DensityClassicLinear}, the basis functions are $\wedge$-shaped in all three directions (radial and two angular), with amplitude 1 reached at a single vertex of the grid and linearly tapering to zero at adjacent vertices. At any point, there are several (up to 8) overlapping basis functions, with their amplitudes summing up to unity. The density integrated with these weight functions still has the meaning of mass, but is associated with a grid vertex rather than grid segment, as shown in the above figure.

In the cylindrical scheme, the density is first expanded into Fourier series in the azimuthal angle $\phi$, and then each $m$-th term ($m=0, 2, \dots, m_\mathrm{max}$) is represented with a 2d B-spline basis set on the orthogonal grid in $R,z$, as shown in Figure~\ref{fig:DensityGrid}, right panel. Similarly to the previous scheme, \ppp{DensityCylindricalTopHat} has basis functions that are non-overlapping and have amplitude 1 inside each cell of the 2d grid, while \ppp{DensityCylindricalLinear} has $\wedge$-shaped basis functions associated with grid vertices rather than segments. There is an additional factor $2\pi R$ in the integration of density times the basis function, so that the constraint values for the $m=0$ term have the meaning of the mass in each grid cell or vertex (hence $\sum_{n=1}^{N_\mathrm{cons}} U_n$ is the total mass within the entire grid, similarly to the classic scheme). The constraint values for higher Fourier terms ($m>0$) can have positive or negative sign, and should not be summed up to get the total mass. Since these higher-$m$ terms must be zero on the $z$ axis from regularity conditions, 1st-degree basis functions at the leftmost vertex in $R$ are excluded from the basis set.

Finally, the \ppp{DensitySphHarm} scheme represents one radial coordinate with a B-spline basis set, and two angular coordinates with the spherical-harmonic basis set with order $l_\mathrm{max},m_\mathrm{max}$. Both this and the previous schemes are conceptually similar to the \ttt{Density\-SphericalHarmonic} (Section~\ref{sec:PotentialMultipoleDetails}) and \ttt{DensityCylGrid} (Section~\ref{sec:PotentialCylSplineDetails}) classes, which represent a 3d density profile as a corresponding interpolated function. The difference is that in those classes, the free parameters are the values of Fourier or multipole coefficients of density expansion at grid points, while in the target classes discussed in this section, the free parameters have the dimension of mass (density integrated over some volume).

The choice of a particular discretization scheme should be tailored to the density profile that is being represented: in the case of a spheroidal profile, classic or spherical-harmonic schemes work best, while for a disky profile (possibly with a non-axisymmetric bar), cylindrical grid is preferred. In all cases, the radial (and vertical, in the cylindrical scheme) grids are defined by the user, typically with uniform or exponential spacing, and enclosing $\gtrsim 90-99\%$ of the total mass.

\paragraph{Kinematic targets} come in two flavors. One is \ppp{KinemShell}, which represents the density-weighted radial and tangential velocity dispersions $\rho\sigma_{r,t}^2$ as functions of radius, projected onto the basis set of B-splines of \ppp{degree=0..3} defined by \ppp{gridr} in spherical radius. 
It is useful to constrain the velocity anisotropy profile $\beta\equiv 1 - \frac12 \sigma_t^2/\sigma_r^2$: if $U_n^r$ and $U_n^t$ are two equal-length arrays of these projections, then the constraints to be satisfied are written as $0 = 2(1-\beta_n)\,U_n^r - U_n^t$, where $\beta_n$ is the value of $\beta$ associated with $n$-th radial basis element. This can be used in the context of ``theoretical'' Schwarzschild models, when the goal is to construct a dynamically self-consistent model with the given density profile and have some control on its kinematic structure by assuming some functional form of $\beta(r)$.

More important in practice is the \ppp{LOSVD} target, which is used to constrain the kinematic structure of the model by observed line-of-sight velocity distributions in the image plane. There are several related ways of representing a LOSVD, and the relation between them is explained below.

The orientation of the image plane in the intrinsic coordinate system associated with the galaxy model is specified by three Euler angles $\alpha,\beta,\gamma$, see Section~\ref{sec:CoordinateDetails} for the definition, and Figure~\ref{fig:EulerAngles} for an illustration. The intrinsic (model) coordinate system is denoted as $xyz$, and the observational coordinate system -- as $XYZ$, with $Y$ axis pointing up in the image plane, $X$ axis pointing left (note the opposite of the usual convention!), and $Z$ axis pointing perpendicular to the image plane (along the line of sight) away from the observer. This unusual sign convention for the $X$ axis is a consequence of the right-handedness of the coordinate frame. $\beta$ is the usual inclination angle; $\gamma$ is the angle between the line of nodes (intersection of $xy$ and $XY$ planes) and the $X$ axis, and $\alpha$ is the angle between the line of nodes and the $x$ (major) axis of the galaxy (it is relevant only for non-axisymmetric systems). If there are several observational datasets, each one needs a separate instance of a \ppp{LOSVD} target, with the parameters \ppp{gamma} possibly different between instances, and the other two angles \ppp{alpha}, \ppp{beta} being identical.

The LOSVDs are recorded in several spatial regions (apertures), which are arbitrary polygonal regions $\Omega_a$ in the image plane. The \ppp{apertures} parameter should contain a list of $N_a\times2$ two-dimensional arrays specifying the $N_a$ coordinates $X,Y$ of $a$-th boundary polygon. Often these apertures come from binning up pixels in a regular two-dimensional grid in the image plane, e.g., using the Voronoi binning approach \cite{CappellariCopin2003} or some other scheme. There is a \Python routine \ttt{getBinnedApertures} that reconstructs the boundary polygons from an array describing the correspondence between bin indices and pixel coordinates. But any alternative way of defining apertures (e.g. circular fibers, sectors in polar coordinates, or spaxels of a long slit) is equally well possible to describe with arbitrary boundary polygons.

The effect of finite spatial resolution in the image plane is encoded in the point-spread function (PSF) of the instrument. The parameter \ppp{psf} can be specified either as a single number (the width of a circular Gaussian), or as a 2d array of several Gaussian components (the first column is the width and the second is the relative fraction of this component, which should sum up to unity).

The kinematic datacube is three-dimensional: two image-plane coordinates $X,Y$ and the line-of-sight velocity $V_Z$. Accordingly, it is first recorded on a rectangular 3d grid in these variables, and represented internally as a 3d tensor-product B-spline: 
\begin{align}  \label{eq:LOSVDinternal}
\mathfrak{f}^\mathrm{(int)}(X,Y,V_Z) = \sum_{i,j,k} A_{ijk}\;B^{(X)}_i(X)\;B^{(Y)}_j(Y)\;B^{(V)}_k(V_Z),
\end{align}
where $A_{ijk}$ are the amplitudes and $B^{(X)},B^{(Y)},B^{(V)}$ are basis functions in each of the three directions. Then the two spatial directions are convolved with the PSF and rebinned onto the array of apertures. The output functions to be represented are the integrals of the LOSVD, convolved with the spatial PSF, over the area of each aperture $\Omega_a$:
\begin{subequations}  \label{eq:LOSVDoutput}
\begin{align}
\mathfrak{f}_a(V_Z) &= \sum_{k} A_{a,k}\;B^{(V)}_k(V_Z), \\
A_{a,k} &= \sum_{i,j} \iint_{X,Y\in \Omega_a} 
\d X\, \d Y  \iint \d X'\, \d Y' \label{eq:LOSVDoutputAmpl} \\
&\times A_{ijk}\;B^{(X)}_i(X')\;B^{(Y)}_j(Y')\;\; PSF\big( \sqrt{ (X-X')^2 + (Y-Y')^2 } \big)  \nonumber.
\end{align}
\end{subequations}
The amplitudes $A_{a,k}$ of B-spline expansion in the velocity dimension (indexed by $k=1..N_V$) for each aperture (indexed by $a$) are stored in the flattened one-dimensional output array: each consecutive $N_V$ numbers refer to one aperture. The conversion between the internal and the output representations is performed transparently to the user, using a single matrix multiplication, for which the matrix is pre-computed in advance and combines the spatial convolution and rebinning steps. The parameters provided by the user are: the grids in the image plane \ppp{gridx}, \ppp{gridy} and velocity \ppp{gridv}, and the degree of B-spline basis set ranging from 0 to 3 (although \ppp{degree=2} or \ppp{3} is strongly recommended for a much greater accuracy at the same grid size, as illustrated in the appendix of \cite{VasilievValluri2020}). The image-plane grid should cover all apertures, and preferrably have an extra margin of $2-3$ times the PSF width for a more accurate convolution, but needs not be aligned with any of the apertures: the integration over $\Omega_a$ is performed exactly no matter what is the overlap between grid segments and aperture polygons. The spatial size of the grid should be comparable with the PSF width or the aperture size, whichever is larger. For instance, in the typical case that an adaptive binning scheme is employed, the apertures may consist of a single spaxel of the detector in the central area, typically smaller than the PSF width, and contain many such spaxels in the outer parts of the image. Then one may define a non-uniform \ppp{gridx} with smaller segments $\simeq$PSF width in the central part, which gradually become comparable to sizes of outermost apertures towards the endpoints in $X$. The parameter \ppp{gridy} may be omitted if it is identical to \ppp{gridx}.

One may also perform smoothing along the velocity axis by providing a nonzero \ppp{velpsf} parameter. Since the integrated-light LOSVDs  are usually produced by a spectral fitting code in a velocity-deconvolved form, this is not needed (although won't hurt if the smoothing width is set to a significantly smaller value than the velocity dispersion). However, if the LOSVD is computed from individual stellar velocities, this parameter may represent the typical observational error (unfortunately, it is not easy to account for variable error bars between individual measurements).

Finally, another important parameter is \ppp{symmetry}, which determines how the model LOSVDs are symmetrized before producing the output array. Possible values are: \ppp{symmetry=} \ppp{'t'} for the triaxial geometry, in which a fourfold discrete symmetry $\{x,y\}\leftrightarrow \{-x,-y\}, z \leftrightarrow -z$ holds even in the case of figure rotation; \ppp{symmetry='a'} for axisymmetric systems, which is approximately enforced by randomizing the azimuthal angle $\phi$ for each recorded point; or \ppp{symmetry='s'} for spherical systems, when both angles are randomized. This is performed in the intrinsic coordinate system $xyz$ before projection: for instance, two out of four possible identical points in the triaxial case correspond to a reflection symmetry $\mathfrak{f}(X,Y,V_Z) = \mathfrak{f}(-X,-Y,-V_Z)$, but two other points project to coordinates unrelated to $X,Y$. This parameter should be in agreement with the symmetry of the potential, but needs to be provided separately, as the \ttt{Target} object has no knowledge of the potential.

\paragraph{Surface density} is the integral of $\mathfrak{f}(X,Y,V_Z)$ along the velocity axis. When the \ttt{Target} is applied to a \ttt{Density} object, it produces an array of aperture masses -- integrals of PSF-convolved LOSVDs over the area of each aperture and over velocity:
\begin{align}  \label{eq:ApertureMass}
\mathfrak M_a \equiv \int \mathfrak{f}(V_Z)\; \d V_Z .
\end{align}
In terms of the B-spline representation of the LOSVD, it can be expressed as the dot product of the vector of amplitudes $A_{a,k}$ (\ref{eq:LOSVDoutputAmpl}) by the vector of integrals of basis functions
\begin{align}  \label{eq:BsplineIntegrals}
\mathcal I_k \equiv \int B_k^{(V)}(V_Z)\; \d V_Z .
\end{align}

\paragraph{Observational constraints} on the LOSVD do not come in the form of B-splines, therefore one needs to convert the coefficients $A_{a,k}$ (output by the \ttt{Target} object as  a 1d flattened array $U_n$ for the entire model, or a 2d array $u_{i,n}$ of coefficients for each $i$-th orbit) into a form suitable for comparison with observations.

The observed LOSVDs are usually not normalized, i.e., they provide the distribution of stars in velocities, but not the total luminosity in the given aperture. This quantity needs to be computed from the model, by applying the LOSVD \ttt{Target} to the \ttt{Density} object.

Typically, the mass-to-light ratio of stars $\Upsilon$ is a free parameter in the models. When changing the mass normalization of all galactic components (including dark matter, central black hole, etc.) by the same factor $\Upsilon$, one can reuse the same orbit library, but rescale the model velocities by a factor $\sqrt\Upsilon$ before comparing the model LOSVDs to the observed ones. The rescaled LOSVD is represented by a B-spline: 
$\mathfrak{f}'_a(V_Z) = \sum_k A'_{a,k}\;B_k\big(\sqrt\Upsilon\,V_Z\big)$, where the new set of basis functions is defined by the velocity grid multiplied by $\sqrt\Upsilon$, and the new amplitudes are $A'_{a,k} = A_{a,k}/\sqrt\Upsilon$.

The observed LOSVD in $a$-th aperture is usually represented by some sort of basis-set expansion 
$\mathfrak{f}^\mathrm{(obs)}_a(V_Z) = \sum_l C_{a,l}\,F_{a,l}(V_Z)$, with a vector of coefficients $\boldsymbol C_a \equiv C_{a,l}$ and the set of basis functions $F_{a,l}(V_Z)$. This could be a B-spline basis, e.g., a velocity histogram (0th-degree B-spline), in which case the basis functions are the same for all apertures, or a Gauss--Hermite (GH) expansion (Section~\ref{sec:MathGaussHermiteDetails}), in which case the basis functions depend on three additional parameters in each aperture -- amplitude $\Xi_a$, center $\mu_a$ and width $\sigma_a$ of the Gaussian function, which is the zeroth term in the expansion. In either case, the model LOSVDs can be \textit{reinterpolated} onto the observational basis set(s), as explained in the \hyperref[sec:MathBasisSetChange]{last paragraph} of Section~\ref{sec:MathBasisSetDetails}. The transformation between the vector of (rescaled) B-spline amplitudes of the model LOSVD $\boldsymbol A'_a$ and the vector of expansion coefficients in the observational basis set $\boldsymbol C_a$ is described by a matrix multiplication: $\boldsymbol C_a = \mathcal G^{-1}\,\mathsf H\, \boldsymbol A'_a$, where $H_{lk} \equiv \langle F_{a,l}, B_k \rangle$ and $\mathcal G_{lm} \equiv \langle F_{a,l}, F_{a,m} \rangle$ are the matrices of inner products of corresponding basis functions.

\paragraph{Example} \label{sec:SchwarzschildExample} of all these steps is provided below (a complete \Python script is given in \texttt{example_forstand.py}). The apertures are Voronoi-binned as described in the file \texttt{voronoi_bins.txt} containing $N_\mathrm{aper}$ rows and 3 columns: $X$ and $Y$ coordinates of bin centers, and bin index. The kinematic measurements are provided in two alternative forms: (1) LOSVD histograms in the file \texttt{losvd_histograms.txt}, containing $N_\mathrm{aper}$ rows and $2\,N_\mathrm{bins}$ columns, each pair of consecutive columns giving the amplitude of LOSVD in a given bin of velocity grid and its error estimate; (2) Gauss--Hermite moments in the file \texttt{losvd_ghmoments.txt}, containing $N_\mathrm{aper}$ rows and 12 columns -- values and error estimates of $\mu, \sigma, h_{3..6}$.\\[2mm]
\texttt{vorbins~~~= numpy.loadtxt("voronoi_bins.txt")\\
apertures~= agama.schwarzlib.getBinnedApertures(\\
\mbox{}~~~~xcoords=vorbins[:,0], ycoords=vorbins[:,1], bintags=vorbins[:,2])\\
histfile~~= numpy.loadtxt("losvd_histograms.txt")\\
obs_gridv~= numpy.linspace(-v_max, v_max, 16)}
\textit{\color{Sepia} \ \# observational vel.~grid, $N_\mathrm{bins}=15$}\\
\texttt{obs_degree= 0}
\textit{\color{Sepia} \ \# histograms are 0th-degree B-splines}\\
\texttt{hist_val~~= histfile[:,0::2];~~hist_err = histfile[:,1::2]}
\textit{\color{Sepia} \ \# odd/even columns}\\
\texttt{ghmfile~~~= numpy.loadtxt("losvd_ghmoments.txt")\\
ghm_val~~~= ghmfile[:,0::2];~~ghm_err = ghmfile[:,1::2]\\
num_aper~~= len(apertures)} 
{\color{Sepia}\textit{\ \# number of apertures}\texttt{ = len(histfile) = len(ghmfile)}}\\[2mm]
Define the density and LOSVD \ttt{Target} objects (only the necessary parameters are provided):\\[2mm]
\texttt{mod_gridx~= numpy.linspace(-r_max, r_max, 50)}
\textit{\color{Sepia} \ \ \# grid should cover all apertures}\\
\texttt{mod_gridv~= numpy.linspace(-v_max, v_max, 25)}
\textit{\color{Sepia} \ \ \# and all velocities in the model}\\
\texttt{mod_degree= 2}
\textit{\color{Sepia} \ \ \# degree of B-splines for representing model LOSVDs (use 2 or 3)}\\
\texttt{tar_los~~~= agama.Target(type="LOSVD", gridx=mod_gridx, gridv=mod_gridv,\\
\mbox{}~~~~degree=mod_degree, apertures=apertures, symmetry="s", psf=psf)\\
mod_gridr~= agama.nonuniformGrid(30, 0.01*r_max, 5.*r_max)\\
tar_den~~~= agama.Target(type="DensitySphHarm", lmax=0, gridr=mod_gridr)
}\\[2mm]
Assume we have a model potential \texttt{pot}, which also doubles as the density profile of stars, and have constructed initial conditions for the orbit library in \texttt{ic} (2d array of shape \texttt{num_orbits}${}\times 6$). Integrate the orbits while collecting the matrices $u_{i,n}^{(t)}$ containing the contribution of $i$-th orbit to $n$-th discretization element of $t$-th target:\\[2mm]
\texttt{mat_den, mat_los = agama.orbit(potential=pot, ic=ic, time=100.*pot.Tcirc(ic),\\
\mbox{}~~~~targets=[tar_den, tar_los])}\\[2mm]
Next we compute the required values of density and kinematic constraints. As explained above, the observational kinematic profiles are not normalized, so we compute the overall scaling factors $\mathfrak M_a$ (\ref{eq:ApertureMass}) from the model density profile. For GH moments, a couple of extra steps are needed. First, the normalization is translated into the amplitude of the base Gaussian, summing the contributions from all even GH moments (\ref{eq:GHnormalization}). Second, the uncertainties on the mean and width of the Gaussian are converted into (approximate) uncertainties on $h_1,h_2$, which can be incorporated into the linear equation system.
Finally, we may need to constrain the PSF-convolved surface density profile (aperture masses) separately from the 3d density profile, at least when working with GH moments (alternatively, one may add a set of constraints that $h_0$ be close to unity). This is expressed by a separate matrix constructed by dot-multiplying the LOSVD matrix of B-spline amplitudes of each orbit by the vector of B-spline integrals $\mathcal I$ (\ref{eq:BsplineIntegrals}). When fitting to LOSVD histograms directly, they will be already normalized, so separate aperture mass constraints are not necessary.\\[2mm]
\texttt{cons_den~~= tar_den(pot)}
\textit{\color{Sepia} \ \ \# required values of 3d density constraints}\\
\texttt{cons_sur~~= tar_los(pot)}
\textit{\color{Sepia} \ \ \# aperture masses (surface density integrated over apertures)}\\[1mm]
\textit{\color{Sepia}\# row-normalize the provided histograms and multiply by aperture masses}\\
\texttt{obs_bsint~= agama.bsplineIntegrals(degree=obs_degree, grid=obs_gridv)}\\
\texttt{num_obs_bs= len(obs_bsint)}
\textit{\color{Sepia}\ \ \# number of bins in observed velocity histograms}\\
\texttt{hist_norm~= hist_val.dot(obs_bsint)}
\textit{\color{Sepia} \ \ \# $\int \mathfrak f_a^\mathrm{(obs)}(V_Z)\, \d V_Z$ from the provided histograms}\\
\texttt{hist_val *= (cons_sur / hist_norm).reshape(num_aper, 1)}\\
\texttt{hist_err *= (cons_sur / hist_norm).reshape(num_aper, 1)}\\[1mm]
\textit{\color{Sepia}\# normalization of the GH series in each aperture with contributions from $h_4$, $h_6$}\\
\texttt{ghm_norm~~= 1 + ghm_val[:,3] * (24**0.5 / 8) + ghm_val[:,5] * (720**0.5 / 48)}\\
\textit{\color{Sepia}\# parameters of GH series (amplitude, center, width)}\\
\texttt{gh_params~= numpy.array([ cons_sur/ghm_norm, ghm_val[:,0], ghm_val[:,1] ]).T}\\
\texttt{ghm_err[:,0:2] /= 2**0.5 * ghm_val[:,1:2]}
\textit{\color{Sepia}\ \ \# $\delta h_1 \approx \delta v / \sqrt{2}\sigma$, $\delta h_2 \approx \delta \sigma / \sqrt{2}\sigma$}\\
\texttt{ghm_val[:,0:2] *= 0} \textit{\color{Sepia}\ \ \# set $h_1=h_2=0$}\\
\texttt{num_obs_gh= ghm_val.shape[1]} 
\textit{\color{Sepia}\ \ \# order of GH expansion (6 in our case)}\\[1mm]
\textit{\color{Sepia}\# matrix of orbital contributions to aperture masses}\\
\texttt{mod_bsint = agama.bsplineIntegrals(degree=mod_degree, grid=mod_gridv)}\\
\texttt{num_mod_bs= len(mod_bsint)}
\textit{\color{Sepia}\ \ \# number of velocity basis functions in each aperture $N_V$}\\
\texttt{mat_sur~~~= mat_los.reshape(num_orbits, num_aper, num_mod_bs).dot(mod_bsint)}\\[2mm]
Now construct a Schwarzschild model for a particular value of mass-to-light ratio $\Upsilon$, scaling the velocity grid of orbit LOSVDs by $\sqrt{\Upsilon}$ and their amplitudes by $1/\sqrt{\Upsilon}$ before converting them into the form compatible with the observational constraints.\\[2mm]
\texttt{mod_gridv_scaled = mod_gridv * Upsilon**0.5}\\[2mm]
If using LOSVD histograms, transform the matrix of B-spline amplitudes of orbit LOSVDs into the matrix of histogram values (amplitudes of 0th-degree B-spline) defined by the observational velocity grid. This is achieved by the following conversion matrix:\\[2mm]
\texttt{
conv~~~~~~= numpy.linalg.solve(\\
\mbox{}~~~~agama.bsplineMatrix(obs_degree, obs_gridv),}
{\color{Sepia}\textit{\ \ \# same as }\texttt{obs_bsint}}\\
\texttt{
\mbox{}~~~~agama.bsplineMatrix(obs_degree, obs_gridv, mod_degree, mod_gridv_scaled))\\
mat_kin~~~= mat_los.reshape(num_orbits * num_aper, num_mod_bs). \textbackslash\\
\mbox{}~~~~dot(conv.T). \textbackslash\\
\mbox{}~~~~reshape(num_orbits, num_aper * num_obs_bs) * Upsilon**-0.5\\
cons_kin~~= hist_val.reshape(num_aper * num_obs_bs)\\
err_kin~~~= hist_err.reshape(num_aper * num_obs_bs)}\\[2mm]
If using GH moments, the routine \ttt{ghMoments} transforms the matrix of B-spline amplitudes of orbit LOSVDs into the matrix of GH moments computed in the observed GH basis defined by parameter \texttt{ghbasis} (different in each aperture). This matrix has moments $h_0..h_6$ (in this example \texttt{num_obs_gh=6}), but we don't use $h_0$ because it is not available observationally (instead, we constrain the aperture mass), so we reshape the matrix and eliminate the 0th column:\\[2mm]
\texttt{
mat_kin~~~= agama.ghMoments(degree=mod_degree, gridv=mod_gridv_scaled, \\
\mbox{}~~~~matrix=mat_los, ghorder=num_obs_gh, ghbasis=gh_params). \textbackslash\\
\mbox{}~~~~reshape(num_orbits, num_aper,~ num_obs_gh+1)[:,:,1:].~~ \textbackslash\\
\mbox{}~~~~reshape(num_orbits, num_aper * num_obs_gh) * Upsilon**-0.5\\
cons_kin~~= ghm_val.reshape(num_aper * num_obs_gh)\\
err_kin~~~= ghm_err.reshape(num_aper * num_obs_gh)}\\[2mm]
Finally, solve the quadratic optimization problem to determine orbit weights. In this example, we require that the 3d density and 2d aperture mass constraints be satisfied exactly (set an infinite penalty for them), while the observational kinematic constraints should be satisfied as closely as possible, with the penalty proportional to the inverse squared observational error:\\[2mm]
\texttt{weights = agama.solveOpt(\\
\mbox{}~~~~matrix=[mat_den.T, mat_sur.T, mat_kin.T],}
\textit{\color{Sepia} \ \ \# list of matrices}\\
\texttt{\mbox{}~~~~rhs=[cons_den, cons_sur, cons_kin],~~~~~~\mbox{}}
\textit{\color{Sepia} \ \ \# list of RHS vectors (constraints)}\\
\texttt{\mbox{}~~~~rpenq=[cons_den*numpy.inf, cons_sur*numpy.inf, 2*err_kin**-2])}
\textit{\color{Sepia}\# penalties}\\[3mm]
Many of the above steps are generally applicable to all observationally-constrained Schwarzschild models. The relevant routines and classes reside in the submodule \texttt{agama.schwarzlib}, and the user- and model-specific tasks may be kept in a separate script adapted from \texttt{example_forstand.py}.

\newpage
\addcontentsline{toc}{section}{References}

%%%%%%%%%%%%%%%%%%%%%%%%%%%

\end{document}